\documentclass{article}
\usepackage{fix-cm}

\usepackage[fontsize=11.4pt]{fontsize}
\usepackage{multirow}
\usepackage[utf8]{inputenc}
\usepackage{amsfonts,amsthm,amsmath,amssymb,bm}
\usepackage{hyperref}
\usepackage{cite}
\hypersetup{hidelinks}
\usepackage[margin = 2.5cm]{geometry}

\usepackage[section]{placeins}
\usepackage{xcolor}
\usepackage[mathscr]{euscript}
\usepackage{bm}
\usepackage{physics}
\usepackage{tikz}
\usepackage{mathtools}
\usetikzlibrary{decorations.pathmorphing}
\usetikzlibrary{arrows.meta}
\usepackage{dsfont}
\usepackage{simplewick}
\usepackage{comment}
\usepackage{float}
\usepackage{hyperref}

\graphicspath{{too_many_diagrams/}}

\renewcommand{\tilde}{\widetilde}
\renewcommand{\d}{\mathrm{d}}

\numberwithin{equation}{section}

\def\Tr{\text{Tr}}

\begin{document}
\thispagestyle{empty}

\vspace*{2.5cm}
\begin{center}

     {\LARGE \bf A backreacted bouncing geodesic in 2D}

\vspace{0.5in}

{\bf Gauri Batra$^1$, Adam Levine$^{2}$ \& Stephen H. Shenker$^1$}

\vspace{0.3in}

$^1$Leinweber Institute for Theoretical Physics, Stanford University, Stanford, CA 94305, USA

\vspace{0.1in}

$^2$Center for Theoretical Physics – a Leinweber Institute, Massachusetts Institute of Technology,
Cambridge, MA 02139, USA 
                
    \vspace{0.5in}

    \vspace{0.5in}
    
\end{center}

\vspace{0.5in}

\begin{abstract}
    In holography, analytically continued finite temperature correlation functions $G(t)$ exhibit signatures of the black hole singularity. These take the form of singularities at certain values of $t$ whose bulk origin is a null geodesic worldline ``bouncing off" the black hole singularity. We study certain quantum corrections to this effect in 2D dilaton gravity that are captured by gravitational backreaction of the worldline onto the black hole geometry. These corrections smooth out the bouncing geodesic singularities. By following the backreacted geodesic saddle in a steepest descent analysis we find instead that  the black hole singularity physics gets encoded in a subtle way involving a Stokes phenomenon near the coincident point singularity of $G(t)$ at $t=0$ (and its thermal image). We point out a parallel between the structure of Stokes lines uncovered in this analysis and the structure present in the analysis of forbidden singularities in large $c$ 2D conformal blocks. Finally we make some preliminary remarks about the significance of quantum corrections in these models beyond backreaction.

\end{abstract}

\vspace{1in}

\pagebreak

\setcounter{tocdepth}{3}

\tableofcontents

\newpage

\section{Introduction and summary}
\label{sec:intro}
\subsection{Introduction}
The black hole singularity is a place where the predictive power of physics fails. An important and longstanding  question is how a consistent theory of quantum gravity resolves this singularity. Gauge/gravity duality, in particular the AdS/CFT correspondence, is our best understood theory of quantum gravity and so it is natural to ask about the fate of this singularity in this context.   This question is difficult to address in general because of the difficulty in expressing bulk observables behind the black hole horizon in terms of boundary observables.\footnote{A sampling of work exploring the black hole singularity in gauge/gravity duality includes \cite{Horowitz:2003he,Silverstein:2005qf,Horowitz:2006mr,Horowitz:2009wm,Itzhaki:2018rld,Frenkel:2020ysx,Grinberg:2020fdj,Leutheusser:2021qhd,Emparan:2021ewh,Bousso:2022tdb,deBoer:2022zps,DeClerck:2023fax,Shahbazi-Moghaddam:2025fxi,Chakravarty:2025ncy,Blacker:2026poc}.} Answering the most interesting questions, like what is the experience of an infalling observer near the singularity, is still beyond our reach.      
Nevertheless it might still be possible to exploit the special properties of certain states to find a simple signature of some aspects of the  singularity in boundary quantities. In this paper we discuss one such signature, the ``bouncing geodesic" singularity.

\begin{figure}[t]
    \centering
    \includegraphics[width=0.4\linewidth]{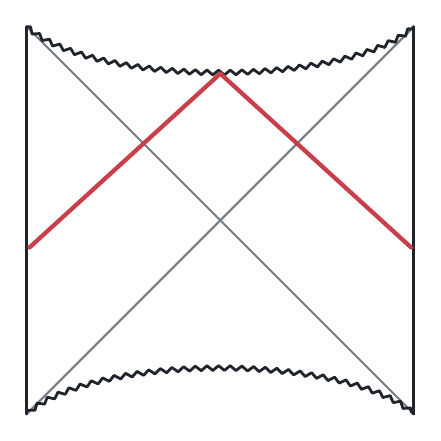}
    \caption{Penrose diagram for the dilaton gravity black hole solution discussed in Section \ref{sec:geometry}. The bouncing geodesic in red is responsible for the $t_c$ singularity in the correlator $G(t)$. A similar diagram describes the $D>3$ AdS-Schwarzschild geometry.}
    \label{fig:penrose}
\end{figure}

In \cite{Maldacena:2001kr} Maldacena pointed out that two-sided correlators in the thermofield double state should be sensitive to the region behind the horizon of the corresponding AdS-Schwarzschild eternal black hole.    For large mass bulk particles, correlators are determined by geodesics which lie on the appropriate steepest descent contour.   For BTZ black holes in AdS$_3$ the authors of \cite{Louko:2000tp,Kraus:2002iv} showed that the geodesics for two-sided correlators pass through the region behind the horizon, tying the correlators directly to geometry behind the horizon.  Building on these
insights the authors of \cite{Fidkowski:2003nf} identified a nearly null spacelike geodesic in the AdS-Schwarzschild geometry for bulk spacetime dimensions $D > 3$ connecting the Left (L) and Right (R) boundaries that ``bounces" off the black hole singularity -- see Figure \ref{fig:penrose}. The properties of such a geodesic should yield some information about the spacetime region near the singularity.

In this paper we will consider the thermal Wightman correlator 
\begin{align}\label{eq:wightcorr}
G(t)=\frac{1}{Z}\Tr\left[e^{-\beta H} O\left(t\right) O(0)\right] .
\end{align}
(We have suppressed the spatial coordinates in \eqref{eq:wightcorr} and for now discuss momentum space correlators at zero spatial momentum.) 
If the nearly  null bouncing geodesic contributed to the thermal Wightman correlator, continued by $\frac{i\beta}{2}$ to be  two-sided,
\begin{align}\label{eq:lrcorr}
    G_{LR}(t)=\frac{1}{Z}\Tr\left[e^{-\beta H} O\left(t-\frac{i\beta}{2}\right) O(0)\right] = \langle O_L(t) O_R(0) \rangle_{\beta},
\end{align}
 it would produce a lightcone singularity of the form 
 \begin{align}\label{eq:lrsing}
 \langle O_L(t) O_R(0) \rangle_{\beta} \sim 1/(t-{t_b})^{2 \mu} .
 \end{align}
 Here ${t_b}$ is the boundary time at which the geodesic becomes null.\footnote{In momentum space, the integral over real space implies that the exponent in \eqref{eq:lrsing} is $2\Delta-d+1$, where $\Delta$ is the conformal dimension of $O$. Here we used the fact that this power is approximately given by the mass $\mu$ for bulk particles of large mass.}  It is not difficult to show that such a singularity is impossible for $t$ in the physical thermal strip $0 \ge \Im t \ge -\beta$ , and in fact this geodesic is not on the steepest descent contour there \cite{Fidkowski:2003nf}.   The question then is to what quantity, if any, this geodesic does contribute.   After a significant effort \cite{Festuccia:2005pi,Festuccia2007BlackHoleSingularities,Dodelson:2023vrw,Ceplak:2024bja,Dodelson:2025jff,Afkhami-Jeddi:2025wra,AliAhmad:2026wem}, a number of quantities have been identified.  In early work Festuccia and Liu \cite{Festuccia:2005pi} argued that the bouncing geodesic determines the behavior of the Fourier transform of \eqref{eq:wightcorr} at large imaginary frequency.    
 
\begin{figure}[t]
    \centering
    \includegraphics[width=0.7\linewidth]{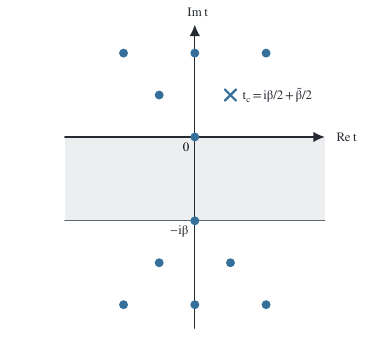}
    \caption{In the $G_N\to 0$ limit, the Wightman function $G(t)$ exhibits a lattice of singularities (see \cite{Dodelson:2025jff}) in its analytic continuation outside the physical strip (pictured in grey). The function $G(t)$ may be written as a sum of integrals over the energy $\tilde\omega$; each singularity is generated by a saddle of any one of these integrals running off to infinite $|\tilde\omega|$.}
    \label{fig:lattice}
\end{figure}

 Here we will focus on a more recent identification: the bouncing geodesic causes singularities in \eqref{eq:wightcorr} directly in the time domain when time is continued {\it outside} of the physical strip $0 \ge \Im t \ge - \beta$ \cite{Dodelson:2025jff,Afkhami-Jeddi:2025wra,AliAhmad:2026wem}.\footnote{As a reminder, the momentum space Wightman function $G(t)$ is periodic in imaginary time, but its analytic continuation outside the physical strip is not.  The full $G(t)$ is only analytic in strips of width $i \beta$.  In position space $G(x, t)$ has a periodic analytic continuation containing standard lightcone singularities with branch points emanating from them \cite{Fulling:1987otn}.  In this case the bouncing geodesic singularities are found by analytically continuing through these branch cuts. Geodesic analysis indicates that we have to set $x=0$ to see these singularities (see also \cite{Grozdanov:2026cut}), although bulk cone-like singularities exist in the stress tensor sector for $x\neq 0$ \cite{Araya:2026shz}. We thank Mukund Rangamani for discussions on this topic.}  These singularities are illustrated in Figure \ref{fig:lattice}.   
 They occur at a lattice of points determined by the basis vectors $t=t_c = i \beta/2 + {\tilde \beta}/2$ and $t=i\beta$, where $t_b$ in \eqref{eq:lrsing} is given by ${\tilde \beta}/2$.\footnote{Singularities at general lattice points correspond to geodesics that bounce multiples times off the singularity, or their thermal images \cite{Dodelson:2025jff,Afkhami-Jeddi:2025wra,AliAhmad:2026wem}.} We refer to these singularities as ``$t_c$" singularities or ``bouncing geodesic" singularities.\footnote{Additional work on these singularities includes \cite{Kaplan:2004qe,Hartnoll:2020fhc,Rodriguez-Gomez:2021pfh,Horowitz:2023ury,Giombi:2026kdz,Jia:2026ryl,Grozdanov:2026ktq,Barrat:2026jfg,Arnaudo:2026axe,Grozdanov:2026lnc,Hartnoll:2026vhu}.} 
 
 The Wightman function $G(t)$ for real $t$ is a one-sided correlator and is determined solely by bulk physics outside of the black hole horizon.   How can it contain any information about the singularity?    The reason is that the bulk geometry in this case, eternal AdS-Schwarzschild, is {\it analytic}.    If we know its behavior outside the horizon well enough we can analytically continue through the horizon to learn about the singularity. We also expect this analyticity to  hold for various kinds of corrections and so should be able to learn about them as well. 
 
 So far we have discussed the bulk in the limit where stringy and quantum corrections are absent. We would like to understand how such corrections affect the $t_c$ singularities. Stringy corrections are certainly interesting as tidal effects are expected to be large near the singularity \cite{Horowitz:1989bv,Horowitz:1990sr}. Whether or not they smooth out the $t_c$ singularity is an open question --- there are indications that both possibilities could be true. In certain simple large $N$ systems like the SYK model \cite{Dodelson:2024atp, Dodelson:2025jff,Buric:2026qsp} and few matrix quantum mechanics \cite{Dodelsonetal:WIP}, $t_c$-like singularities persist away from a gravitational regime.   This suggests that such singularities might be a generic feature of large $N$ (or classical) chaotic systems.  On the other hand, in the case of the $SL(2,\mathbb R)_k/U(1)$ black hole, corrections to the reflection coefficient due to non-perturbative in $\alpha'$ effects \cite{Itzhaki:2018rld} point to the smoothing out of the $t_c$-like singularity.\footnote{We thank Simon Caron-Huot and Douglas Stanford for pointing this out to us.  See \cite{Afkhami-Jeddi:2025wra} for the relation between the reflection coefficient and the $t_c$ singularity.} In any event, in this paper we focus on quantum corrections to the $t_c$ singularities.   We do so in a very simple context -- two-dimensional dilaton gravity -- where stringy degrees of freedom are absent.   
 
 Classical black holes in the  simplest holographic version of two dimensional dilaton gravity, JT gravity,\footnote{For reviews see \cite{Sarosi:2017ykf,Mertens:2022irh,Turiaci:2024cad}.} do not have curvature singularities. 
 But theories with a more complicated dilaton potential\footnote{For reviews see \cite{Grumiller:2002nm,Witten:2020wvy,Mertens:2022irh} and references therein.} can have black holes with such singularities and geodesics that bounce off them.  In particular the kinds of exponential dilaton potentials that correspond to conical defect insertions \cite{Maxfield:2020ale,Witten:2020wvy} can have these features \cite{Kruthoff:2024gxc}.     Coupling a heavy spectator matter field  to the metric allows us to study the $t_c$ singularity in its correlator $G(t)$.    
 
 In this paper we discuss a certain class of quantum corrections to $G(t)$ and their effect on its singularity structure. 
 In particular, we evaluate the backreaction of the black hole geometry due to the high energy flowing in the bouncing geodesic worldline.  Our central result is that this effect is sufficient to smooth out the sharp $t_c$ singularities, but results in a more subtle imprint of the black hole singularity at $t=0$ where the boundary operators coincide. In other words, the place where we may expect to find information about quantum gravity effects very close to the black hole singularity is not near $t=t_c$, but instead is the vicinity of $t=0$.  We now give an outline of our paper and a summary of our main results. 

 \subsection{Outline and summary}
 In {\bf Section \ref{sec:setup}}, we introduce the technology from the bulk gravitational theory needed to analyze $G(t)$. Using a boundary spectral decomposition, we write $G(t)$ as an integral over two energy variables $\bar E$ and $\tilde\omega$.  We then approximate the integrand using the bulk gravitational path integral with fixed energy boundary conditions evaluated semiclassically, following the JT analysis in \cite{Bah:2022uyz}. It is proportional to $e^{\mathcal{I}}$, where $\mathcal{I}$ is the action of a classical geometry formed by patching two black hole geometries with masses $2\bar E-G_N\tilde\omega$ and $2\bar E+G_N\tilde\omega$ across the worldline of the particle using Israel junction conditions.\footnote{Note that $G_N$ is dimensionless in 2D.} This particle carries stress-energy ${\tilde \omega}$, and the difference of black hole energies is due to its gravitational backreaction on the geometry. The rest of the section is devoted to computing and analyzing this action.   
 
 We focus on two examples of dilaton potential, a power law and an exponential potential.   The power law potential is simpler in some respects and unless further specified our results refer to it.  The exponential potential is discussed in detail in Appendix \ref{sec:exp_potential}.

 In {\bf Section \ref{sec:probe_singularities}} we show how the resulting integral over $\bar E$ and $\tilde\omega$ gives rise to the lattice of singularities in Figure \ref{fig:lattice} as we take $G_N\to 0$.  We sometimes refer to this as the ``probe" limit.   Every point in the complex $(\bar E,\tilde\omega)$ plane corresponds to a geometry with a geodesic. The $t_c$ singularity is caused by a saddle point of the integrand that approaches infinite $\tilde\omega$ and corresponds to a geodesic with turning point close to the singularity. We call this the ``bouncing geodesic saddle". We emphasize that this null geodesic is approached via a limit of complex geodesics \cite{AliAhmad:2026wem}.\footnote{This family differs from that considered in \cite{Fidkowski:2003nf, Festuccia2007BlackHoleSingularities}, which approach the null bouncing geodesic through real spacelike geodesics.}  The large $\tilde\omega$ part of the integral describing the correlator takes the form
 \begin{align}
 \label{eq:probecorr}
     G(t) \sim \int^\infty \d \tilde\omega\, \exp\left[-i\tilde\omega t + i \tilde\omega t_c +2\mu \log \tilde\omega \right]
     .
 \end{align}
 For large particle mass $\mu$ we can evaluate this integral by saddle point, giving a saddle point value $\tilde \omega_* = -\frac{ 2 i\mu}{t-t_c}$.  This yields the probe approximation $t_c$ singularity:
 \begin{align}
 \label{eq:probecorrt}
     G(t) \sim 1/(t-t_c)^{2 \mu} .
 \end{align}

 In {\bf Section \ref{sec:smoothing}} we study the smoothing of the $t_c$ singularity at finite $G_N$ in detail. In {\bf Subsection \ref{sec:first_correction}} we find that in the vicinity of $t_c$, the leading $G_N$ correction to the backreacted action is a term $\sim G_N {\tilde \omega}^2$.  The correlator  is now given by a Gaussian integral:
 \begin{align}
 \label{eq:correct}
     G(t)\sim \int^\infty \d \tilde\omega\, \exp\left[-i\tilde\omega t + i \tilde\omega t_c +2\mu \log \tilde\omega + G_N \tilde\omega^2 \gamma(\beta) \right].
 \end{align}
 The correction becomes significant when $t-t_c \sim \sqrt{G_N}$, smoothing out the power law $t_c$ singularity into a sharp Gaussian.\footnote{This kind of smoothing of the bouncing geodesic singularity has also been noted in \cite{Ruan:2026talk}.}
 This Gaussian smoothing mechanism for small $G_N$ is quite generic and applies to a large class of dilaton potentials. More generally, it smooths out many kinds of singularities in AdS/CFT. These include forbidden singularities in 2D CFTs \cite{Fitzpatrick:2016ive}, bulk cone singularities \cite{Batra:WIP}, and the fake disk singularity in SYK \cite{StanfordTangPrivateCommunication}.   Stringy effects lead to such a Gaussian smoothing for the bulk point singularity \cite{Maldacena:2015iua}. In the case of bulk cone singularities, the physical mechanism behind the smoothing is the backreaction of the null geodesic responsible for the singularity on the spacetime, which is similar to the mechanism we consider here. 

In {\bf Subsection {\ref{sec:stokes}}} we study what happens to the bouncing geodesic saddle at small but finite $G_N$ by studying the global structure in the complex $t$ plane of certain saddles contributing to the correlator.
 At finite $G_N$ at $t=t_c$, the bouncing geodesic saddle represents a geodesic whose turning point is at a proper distance of $\mathcal{O}(G_N)$ from the singularity. By studying it and the other saddles that control the behavior of \eqref{eq:correct} near $t_c$ as a function of complex $t$ we are able to track the bouncing geodesic's fate as we leave the region near $t_c$ and move towards $t=0$. In particular we numerically locate the structure of Stokes lines of $G(t)$ involving these saddles.\footnote{These are not the only saddles that contribute to $G(t)$ but they are the ones that describe the leading behavior near $t_c$. They are the only ones that are relevant for studying the bouncing geodesic saddle  and for describing the singular behavior near $t=0$ or $-i\beta$. Other saddles are discussed in Appendix \ref{sec:additionalsaddles}.}

 The answer is path-dependent---along one path this saddle represents a geodesic whose turning point moves closer to the singularity as $t$ approaches zero. Along another path it turns into a saddle for which the turning point is far from the singularity, which then participates in a Stokes phenomenon with the near-singularity turning point saddle at a Stokes line emanating from $t=0$. Upon crossing this line, the latter saddle subsequently starts to contribute. 
 
 Regardless of path,  it is the coincident point singularity whose vicinity encodes information about the region very close to the black hole singularity at finite $G_N$, not $t=t_c$.\footnote{The correlator is large in a region extending from $t=t_c$ to $t=0$, where the bouncing geodesic saddle is the dominant saddle.} 
The basic reason that the signature of the singularity is encoded near $t=0$ as opposed to $t=t_c$ is the fact that the inverse temperature of the highly energetic black hole on one side of the Israel junction goes to zero, driving the effective $t_c$ to $0$.

 In {\bf Section \ref{sec:OPE_sing}} we analyze the behavior near $t=0$ analytically. In addition to the Stokes phenomenon we also find that backreaction turns the point $t=0$ into an essential singularity. This essential singularity is locally described by the integral 
\begin{align}
\label{eq:intro_Gt}
    G(t)\sim \int^\infty \d s_+\, \exp\left[-\frac{it}{2 G_N}s_+^2 -\frac{\pi}{G_N}s_++2\mu \log(s_+^2)\right].
\end{align}
The bouncing geodesic saddle is described by a saddle point of this integral which is picked up across a Stokes line emanating from  $t=0$. In this regime, the saddle gives a contribution to the correlator  that behaves like\footnote{Note that saddle points representing high energy geodesics with turning point near the boundary are also picked up in a similar way, and have similar $t\to 0$ behavior. For this reason, this essential singularity appears in pure JT gravity as well. The exact numerical coefficients and perturbative corrections around the bouncing geodesic saddle distinguish it from the saddle with turning point near the boundary.} 
\begin{align}
\label{eq:exppottime}
    e^{\mathcal{J}_{\tilde\Lambda}}\sim t^{-4\mu}\exp\left[-\frac{i\pi^2}{2 G_N t}\right].
\end{align}
This small $t$ behavior changes for the exponential potential.  In this case  $-\frac{2i}{G_N} \log^2s_+$ is added to the argument of the exponential in \eqref{eq:intro_Gt} giving rise to an additional $\frac{2i}{G_N} \log^2(1/t)$ term in the exponential of \eqref{eq:exppottime} (see Appendix \ref{sec:exp_potential}).

 In {\bf Section \ref{sec:windings}} we discuss the fate of the full set of the singularities on the lattice displayed in Figure \ref{fig:lattice}. We find a similar structure of Stokes lines dictating the dynamics of the saddles causing these other singularities. An infinite number of Stokes lines extend to the point $t=0$, across which an infinite number of saddles are picked up. These saddles add up to form a natural boundary of analyticity that is a curve extending from $t=0$.\footnote{Assuming the validity of the saddle point approximation.} This natural boundary is not a distinctive signal of the curvature singularity --- in Appendix \ref{sec:JT} we show that it exists in pure JT gravity as well.

 In {\bf Section \ref{sec:cft}} we find a similar structure of Stokes lines in the context of the resolution of  forbidden singularities in 2D CFTs. As a concrete example we study the Virasoro vacuum block contributing to a four-point function involving two heavy degenerate operators with large negative scaling dimension and two light operators. We find that the Stokes phenomenon of Sections \ref{sec:stokes} and \ref{sec:OPE_sing} also occurs near the heavy-light OPE point in this case. This shows that this phenomenon is consistent with the presence of a convergent OPE-like expansion near the coincident points, and hence is a possibility for higher-dimensional CFTs as well. It points to the possible universality of this phenomenon --- we outline some future directions for probing it further in conformal field theories.

 In {\bf Section \ref{sec:diagrams}} we give a preliminary discussion of loop corrections in $G_N$ that go beyond the effects of backreaction.  This tentative analysis gives evidence that the $G_N\tilde\omega^2$ correction in \eqref{eq:correct} is robust to loop corrections. For power law potentials these loop corrections may go out of control near $t=0$. For the exponential dilaton potential, we give evidence that these corrections remain small and the behavior near $t=0$ outlined above seems to be robust as well. We plan to study this more carefully in future work \cite{BatraShenker:WIP}.

\section{Backreacted geometries from junction conditions}
\label{sec:setup}

\subsection{The correlator as an integral over energies}
\label{sec:integral}

To set up the bulk calculation let us begin by rewriting the boundary correlator, inserting complete sets of energy eigenstates to write the thermal two-point function as an integral over energies. Focusing on the thermal Wightman function, we have
\begin{align}
\nonumber
    G(t)&=\frac{1}{Z}\Tr [ e^{-\beta H} O(t) O(0)]\\
    &=\frac{1}{Z}\sum_{m,n}\bra{E_n} e^{-\beta H} e^{iHt} O(0) e^{-iHt} \ket{E_m}\bra{E_m}O(0)\ket{E_n} \nonumber\\
    &=\frac{1}{Z}\sum_{m,n} e^{-\beta E_n} e^{i(E_n-E_m)t}|\bra{E_n}O(0)\ket{E_m}|^2\nonumber\\
    &\approx \frac{1}{Z} \int_0^\infty \d E_n \int_0^\infty \d E_m \,\rho(E_n)\rho(E_m) e^{-\beta E_n} e^{i(E_n-E_m)t}|\bra{E_n}O(0)\ket{E_m}|^2.
    \label{eq:int_correlator}
\end{align}
In the last line, we have replaced the discrete spectrum by a smooth density of
states $\rho(E)$.
Note that because of the exponential suppression of high energy states, $G(t)$ is analytic in the strip defined by $-\beta<\Im(t)<0$.  Here the integral over energies converges absolutely. In
particular, no singularities can occur in its interior.

\begin{figure}[t]
    \centering
    \includegraphics[width=0.47\linewidth]{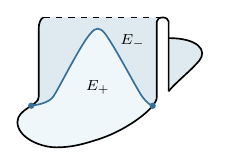}
    \includegraphics[width=0.4\linewidth]{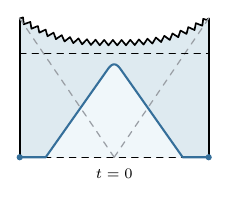}
    \caption{We patch two black hole geometries with masses $E_+$ and $E_-$ across a geodesic of a given energy. 
    The geodesic in this figure is a complex geodesic obtained by the continuation described in Section \ref{sec:geodesics} (see also Figures \ref{fig:complex_geodesic_contour} and \ref{fig:P_path}). It starts at a complex time on the boundary (see left figure), and moves on a geometry involving forward and backward Lorentzian evolution. As its energy increases, the turning point of this geodesic gets closer to the black hole singularity from a complex direction in the radial coordinate $\phi$.  On the right, we picture the projection of this geodesic onto the real $\phi$ Lorentzian slice. The black dashed line on the top indicates the location of the fold from forward to backward time evolution. The singularity lies beyond this dashed line.}
    \label{fig:complex_geodesic}
\end{figure}

To incorporate the backreaction of the bouncing geodesic, we approximate the product of the density of states factors and the matrix elements squared by the gravitational path integral with fixed energy boundary conditions (following the treatment of JT in \cite{Bah:2022uyz}). We treat the matter field sourced by $O$ as a spectator field, so no matter loops are included.  We restrict ourselves to the simplest, disk, topology. In the semiclassical limit, this path integral decomposes
into a sum over saddles, so that this product is given by the sum over multiple exponentials:
\begin{align}\label{eq:saddle_sum}
\rho(E_n)\rho(E_m)|\bra{E_n}O(0)\ket{E_m}|^2\approx \sum_i e^{\mathcal{I}_i}.
\end{align}
The argument of each exponential is the on-shell action of the $i$th geometry, with each geometry obtained by gluing two black hole regions across the geodesic of a particle, see Figure \ref{fig:complex_geodesic}.
The two different black holes have (scaled) masses $E_+=2 G_N  E_m$ and $E_-=2 G_N E_n$. The distinct saddles then correspond to
different complex geodesic contours (we will soon justify why this is the case). Some of these geodesics have turning point close to the boundary, and some of these geodesics have turning point close to the black hole singularity.   Where necessary we will take the mass of the particle $\mu$ large to justify the geodesic approximation.\footnote{We discuss finite $\mu$ corrections in Appendix \ref{sec:feynman_rules}.}

For the analysis in this paper, it is convenient to write the above integral in terms of the average energy $\bar E$ and energy difference $\tilde\omega$ measured at the boundary:
\begin{align}
\label{eq:correlator_omega}
    G(t)=\frac{1}{Z}\sum_i\int_0^\infty \d \bar E \int_{-2\bar E/G_N}^{2\bar E/G_N} \d \tilde\omega \, e^{-\frac{\beta \bar E}{G_N}-i\tilde\omega\left(t+\frac{i\beta}{2}\right)+\mathcal{I}_i},
\end{align}
where we defined
\begin{align}
\label{eq:energy_def}
    \frac{\bar E}{G_N}=\frac{E_m+E_n}{2},\quad \tilde\omega=E_m-E_n
\end{align}
and suppressed overall constant factors. In the $G_N\to 0$ limit, we can perform the $\bar E$ integral by saddle point. This picks out a certain $\bar E_0$ corresponding to the mass of the black hole. The individual saddles above then correspond to different complex geodesics with energy\footnote{Why the energy is given by $\tilde\omega$ will become clearer in Section \ref{sec:backreaction}.} proportional to $\tilde\omega$ moving on this fixed black hole geometry. The geodesic saddles included in the sum in \eqref{eq:saddle_sum} are determined by matching to the expression for $G(t)$ for $G_N\to 0$ obtained via the wave equation; we explain this further in Appendix \ref{sec:qnms}.

\subsection{Description of the geometry}
\label{sec:geometry}
Consider a two-dimensional dilaton gravity theory coupled to a massive point particle \cite{Grumiller:2002nm,Witten:2020wvy,Mertens:2022irh}. In Euclidean signature the action is
\begin{align}\label{eq:action}
\mathcal{I} = -I_E = \frac{1}{2 G_N} \int \d^2 x \sqrt{g} \left( \phi R + U(\phi) \right)-\mu \int \sqrt{h}\, \d s+\frac{\phi_b}{G_N}\int \d v\,\sqrt{h} K . 
\end{align}
Here $\mu$ is the mass of the bulk field dual to the boundary operator of interest, $\phi_b$ is the boundary value of the dilaton, and the last term is the usual Gibbons-Hawking boundary term at the cutoff surface. The integral over $s$ is taken along the particle worldline, while the integral over $v$ is taken along the asymptotic boundary. We have suppressed any counterterms. We first review the classical black hole solutions in the absence of the particle. Varying with respect to the dilaton gives
\begin{align}\label{eq:dilaton_eom}
    R=-U'(\phi),
\end{align}
while the metric variation gives
\begin{align}
\label{eq:metric_eom}
    \nabla_\mu\nabla_\nu \phi - g_{\mu\nu} \nabla^2\phi +\frac{1}{2} g_{\mu\nu} U(\phi)=0.
\end{align}
In dilaton gauge, where the dilaton itself is used as the radial coordinate, these equations admit the one-parameter family of solutions (in Euclidean signature)
\begin{align}\label{eq:metric}
ds^2 = f(\phi) \d\tau^2 + \frac{\d \phi^2}{f(\phi)},\quad f(\phi)=W(\phi)-E,\quad W'(\phi)=U(\phi),
\end{align}
where an additive constant in $W$ has been absorbed into $E$, which labels the black hole mass. We will focus on asymptotically AdS solutions, for which $W(\phi)\sim \phi^2$ at large positive $\phi$. 

For suitable choices of the dilaton potential, the Lorentzian continuation of these geometries contains a curvature singularity \cite{Kruthoff:2024gxc}. This is immediate from \eqref{eq:dilaton_eom}: singularities occur when $U'(\phi)$ diverges at some point in the continuation. In this work we mainly focus on the two potentials
\begin{align}\label{eq:potentials}
    U(\phi)=2\phi+\frac{2}{\phi^3},\quad U(\phi)=2\phi+e^{-\phi}.
\end{align}
They exhibit curvature singularities at $\phi=0$ and $\phi=-\infty$, respectively.  The first potential gives rise to a geometry that is the 2D reduction of the higher-dimensional AdS$_5$-Schwarzschild black hole \cite{Fidkowski:2003nf,Festuccia:2005pi,AliAhmad:2026wem}, while the second potential admits a simple defect expansion \cite{Maxfield:2020ale,Witten:2020wvy}. As we will see in Section \ref{sec:diagrams}, there are indications that the loop corrections for the second potential are better controlled.
The main body of the paper focuses on the $1/\phi^3$ potential, while the exponential potential is treated in Appendix \ref{sec:exp_potential}. For the first potential, $W(\phi)=\phi^2-\phi^{-2}$, and the horizon is the positive root of $W(\phi)=E$:
\begin{align}\label{eq:phi_h}
    \phi_h(E)=\frac{\sqrt{E+\sqrt{4+E^2}}}{\sqrt{2}},
\end{align}
with the inverse temperature of the black hole given by
\begin{align}\label{eq:beta}
    \beta(E)=4\pi \phi_h'(E).
\end{align}
The equation $W(\phi)=E$ also has other solutions. Let us define the quantities $\tilde\phi_h(E)$ and $\tilde\beta(E)$ as
\begin{align}\label{eq:tphi_h}
    \tilde\phi_h(E)&=\frac{\sqrt{E-\sqrt{4+E^2}}}{\sqrt{2}},\\
    \label{eq:tbeta}
     \tilde\beta(E)&=4\pi i\,\tilde\phi_h'(E).
\end{align}
The four solutions of $W(\phi)=E$ for this potential are then given by
$\pm \phi_h(E)$ and $\pm \tilde\phi_h(E)$.

An important feature of these models is the ``bending in'' of the singularity in the Penrose diagram, as illustrated in Figure \ref{fig:penrose}. This bending is controlled by the nonzero quantity $\tilde\beta$ \cite{Fidkowski:2003nf,Festuccia:2005pi}. As emphasized in \cite{Fidkowski:2003nf}, this geometry admits a geodesic saddle that approaches the singularity and bounces before returning to the asymptotic region. The rest of this section first describes the allowed geodesics on the above geometry, then constructs the backreacted versions of these saddles.

\subsection{Complex solutions to the geodesic equation}
\label{sec:geodesics}
In the $G_N\to 0$ limit, the matter particle coupled to the metric and dilaton acts as a probe.
Its equations of motion dictate that it follow geodesics on the above background geometry. In Euclidean signature, the relevant geodesics are spacelike, and are solutions to the geodesic equation given by
\begin{align}\label{eq:gdeq}
\left(\frac{\d\phi}{\d u}\right)^2-W(\phi) + E + P^2 = 0.
\end{align}
Here $u$ is the proper length along the geodesic, and $P$ is the conserved energy conjugate to Euclidean time translations:
\begin{align}\label{eq:energy}
    P=f(\phi) \frac{\d\tau}{\d u}.
\end{align}
As mentioned in Section \ref{sec:integral}, to satisfy the fixed energy boundary conditions described below \eqref{eq:int_correlator}, we have to sum over geodesics with a given energy in the above background. Each geodesic is parameterized by some trajectory $(\phi(u),\tau(u))$, where $\tau(u)$ is determined by \eqref{eq:energy} once the endpoint of the geodesic in $\phi$ is specified. Using \eqref{eq:gdeq} and \eqref{eq:energy}, the coordinate time elapsed along a geodesic that starts and ends at the Euclidean boundary is given by
\begin{align}\label{eq:int_tau}
\Delta \tau = 2 \int^{\infty}_{\phi_t} \d\phi \frac{P}{(W(\phi) -E)\sqrt{W(\phi) - E - P^2}},
\end{align}
where $\phi_t$ is the turning point of the geodesic. 

\begin{figure}[t]
    \centering
    \includegraphics[width=\linewidth]{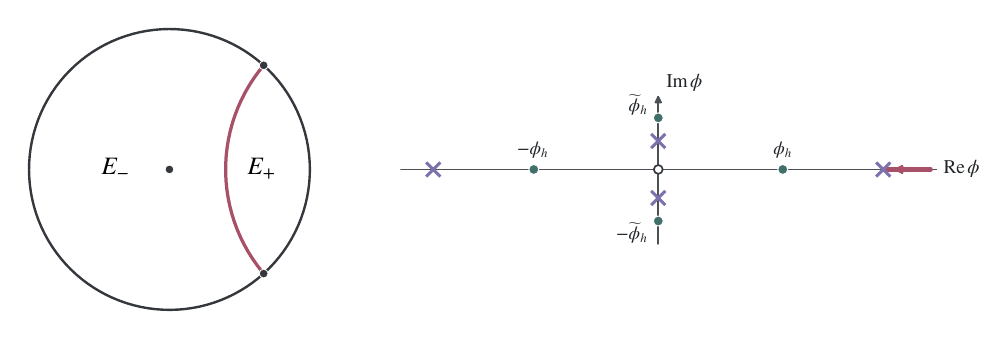}
    \caption{The left figure pictures the geodesic that gives the coincident point singularity at large energies. To compute its backreaction in Section \ref{sec:backreaction}, we patch two black hole solutions with different masses $E_+$ and $E_-$ across this geodesic.
    On the right we depict the contour in the complex $\phi$ plane that corresponds to this geodesic, along with the turning points (labelled by crosses). 
}
    \label{fig:OPE}
\end{figure}

Different geodesics with the same energy $P$ are then defined by their turning point $\phi_t$ along with the contour from the boundary to $\phi_t$. Turning points occur when $\d\phi/d u=0$, or equivalently when the square root in \eqref{eq:int_tau} vanishes:
\begin{align} \label{eq:tp_eq}
    W(\phi_t)=E+P^2.
\end{align}
This equation has multiple solutions. The first set of solutions has turning points that occur in the asymptotic AdS region, where $W(\phi)\approx \phi^2$ and hence $\phi_t\approx \pm \sqrt{E+P^2}$. We call these the ``OPE geodesics". As we increase $|P|$, the resulting geodesic approaches the AdS boundary, and the geodesic probes the coincident-point region of the boundary correlator. One of these solutions is displayed in Figure \ref{fig:OPE}.

\begin{figure}[t]
    \centering
    \includegraphics[width=0.6\linewidth]{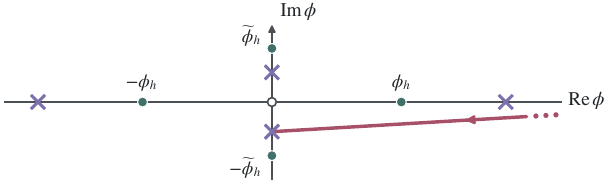}
    \caption{The contour in the complex $\phi$ plane that corresponds to the geodesic in Figure \ref{fig:complex_geodesic}. The turning points are labelled by crosses, with the turning point of this geodesic approaching the singularity at $\phi=0$ from a complex direction.}
    \label{fig:complex_geodesic_contour}
\end{figure}

Another set of solutions has turning points that occur in the region near the singularity, where $W(\phi)$ goes to $-\infty$. For the $1/\phi^3$ potential in \eqref{eq:potentials}, these turning points are given by 
\begin{align}
    \phi_t\approx \pm i/\sqrt{E+P^2}
\end{align}
at large $|P|$.
They approach the black hole singularity from a (generically) complex direction as $|P|\to\infty$, limiting to a null bouncing geodesic at infinite energy (see Figure \ref{fig:complex_geodesic} and the contour in Figure \ref{fig:complex_geodesic_contour}). In Section \ref{sec:probe_singularities} we will see that such a saddle gives rise to the singularity at $t=t_c$. Both kinds of saddles above are responsible for the full lattice of singularities of the thermal two-point function in Figure \ref{fig:lattice}.

\begin{figure}[t]
    \centering
    \includegraphics[width=0.5\linewidth]{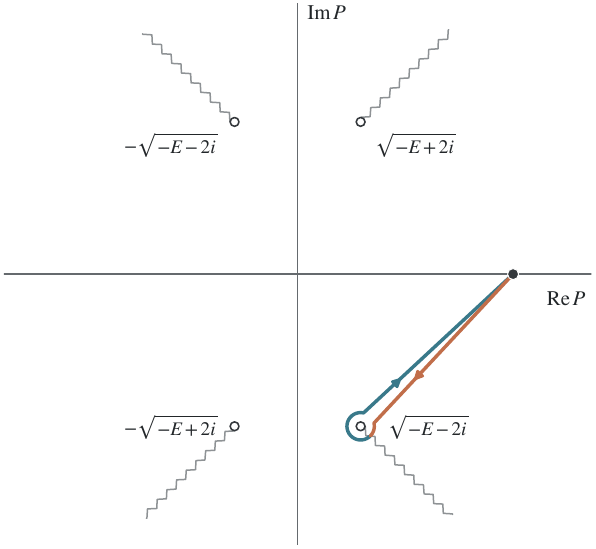}
    \caption{A path in the complex $P$ plane that analytically continues the OPE geodesic to a complex geodesic with a turning point near the singularity. The path changing color implies that it has changed sheets.}
    \label{fig:P_path}
\end{figure}

The relation between these two kinds of solutions is particularly transparent in the complex $P$ plane. The solution $\phi_t(P)$ to the turning point equation \eqref{eq:tp_eq} is branched, and one solution may be analytically continued to another by winding $P$ around the branch points of $\phi_t(P)$ (this was pointed out in \cite{AliAhmad:2026wem}). For the $1/\phi^3$ potential, the turning points are given by
\begin{align}\label{eq:tp}
    \phi_{t,\pm}^2=\frac{E+P^2\pm\sqrt{(E+P^2)^2+4}}{2},
\end{align}
with the relevant branch points being
\begin{align}\label{eq:probe_branches}
    P_*=\pm\sqrt{-E\pm 2i}
\end{align}
Analytically continuing $P$ around these points switches the sign of the square root in \eqref{eq:tp} and exchanges the near-boundary and near-singularity turning points. See Figure \ref{fig:P_path} for a path in $P$ that performs this exchange.

Because the different geodesics are related by analytic continuation in $P$, their coordinate times $\Delta\tau$ are also related by monodromies around these branch points.
To see this, consider the OPE geodesic of Figure \ref{fig:OPE} at a given energy $P$, and follow the path in Figure \ref{fig:P_path} by winding $P$ clockwise around $P_*=\sqrt{-E-2i}$ before returning to the same value of $P$. As discussed above, the result is a complex geodesic with turning point close to the singularity. Based on the monodromy of $\Delta\tau(P)$ under this path, the coordinate times of the OPE geodesic and this complex geodesic satisfy the relation
\begin{align}\label{eq:time_reln}
   \Delta\tau_{(1,0)}= \Delta\tau_{(0,0)}-\frac{\beta}{2}+\frac{i\tilde\beta}{2}.
\end{align}
Here we have labelled the geodesic with turning point close to the singularity as $(1,0)$ and the OPE geodesic as $(0,0)$, 
based on the singularities they correspond to in the lattice in Figure \ref{fig:lattice}; a geodesic labelled $(n,m)$ with $n,m\geq 0$ gives rise to a singularity at $t=n(i\beta/2+\tilde\beta/2)+m(i\beta/2-\tilde\beta/2)$. Each geodesic labelled by $(n,m)$ corresponds to a classical solution contributing to the sum over gravitational saddles that specifies the integrand in \eqref{eq:saddle_sum} --- we explain this further in Section \ref{sec:probe_singularities}.

 To see the above monodromy,  note that the integral for $\Delta \tau$ in \eqref{eq:int_tau} evaluates to
\begin{align}
\label{eq:tau_formula}
\Delta \tau
=
&-\frac{i\beta}{4\pi}
\log
\left[
\frac{
\left(P+\sqrt{-E+2i}\right)
\left(P-\sqrt{-E-2i}\right)
}{
\left(P+\sqrt{-E-2i}\right)
\left(P-\sqrt{-E+2i}\right)
}
\right] \nonumber\\
&+
\frac{\tilde\beta}{4\pi}
\log
\left[
\frac{
\left(P+\sqrt{-E-2i}\right)
\left(P+\sqrt{-E+2i}\right)
}{
\left(P-\sqrt{-E-2i}\right)
\left(P-\sqrt{-E+2i}\right)
}
\right],
\end{align}
where we used the relations in \eqref{eq:beta} and \eqref{eq:tbeta}.
The monodromies of the two logarithms above then give \eqref{eq:time_reln}. We can similarly find a relation between the regulated lengths of these geodesics:
\begin{align}
    \label{eq:L_reln}\mathcal{L}_{(1,0)}=\mathcal{L}_{(0,0)}+i\pi,
\end{align}
where
\begin{align}
    \mathcal{L}
&\equiv
\lim_{\phi_c\to\infty}
\left[
2\int_{\phi_t}^{\phi_c}
\frac{\d\phi}
{\sqrt{W(\phi)-E-P^2}}
-2\log(2\phi_c)
\right]
\nonumber\\
&=
-\frac{1}{2}
\log\left[
\left(P-\sqrt{-E-2i}\right)
\left(P+\sqrt{-E-2i}\right)
\left(P-\sqrt{-E+2i}\right)
\left(P+\sqrt{-E+2i}\right)
\right].
\label{eq:regulated_length}
\end{align}
More generally, winding $P$ multiple times around the four branch points generates an infinite family of complex geodesic saddles, with different coordinate times and geodesic lengths. In the complex $\phi$ plane, these saddles correspond to different integration contours connecting the boundary to the appropriate turning point.

Note that the bouncing geodesic family we have identified in this section (based on the analysis in \cite{AliAhmad:2026wem}) differs from the one considered in \cite{Fidkowski:2003nf,Festuccia:2005pi}, which is a limit of real spacelike geodesics. The latter appears as a solution to the geodesic equation when $P$ is the energy conjugate to Lorentzian time translations, and hence imaginary. At large imaginary $P$, \eqref{eq:tau_formula} reproduces the coordinate time elapsed along those real spacelike geodesics \cite{Fidkowski:2003nf,Festuccia:2005pi}.

\subsection{Backreaction of the particle and the on-shell action}
\label{sec:backreaction}
We want to find how a particle following the geodesics described above backreacts on the geometry. The relevant combined solutions to the particle-metric-dilaton system may be found using the Israel junction conditions (see \cite{Bah:2022uyz} for a similar discussion in pure JT gravity). Given a solution to the geodesic equations of motion, the backreacted geometry is given by patching together two black hole metrics on the two sides of the geodesic, see Figure \ref{fig:complex_geodesic}.

Let the masses of the two geometries we patch together be $E_+$ and $E_-$.  First note that the induced metric computed from either side of the geodesic should be the same, so that we may use the same proper length variable for both geometries. This implies 
\begin{align}
\left(\frac{\d\phi}{\d u}\right)^2 - W(\phi) + E_{\pm} + P_{\pm}^2 = 0,\ \ P_{\pm} = (W(\phi) - E_{\pm}) \frac{d\tau_{\pm}}{\d u}.
\end{align}
Note that the conserved energy of the geodesic depends on whether we use the $E_+$ or $E_-$ geometry. The first junction condition says that the dilaton is continuous across the particle, implying the relation
\begin{align}
\label{eq:continuity}
    E_+ + P_+^2 = E_- + P_-^2.
\end{align}
The second junction condition determines the discontinuity of the normal derivative of the dilaton across the particle worldline. This discontinuity is obtained from the metric equation of motion \eqref{eq:metric_eom} with the particle source included, which gives
\begin{align}
\Delta \partial_n \phi = \mu G_N. 
\end{align}
Now, the unit normal to the geodesic on either side is given by $(\dot \phi,-\dot\tau_\pm)$, where the derivative is with respect to the proper length $u$. Using the fact that $\partial_\mu\phi=(0,1)$, this implies that the derivative $\partial_n\phi$ on either side equals $n^\mu_\pm \partial_\mu\phi=-\dot\tau_\pm f_\pm=-P_\pm$. We therefore obtain the relation
\begin{align}\label{eq:Pdiff}
    P_+ - P_- = -\mu G_N.
\end{align}
Using \eqref{eq:continuity}, we can then solve for $P_+$ and $P_-$ in terms of $E_+$, $E_-$ and $\mu$, fully determining the backreaction. The physical picture of this backreaction is that the particle injects energy into the black hole and hence changes its mass.

To use the geometry derived here for computing the thermal two-point function as in \eqref{eq:saddle_sum}, we need to compute its on-shell action. We will first compute the on-shell action for the geodesic on the Euclidean disk pictured in Figure \ref{fig:OPE}.\footnote{This is along the lines of the computation in Appendix B of \cite{Bah:2022uyz}.} The on-shell action for solutions involving the other geodesics described in Section \ref{sec:geodesics} can then be obtained via appropriate analytic continuation, as we will explain. 

Using the dilaton equation of motion $R=-U'(\phi)$, the on-shell action in \eqref{eq:action} becomes 
\begin{align}
\mathcal{I}_{(0,0)} = \frac{1}{2G_N} \int \d\tau\, \d\phi \left( - \phi U'(\phi) + U(\phi) \right) + \frac{\phi_b}{G_N} \int \d v \sqrt{h} K +E_{\text{phys}}\int \d \tau - \mu \int  \sqrt{h}\, \d s,
\end{align}
where the $\tau$ and $\phi$ integrals are over the geometry pictured in Figure \ref{fig:OPE}. Here, we also included an extra $E_{\text{phys}}\int \d \tau$ term to impose fixed energy boundary conditions. $E_{\text{phys}}$ is the physical energy measured at the boundary, which is either $E_+/(2G_N)$ or $E_-/(2G_N)$ depending on the segment of the boundary. We can then integrate by parts to reduce the first integral above to pure boundary terms, since
\begin{align}
    - \phi U'(\phi) + U(\phi)=\frac{\d}{\d\phi}\left(2W(\phi)-2E-\phi U(\phi)\right).
\end{align}
Here the $2E$ term is an integration constant that simplifies the answer.
The first contribution to the boundary terms comes from the horizon $\phi=\phi_h(E_-)$, which lies in the $E_-$ part of the geometry (see Figure \ref{fig:OPE}). Using smoothness at the horizon to determine the integral $\Delta_{\rm{o}}$ over the coordinate $\tau$ around this point, it contributes
\begin{align}
    \frac{1}{2G_N} \Delta_{\rm{o}} \phi_h U(\phi_h)=\frac{2 \pi \phi_h (E_-)}{G_N}.
\end{align}
The next contribution is from the integrals along the AdS boundary. 
Along the lines of \cite{Bah:2022uyz}, these terms all add up to zero on adding appropriate counterterms at the boundary. In the remaining expression, only integrals along the particle worldline remain:
\begin{align}
\mathcal{I}_{(0,0)}  = &+\frac{1}{2G_N} \int d\tau_+ \phi(\tau) U(\phi(\tau)) - \frac{1}{2G_N} \int d\tau_- \phi(\tau) U(\phi(\tau)) + \frac{2\pi \phi_h(E_-)}{G_N} \nonumber \\
& -\frac{1}{G_N}\left( \int d\tau_+ (W(\phi(\tau_+)) - E_+) - \int d\tau_- (W(\phi(\tau_-)) - E_-)\right)- \mu \int \sqrt{h}.
\end{align}
Here, the $\tau$ integrals are along the particle worldline, with the $+$ integral on the side of the worldline with black hole mass $E_+$, and the $-$ integral on the side of the worldline with black hole mass $E_-$. Using \eqref{eq:Pdiff} and \eqref{eq:energy}, we can show that the terms in the second line of the above equation cancel one another. 
We can then use the geodesic equation to replace the $\tau$ integral by a $\phi$ integral in the first line, leading to
\begin{align}\label{eq:S_on_shell}
\mathcal{I}_{(0,0)}&=\frac{2\pi \phi_h(E_-)}{G_N} + (I_+^{\rm{OPE}} - I_-^{\rm{OPE}}),\nonumber\\
I_\pm^{\rm{OPE}}&=\frac{1}{G_N} \int_{\phi_t}^{\infty} \frac{d\phi \, \phi\, U(\phi) P_\pm}{(W(\phi) - E_\pm) \sqrt{W(\phi) - E_\pm -P_\pm^2}}. 
\end{align}
This expression for the on-shell action is the main result of this section. For the $1/\phi^3$ potential, the integral above is analytically doable, see Appendix \ref{sec:phi2_formulas}. 

So far our discussion revolved around the OPE geodesic on the Euclidean disk. Our main interest is in the complex geodesic that approaches the singularity at high energies, see Figure \ref{fig:complex_geodesic}. To obtain its on-shell action $\mathcal{I}_{(1,0)}$, we can consider the expression in \eqref{eq:S_on_shell} with the contour for the integrals going along this complex geodesic. However, there is an easier rewriting of $\mathcal{I}_{(1,0)}$. As discussed at the end of Section \ref{sec:geodesics}, this complex geodesic is related to the OPE geodesic by appropriately continuing $P$. This means that to obtain the on-shell action for the geometry in Figure \ref{fig:complex_geodesic}, we may start with $\mathcal{I}_{(0,0)}$ and analytically continue in both $P_+$ and $P_-$ using the path depicted in Figure \ref{fig:P_path}. Similar to the relation for the coordinate time \eqref{eq:time_reln} between the two geodesics, we obtain the relation
\begin{align}\label{eq:S_reln}
    \mathcal{I}_{(1,0)}=\mathcal{I}_{(0,0)}+\frac{\pi \phi_h(E_-)}{G_N} + \frac{\pi \tilde \phi_h(E_-)}{G_N} -\frac{\pi \phi_h(E_+)}{G_N} - \frac{\pi \tilde \phi_h(E_+)}{G_N}-i\pi\mu
\end{align}
between their on-shell actions. This may be seen from the monodromies of the logarithms in the expression for $\mathcal{I}_{(0,0)}$ in Appendix \ref{sec:phi2_formulas}.  This is a simple expression for $\mathcal{I}_{(1,0)}$ in terms of integrals over the contour corresponding to the OPE geodesic.

\begin{figure}[t]
    \centering
    \includegraphics[width=0.7\linewidth]{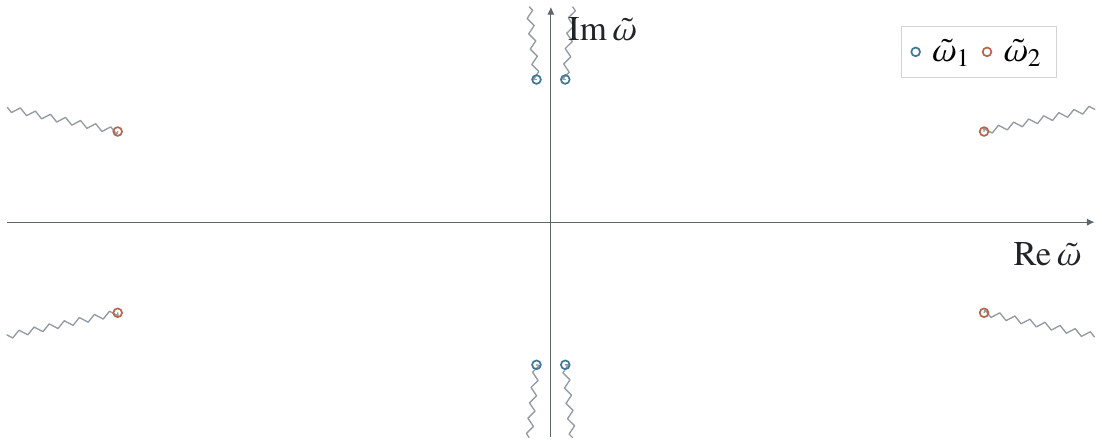}
    \caption{A depiction of the branch points and the conventional choice of branch cuts of the on-shell action $\mathcal I_{(0,0)}$ for the $1/\phi^3$ potential. The $\tilde\omega_2$ branch points move to infinity in the $G_N\to 0$ limit. Large $\tilde\omega$ corresponds to a geometry with a geodesic whose turning point is close to the boundary.
    Winding clockwise around the lower right $\tilde\omega_1$ branch point and moving to large $\tilde\omega$ takes the turning point of this geodesic close to the singularity.}
    \label{fig:branch_cuts}
\end{figure}

In the complex $(\tilde\omega,\bar E)$ variables, this operation of continuing both $P_+$ and $P_-$ is equivalent to continuing $\tilde\omega$ to another sheet. This is because of the relation
\begin{align}
    P_\pm=\frac{\tilde\omega}{\mu}\mp \frac{\mu G_N}{2}.
\end{align}
Since both $|P_+|$ and $|P_-|$ approach $|\tilde\omega/\mu|$ as $G_N\to 0$, this implies that $\tilde\omega$ may be interpreted as the energy per unit mass of the particle.
It is also helpful to understand the analytic structure of the action in the complex $\tilde\omega$ plane. The action has two sets of four branch points (or more precisely, branch curves that depend on $\bar E$), which are depicted in Figure \ref{fig:branch_cuts}. The first set includes logarithmic branch points given by
\begin{align}
\label{eq:bp_1}
    \tilde\omega_1= \pm  \mu\sqrt{-2\bar E-\frac{(\mu G_N)^2}{4}\pm 2i}.
\end{align}
These also survive in the $G_N\to 0$ limit.
The second set includes square root branch points given by
\begin{align}
\label{eq:bp_2}
    \tilde\omega_2=\frac{1}{G_N}\left(\pm 2i\pm 2\bar E\right).
\end{align}
These get pushed off to infinite $\tilde\omega$ on taking $G_N\to 0$. Winding around the $\tilde\omega_2$ branch points exchanges $\phi_h$ with $\tilde\phi_h$.

In this backreacted solution, let us emphasize that as the energy difference $\tilde\omega$ between the two states we are patching together goes to infinity, the turning point of the complex geodesic interpolating between them still goes to the black hole singularity. This may be seen by writing $E_\pm$ and $P_\pm$ in terms of $\bar E$ and $\tilde\omega$. The turning point is given by $W(\phi_t)=E_\pm+P_\pm^2=2\bar E
    +\frac{\tilde\omega^2}{\mu^2}
    +\frac{\mu^2G_N^2}{4}$. For large $\tilde\omega$, the turning point $\phi_t$ approaches $\phi=0$, behaving as
    \begin{align}
        \phi_t\sim \pm\frac{i\mu}{\tilde\omega},\quad  \tilde\omega\to\infty
    \end{align} 
even at finite $G_N$.

\section{Bouncing singularities in the probe limit}
\label{sec:probe_singularities}

In this section we will show how the expression for the on-shell action derived in the previous section leads to singularities in the correlator $G(t)$ in the $G_N\to 0$ or probe limit. The analytic continuation of $G(t)$ exhibits a lattice of singularities (see Figure \ref{fig:lattice} and \cite{Dodelson:2025jff}), with each singularity arising from a geodesic either approaching the singularity or approaching the boundary as it approaches infinite energy. Each geodesic is represented by a contour in the complex $\phi$ plane.

Before taking the $G_N\to 0$ limit, it is useful to integrate by parts and write the on-shell action for a general geodesic as 
\begin{align}
\mathcal{I}
&=
\frac{2\pi \phi_h(E_-)}{G_N}
+
2\mu \nonumber\\
&
+\frac{2}{G_N}
\int_{\phi_t}^{\infty} d\phi
\left[
\tan^{-1}\!\left(
\sqrt{\frac{P_+^2}{W(\phi)-E_+ - P_+^2}}
\right)
-
\tan^{-1}\!\left(
\sqrt{\frac{P_-^2}{W(\phi)-E_- - P_-^2}}
\right)
\right].
\end{align}
Here, the geodesic is specified by the contour from its turning point $\phi_t$ to $\infty$.
Using the relations \eqref{eq:continuity}, \eqref{eq:Pdiff} and \eqref{eq:energy_def} to write all variables in terms of $\bar E$ and $\tilde\omega$, expanding for small $G_N$, and renaming $2\bar E$ to $E$ inside the integrals, we then find
\begin{align}\label{eq:S_probe}
\mathcal{I}
&\approx
\frac{2\pi \phi_h(2\bar E)}{G_N}
-2\pi\phi_h'(2\bar E)\tilde\omega
-
2\mu
\int_{\phi_t}^{\infty}
\frac{d\phi}{\sqrt{W(\phi)-E-P^2}}
\nonumber\\
&\quad
+
2\tilde\omega
\int_{\phi_t}^{\infty}
d\phi\,
\frac{P}
{\left(W(\phi)-E\right)
\sqrt{W(\phi)-E-P^2}}
+
2\mu .
\end{align}
Here, we also used $E_-=2\bar E-G_N\tilde\omega$ to expand the first term. Note $\tilde\omega$ is held fixed as we take $G_N$ to zero.
Now, the first integral above is the expression for half the geodesic length, and the second integral is the expression for half of the coordinate time elapsed along the geodesic \eqref{eq:int_tau}, so that the on-shell action above may be rewritten as 
\begin{align}
\label{eq:S_probe_2}
\mathcal{I}
&\approx
\frac{2\pi \phi_h(2\bar E)}{G_N}
-2\pi\phi_h'(2\bar E)\tilde\omega
-
\mu
\ell(\tilde\omega, \bar E)
+
\tilde\omega \tau(\tilde\omega,\bar E)
+
2\mu .
\end{align}
Here the subtraction involved to obtain a finite geodesic length is implicit.
The next step is to understand the resulting singularities in the correlator $G(t)$, using the expression in terms of the integrals over $\bar E$ and $\tilde\omega$ in \eqref{eq:correlator_omega}:
\begin{align}
     G(t)&=\frac{1}{Z}\sum_i\int_0^\infty \d \bar E \int_{-2\bar E/G_N}^{2\bar E/G_N} \d \tilde\omega \, e^{-\frac{\beta \bar E}{G_N}-i\tilde\omega\left(t+\frac{i\beta}{2}\right)+\mathcal{I}_{\text{on-shell},i}}.
\end{align}
To leading order in $G_N$, the saddle point of the $\bar E$ integral tells us
\begin{align}\label{eq:beta_barE}
    \beta=\frac{\pi \sqrt{\bar E_0+\sqrt{1+\bar E_0^2}}}{\sqrt{1+\bar E_0^2}}=4\pi \phi_h'(2\bar E_0),
\end{align}
where $\bar E_0$ is the saddle point value of $\bar E$.
This is the relation between the mass and the temperature of this black hole, see \eqref{eq:beta}. Plugging in this saddle point\footnote{Note this saddle point is stable for $\bar E_0>1/\sqrt 3$.}, we obtain the expression for $G(t)$ as an integral over $\tilde\omega$:
\begin{align}
    G(t)
    &\propto
    \int_{-\infty}^{\infty}\d\tilde\omega\,
    e^{\mathcal J_0(\tilde\omega,t)}
    +\sum_i\int^{\infty}\d\tilde\omega\,
    e^{\mathcal J_i(\tilde\omega,t)}
    \nonumber
    +\sum_j\int_{-\infty}\d\tilde\omega\,
    e^{\mathcal J_j(\tilde\omega,t)}
    \nonumber\\
    &=G_{0,0}(t)
    +\sum_{\substack{n,m\geq0\\n+m\geq1}}
    \bigl[G_{n,m}(t)+G_{-n,-m}(t)\bigr],
    \label{eq:sum2}\\
    \mathcal J_i(\tilde\omega,t)
    &\equiv -i\tilde\omega t
    -\mu\,\ell_i(\tilde\omega,\bar E_0)
    +\tilde\omega\,\tau_i(\tilde\omega,\bar E_0).
    \nonumber
\end{align}
Here the range of integration over $\tilde\omega$  approaches  $(-\infty,\infty)$ for $G_N\to0$, and we suppress constant prefactors. We have split the sum over saddles into separate sums over $i$ and $j$; these separate the saddles contributing to the positive and negative $\tilde\omega$ parts of the contour. We also then rewrote each term in the sum over $i$ as $G_{n,m}(t)$, and each term in the sum over $j$ as $G_{-n,-m}(t)$.
As explained in Section \ref{sec:integral} and Appendix \ref{sec:qnms}, this is the sum over gravitational saddles consistent with the fixed energy boundary conditions. In particular, the gravitational saddles included above are determined by matching to the expression for $G(t)$ for $G_N\to 0$ obtained via the wave equation.  As then discussed in Section \ref{sec:geodesics}, each saddle corresponds to a geometry with a different geodesic, which may be obtained by starting with the OPE geodesic on the Euclidean disk and winding around branch points of $\tau$ and $\mathcal L$ in the complex $P$ plane. 

In the probe limit, each term gives rise to a singularity in the complex $t$ plane. These singularities come from the respective $\tilde\omega\to\infty$ and $\tilde\omega \to -\infty$ regions of the integrals above. Note that the singularity caused by the $\tilde\omega\to\infty$ region of the winding $G_{n,m}(t)$ is located at
\begin{align}
\label{eq:tnm}
    t_{n,m}=n\left(\frac{i\beta}{2}+\frac{\tilde\beta}{2}\right)+m\left(\frac{i\beta}{2}-\frac{\tilde\beta}{2}\right).
\end{align}
Each such singularity is accompanied by its KMS image from the large $-\tilde\omega$ region of $G_{-n,-m}(t)$, which in the complex $t$ plane is located at $t_{n,m}^{\rm image} =-t_{n,m}-i\beta$.
The sum over all windings then generates the lattice shown in Figure \ref{fig:lattice}.

First consider the $(n,m)=(0,0)$ winding in the sum above, which gives the coincident point singularities at $t=0$ and $t=-i\beta$. These arise from geodesics on the Euclidean disk approaching the boundary at infinite energy, as shown in Figure \ref{fig:OPE}. Let us focus on the singularity at $t=0$, which comes from the large positive $\tilde\omega$ region. In this regime, the regulated proper length $\ell(\tilde\omega,\bar E)$ of the OPE geodesic diverges logarithmically with $\tilde\omega$ as $-2\log\tilde\omega$. The origin of this divergence is the integral over the asymptotically AdS region, where $W(\phi)\sim \phi^2$. Moreover, as $\tilde\omega$ gets larger and the geodesic gets closer to the AdS boundary, the coordinate time $\tau(\tilde\omega,\bar E)$ goes to zero. Therefore for large positive $\tilde\omega$, the contribution from this saddle behaves as 
\begin{align}
\label{eq:00int}
    G_{0,0}(t)\sim \int^{\infty}\d\tilde\omega \exp\left[-i\tilde\omega t+
    2\mu
    \log\tilde\omega
    \right],
\end{align}
which diverges like $1/t^{2 \mu +1}$ as $t$ approaches $0$. We can also say this in the language of the saddle point approximation, which in this case is valid in the large $\mu$ regime.  The singularity at $t=0$ is caused by a saddle point ${\tilde\omega_*}$ of this integrand, given by
\begin{align}
    -i t + \frac{2\mu}{\tilde\omega_*}=0,
\end{align}
which approaches infinity as $t\to 0$.
Note that inserting ${\tilde\omega_*}$ into the action gives  $G_{0,0}(t) \sim 1/t^{2 \mu}$.  The $+1$ shift in the exponent is a one loop effect, which is small at large $\mu$.

Next consider the first winding, which is the contribution of the complex geodesic of Figure \ref{fig:complex_geodesic}. Using its probe limit relations \eqref{eq:time_reln} and \eqref{eq:L_reln} to the OPE geodesic, at large $\tilde\omega$ this saddle instead contributes as 
\begin{align}\label{eq:probe_bg}
    G_{1,0}(t)\sim \int^{\infty}\d\tilde\omega \exp\left[-i\tilde\omega t+i \tilde\omega t_c+
    2\mu
    \log\tilde\omega
    \right],\quad t_c=\frac{i\beta}{2}+\frac{\tilde\beta}{2},
\end{align}
which diverges when $t=t_c$. This is a bouncing geodesic singularity, caused by a saddle of the above integral going to infinite $\tilde\omega$ for $t\to t_c$:
\begin{align}
\label{eq:probe_tc_saddle}
    -i(t-t_c)+\frac{2\mu}{\tilde\omega_*}=0.
\end{align}
This gives the clean signature of the curvature singularity of the black hole in the probe limit.
In the rest of the paper, we discuss how this saddle point analysis is modified at small nonzero $G_N$.

Before proceeding, let us make some more comments about the integrals $G_{0,0}(t)$ and $G_{1,0}(t)$. The integral for $G_{0,0}(t)$ runs from $\tilde\omega=-\infty$ to  $\tilde\omega=\infty$, and is analytic in the physical strip $-\beta<\Im t<0$. The coincident point singularities appear at the edges of this strip. On the other hand, the integral for $G_{1,0}(t)$ only covers positive $\tilde \omega$, and so is analytic for $\Im t<\beta/2$, which extends outside the physical strip. The $t_c$ singularity appears at $\Im t=\beta/2$, at the edge of this region. Similarly, the singularities \eqref{eq:tnm} appear at the edges of the regions where $G_{n,m}(t)$ or $G_{-n,-m}(t)$ stop being analytic.

One could have asked if the analytic continuation of $G_{0,0}(t)$ outside of the physical strip has a singularity at $t=t_c$. This analytic continuation may be performed by rotating the defining contour into the complex $\tilde\omega$ direction by the appropriate amount. This contour has to smoothly pass through the $\tilde\omega_1$ branch cuts of Figure \ref{fig:branch_cuts}, and so goes onto another sheet. As we do so, the asymptotic behavior along this contour varies smoothly and is still given by \eqref{eq:00int}, so that  the large $\tilde\omega$ region again gives a singularity at $t=0$, not $t=t_c$.

\section{Smoothing out the $t_c$ singularity and Stokes phenomena}
\label{sec:smoothing}
The probe limit analysis of Section \ref{sec:probe_singularities} makes the origin of the $t_c$ singularity quite sharp: as $t$ approaches $t_c$, the saddle point of the integral over $\tilde\omega$ is driven to infinity. At finite $G_N$, however, the geodesic backreacts on the geometry, and the large energy region of the integral is no longer described by the probe bouncing geodesic saddle. In this case, we will see that the would-be $t_c$ singularity is smoothed out by the backreaction. The dominant contribution near $t_c$ instead comes from geodesics with energies of $\mathcal{O}(G_N^{-1/2})$, whose turning points remain at a proper distance of $\mathcal{O}(G_N)$ from the singularity.  We analyze the local behavior near $t_c$ by saddle point, and describe the local Stokes behavior of the participating saddles.

We then follow these saddles as we move $t$ away from this region, determining  their Stokes graph.   We find that their Stokes lines end at the coincident point singularities, where the bouncing geodesic actually reaches the black hole singularity. As we will  see in Section \ref{sec:OPE_sing}, the true singularities of each winding contribution $G_{n,m}(t)$ at finite $G_N$ only occur at the locations of these coincident point singularities. 

In this section we focus on the saddle points that dominate near $t_c$ and give the leading singular behavior at the coincident points. There are other saddle points that contribute to $G(t)$ but  are subleading in the regions of interest to us. In particular, they don't affect the behavior of the saddle point corresponding to the bouncing geodesic. Examples of such saddles are discussed in Appendix \ref{sec:additionalsaddles}.

\subsection{The first correction}\label{sec:first_correction}
As a first attempt to understand the region near $t=t_c$, we can include the leading  $\mathcal{O}(G_N)$ correction to the action in a small $G_N$ expansion. As we will see, in the appropriate range of $\tilde\omega$, this first correction adds a term of the form
\begin{align}\label{eq:delta_S}
    \Delta \mathcal{I}_{(1,0)} \sim G_N \tilde\omega^2
\end{align}
to the on-shell action in \eqref{eq:probe_bg}, and
smooths out the $t_c$ singularity.

With all the $G_N$'s made explicit, the expression we are expanding for small $G_N$ has the following form (using \eqref{eq:S_on_shell} and \eqref{eq:S_reln}):
\begin{align}
    \mathcal{I}_{(1,0)}&=\frac{3\pi\phi_h(E_-)}{G_N}+ \frac{\pi \tilde \phi_h(E_-)}{G_N} -\frac{\pi \phi_h(E_+)}{G_N} - \frac{\pi \tilde \phi_h(E_+)}{G_N} +(I_+^{\rm{OPE}}-I_-^{\rm{OPE}}), \nonumber\\
    \label{eq:Phi_Lambda}
    &=\frac{2\pi\phi_h(E_-)}{G_N}+\frac{1}{G_N} \left(f(G_N) - f(-G_N)\right) ,\quad 
    E_\pm = 2\bar E \pm \tilde\omega G_N.
\end{align}
This is the on-shell action for the complex geodesic of Figure \ref{fig:complex_geodesic}, which bounces off the singularity at high energies. 
(Here we omitted additive  constants.) The term $f(G_N) - f(-G_N)$ comes from the integrals along the two different sides of the complex particle worldline. This term is sent to the negative of itself when $G_N$ flips sign, since that operation exchanges $(E_+,P_+)$ with $(E_-,P_-)$. This implies that only odd powers appear in the expansion of $f(G_N) - f(-G_N)$, so that it only contributes to $\mathcal{I}_{(1,0)}$ at $\mathcal O(G_N^k)$, $k$ even,  after dividing by $G_N$.  For now we just keep the $\mathcal{O}(G_N)$ contribution discussed above. 

Only the entropy term contributes at order $G_N$. Taylor expanding this term gives the contribution $\frac{G_N^2}{G_N}2 \pi  \phi_h''(2\bar E) \frac{\tilde\omega^2}{2}$.  The on-shell action in \eqref{eq:probe_bg}, corrected to $\mathcal{O}(G_N)$, then becomes: 
\begin{align}\label{eq:corrected}
\mathcal{I}_{(1,0)}
\approx&\,
\frac{2\pi\phi_h(2\bar E)}{G_N}
-2\pi
\left[2 \phi_h'(2\bar E)+\tilde\phi_h'(2\bar E)\right]\tilde\omega
-
\mu
\ell_{\text{OPE}}(\tilde\omega,\bar E)
+
\tilde\omega \tau_{\text{OPE}}(\tilde\omega,\bar E)
+
\pi G_N \phi_h''(2\bar E) \tilde\omega^2.
\end{align}
Here we again dropped unimportant additive constants. Recall from Section \ref{sec:probe_singularities} that at large $\tilde\omega$, $\ell_{\rm{OPE}}$ has a logarithmic dependence on $\tilde\omega$ and  $\tau_{\rm{OPE}}$ goes to zero. Now, 

Let's first consider a microcanonical ensemble where the total energy $\bar E$ is fixed.  We integrate the exponential of the above action added to $-i\tilde\omega \left(t+\frac{i\beta}{2}\right)$  over $\tilde\omega$ to obtain the two-point function. The added Gaussian term $\sim \tilde\omega^2$  reflects the change in the mass and hence the entropy of the black hole because of the energy injected by the particle into the black hole.  This costs entropy and so the coefficient of $\omega^2$ is negative. 

For the canonical ensemble, the two-point function is obtained by integrating the exponential of the above action added to $-\frac{\beta \bar E}{G_N}-i\tilde\omega \left(t+\frac{i\beta}{2}\right)$ over both $\bar E$ and $\tilde\omega$.
We can first evaluate the $\bar E$ integral by saddle point. We find that the saddle point in $\bar E$ is shifted by an amount proportional to $G_N\tilde\omega$ compared to the probe limit saddle point $\bar E_0$:\footnote{The terms involving $\ell_{\mathrm{OPE}}$ and $\tau_{\mathrm{OPE}}$ also shift the $\bar E$ saddle above. However, their leading large $\tilde{\omega}$ dependence  is independent of $\bar E$ --- they correct \eqref{eq:barE_saddle} only at $\mathcal{O}(G_N/\tilde{\omega}^{2})$.}
\begin{align}\label{eq:barE_saddle}
    \bar E_*
    =
    \bar E_0
    +
    G_N\tilde\omega
    \left[
        1+
        \frac{\tilde\phi_h''(2\bar E_0)}
        {2\phi_h''(2\bar E_0)}
    \right].
\end{align}
Note that the coefficient of $G_N\tilde\omega$ above is complex, so that this saddle represents patching together two complex geometries across the particle worldline.

\begin{figure}[t]
    \centering
    \includegraphics[width=0.75\linewidth]{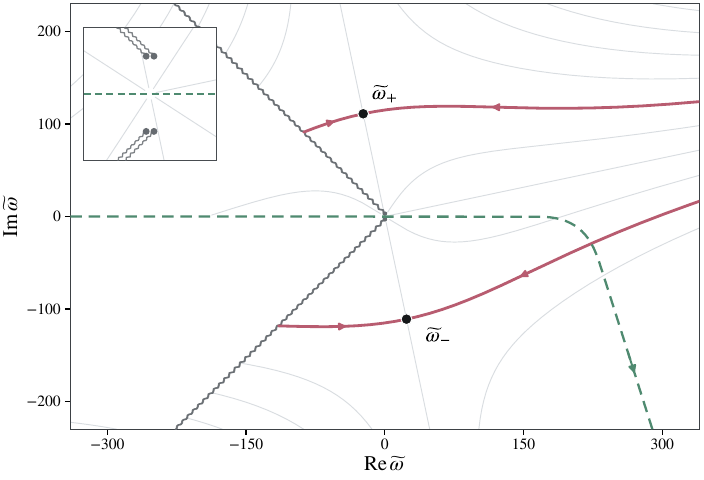}
    \caption{A depiction of two of the saddles of the Gaussian action in \eqref{eq:omega2_model} at $t=t_c$, for $\beta=2$, $G_N=10^{-3}$, and $\mu=1$. The zigzag lines represent branch cuts that arise from branch points in the region where $\tilde\omega$ is $\mathcal{O}$(1). These are the $\tilde\omega_1$ branch points of \eqref{eq:bp_1}, and are depicted in the top left box, which zooms in on the region around $\tilde\omega=0$. In this region, the action in \eqref{eq:omega2_model} receives corrections. In this figure, we have moved the branch cuts compared to Figure \ref{fig:branch_cuts} so that the saddles $\tilde\omega_\pm$ may be seen on the same sheet. 
    The defining contour (dashed green line) is the rotated contour of \eqref{eq:rotated}. The steepest ascent contours are pictured in red. The steepest ascent contour of the $\tilde\omega_+$ saddle does not intersect the defining contour. Instead, it hits a branch cut and goes onto another sheet. This means that it does not contribute as a saddle point. The steepest ascent contour of the $\tilde\omega_-$ saddle does intersect the defining contour, and hence it contributes to the integral. Note that the Gaussian description of \eqref{eq:omega2_model} is valid only on the part of this contour with $\Re \tilde\omega>0$; a different description comes into play for the $\Re \tilde\omega<0$ part of the contour. The $\tilde\omega_-$ saddle is the only saddle that contributes to this $\Re \tilde\omega>0$ part.}
    \label{fig:gaussian_saddles}
\end{figure}

\begin{figure}[t]
    \centering
    \includegraphics[width=0.9\linewidth]{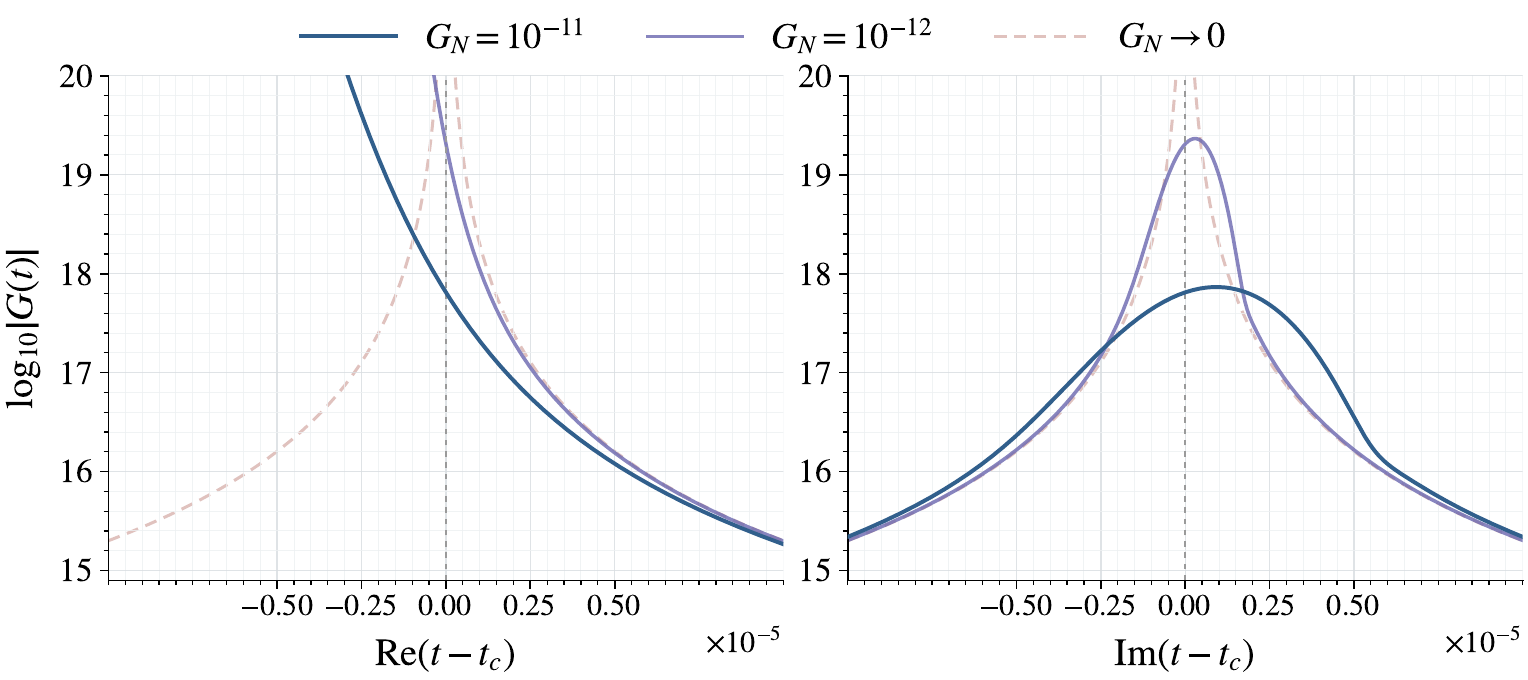}
    \caption{A plot of the correlator along two axes passing through $t=t_c$, for $\beta=2$, $\mu=1$ and different values of $G_N$. This is generated using the Gaussian model of \eqref{eq:omega2_model}, which may be evaluated exactly in terms of parabolic cylinder functions. This model is valid for $t-t_c\sim \mathcal{O}(\sqrt{G_N})$; for the sake of clarity we use it beyond its regime of validity in this figure. As $G_N$ goes to zero, we see the appearance of the $t_c$ singularity, with the peak blowing up as $G_N^{-\left(\mu+\frac{1}{2}\right)}$.}
    \label{fig:resolution}
\end{figure}

When $G_N$ is small, the saddle point approximation in $\bar E$ is accurate.
We can then evaluate $-\beta\bar E/G_N+\mathcal{I}_{(1,0)}$ on this saddle to obtain the model advertised at the start of this section in \eqref{eq:delta_S}.\footnote{Note we could also have expanded around the average energy as $\bar E=\bar E_*+\delta\bar E$ and performed the resulting Gaussian integral over the fluctuations $\delta\bar E$ instead of doing the integral by saddle point.} Indeed, at large positive $\tilde\omega$, using $\ell_{\text{OPE}}\to -2\log\tilde\omega$ and $\tilde\omega\tau_{\text{OPE}}\to 2\mu$ (an unimportant constant), the corrected integral for $G_{1,0}(t)$ becomes
\begin{align}
\label{eq:omega2_model}
    G_{1,0}(t)\sim \int^\infty \d \tilde\omega \exp\,\left[-i\tilde\omega t + i \tilde\omega t_c +2\mu \log \tilde\omega + G_N \tilde\omega^2 \gamma(\beta) \right].
\end{align}
Note that this form of the integral is only valid for the large $\tilde\omega$ part of the integration contour, and changes when we approach its large negative  $\tilde\omega$ part, where the contributions from the corresponding negative windings come into play. When following complex saddles of $G_{1,0}(t)$, its action is continued continuously through any branch cuts encountered. 
Here, the coefficient of the correction term $\gamma(\beta)$ is given by 
\begin{align}
\label{eq:omega_contour}
    \gamma(\beta)=\pi\phi_h''(2\bar E_0)
    \left[
        1-
        \left(
            2+
            \frac{\tilde\phi_h''(2\bar E_0)}
                 {\phi_h''(2\bar E_0)}
        \right)^2
    \right].
\end{align}
where the $\bar E_0 $ dependence on $\beta$ is given implicitly by \eqref{eq:beta}. 
We see that this coefficient gets corrected compared to its microcanonical value due to the fluctuations of the total energy. It is complex; for a sufficiently large mass black hole it has a \textit{positive} real part.\footnote{The real part of this coefficient is negative for $\bar E_0$ below an $\mathcal{O}(1)$ threshold. The analysis for that case may be done in a way that is similar to the positive real part case considered in this section.}  A rotation of the $\tilde\omega$ contour is required to make the integral convergent; in Section \ref{sec:OPE_sing} we will explain this contour rotation from first principles in the full two-dimensional integral.\footnote{Recall that in the $G_N\to 0$ limit, the $G_{1,0}(t)$ integral also becomes analytic in the additional range $0<\Im t<\beta/2$, with the $t_c$ singularity at the edge of this strip. Our analysis in this section indicates that the first $G_N$ correction is sensitive to the failure of analyticity outside the $-\beta<\Im t<0$ strip. A rotation of the contour is required to analytically continue the answer beyond this strip.} In particular, the $\tilde\omega$ contour is rotated so that it asymptotes in the direction
\begin{align}
\label{eq:rotated}
    \theta=-\frac{\left(\pi+\arg\gamma(\beta)\right)}{2}.
\end{align}
The resulting $\tilde\omega$ integral can be done exactly in terms of parabolic cylinder functions.
At large $\mu$ we can evaluate it using the saddle point approximation.
The saddle points of the integrand are located at 
\begin{align}\label{eq:quad_saddles}
    \tilde\omega_\pm=\frac{i(t-t_c)\pm\Delta}{4 G_N \gamma(\beta)}, \quad \Delta=\sqrt{-(t-t_c)^2-16\mu G_N \gamma(\beta)}.
\end{align}
There are infinite number of such saddle points that  lie on different sheets of 
the logarithmic branch point of the action in \eqref{eq:omega2_model}. They all satisfy the same saddle point equations. Two of them are pictured in Figure \ref{fig:gaussian_saddles}.
As explained in the caption of this figure, we see that only one saddle $\tilde\omega_-$ contributes when $t=t_c$. The integral is evaluated by deforming the defining contour into its steepest descent contour. This saddle is the backreacted version of the bouncing geodesic saddle. As we take $G_N$ to $0$, the integral is dominated by the part of its steepest descent contour that grows with $\tilde\omega$. This leads to the $t_c$ singularity. At finite $G_N$, the subsequent decay of the integrand for $\tilde\omega>\tilde\omega_-$ smooths out the singularity. For $t \sim t_c$ this turnover happens at $|\tilde \omega| \sim 1/\sqrt{G_N}$.  To illustrate this smoothing, in Figure \ref{fig:resolution} we plot the correlator in the vicinity of $t_c$ using the above Gaussian approximation.

For small $G_N$ and $t-t_c\sim \mathcal{O}(\sqrt{G_N})$, the Gaussian approximation to the on-shell action controls the full answer for $G_{1,0}(t)$. To see this, note that at large $\tilde\omega$, the corrections to the probe on-shell action organize themselves as 
\begin{align}\label{eq:exp_form}
    \Delta  \mathcal{I}_{(1,0)}^{\text{probe}}\sim G_N\tilde\omega^2 + G_N^2\tilde\omega^3 + G_N^3 \tilde\omega^4+\cdots.
\end{align}
In other words, any term in this expansion is of the form $G_N^K \tilde\omega^L$, with $L \leq K+1$.  It is clear that the expansion of the entropy term in $\mathcal{I}_{(1,0)}$ (which is proportional to $\phi_h(2\bar E - \tilde\omega G_N)$), takes this form. To see that the term involving the difference of integrals along the complex worldline ($\equiv I_+-I_-$) also takes this form, it is helpful to first write the difference as
\begin{align}
\label{eq:I_diff}
I_+-I_-\sim \tilde\omega
\int_{\phi_t}^{\infty} d\phi\,
\phi\, U(\phi)\,
\frac{
2-\frac{\mu^2\left(W(\phi)-2\bar E\right)}{\tilde\omega^2}
}{
\left[\left(W(\phi)-2\bar E\right)^2-G_N^2\tilde\omega^2\right]
\sqrt{
-1+\frac{\mu^2\left(W(\phi)-2\bar E\right)}{\tilde\omega^2}
-\frac{\mu^4G_N^2}{4\tilde\omega^2}
}
}.
\end{align}
The endpoint region of this integral, with $\phi\sim 1/\tilde\omega$, is suppressed in $\tilde\omega$.\footnote{This says something rather interesting, which is that the corrections to \eqref{eq:probe_bg} that smooth out the $t_c$ singularity do not come from the region near the turning point of the bouncing geodesic.}  The leading order in $\tilde\omega$ contribution comes from the $\mathcal{O}(1)$ region in $\phi$. The integral over this region has an expansion in $G_N^2\tilde\omega^2$ and terms suppressed at large $\tilde\omega$, with each term multiplied by $\tilde\omega$. To leading order in $\tilde\omega$, this gives the terms with even powers of $G_N$ in \eqref{eq:exp_form}. This may also be seen from the exact expression for the action in Appendix \ref{sec:phi2_formulas}.

The form of $\Delta  \mathcal{I}_{(1,0)}^{\text{probe}}$ in \eqref{eq:exp_form} then ensures the validity of the Gaussian approximation when $t-t_c$ is $\mathcal{O}(\sqrt{G_N})$. For this range in $t$, the dominant contribution to $G_{1,0}(t)$ comes from $\tilde\omega\sim \mathcal{O}(1/\sqrt{G_N})$. For this range of $\tilde\omega$ all the $\mathcal{O}(G_N^2)$ and higher corrections to the integrand in \eqref{eq:omega2_model} are suppressed compared to the leading $G_N\tilde\omega^2$ term. In fact, these corrections are suppressed relative to the Gaussian term for $\tilde\omega\sim \mathcal{O}(1/G_N^p)$ with $p<1$. For $\tilde\omega \sim \mathcal{O}(1/G_N)$, as we will see in Section \ref{sec:OPE_sing}, the integrand decays exponentially along the steepest descent contour controlling $G_{1,0}(t)$ above, and so doesn't affect the answer from the Gaussian approximation. 

As discussed in Section \ref{sec:intro}, the Gaussian correction $\propto G_N \tilde\omega^2$ is a universal correction that smooths out many kinds of forbidden singularities in AdS/CFT. Despite the universality of this Gaussian correction, the physics that contributes to the $G_N\tilde\omega^2$ is not universal. For the case of the bouncing geodesic singularity considered in this paper, this correction comes from two sources --- the change in the mass and hence entropy of the black hole, and fluctuations of the total energy in the canonical ensemble.\footnote{The Gaussian smoothing of the $t_c$ singularity from the latter source was also pointed out in \cite{Ruan:2026talk}.}
 In the microcanonical ensemble only the first effect leads to the $G_N\tilde\omega^2$ correction.\footnote{Another ensemble of interest is one where we fix one of the energies involved in the matrix elements we are computing. For instance, for the fixed $E_-\equiv 2\bar E-\tilde\omega$ ensemble, we only integrate over the $E_+$ variable. This ensemble is the one relevant for considering heavy-light-light-heavy correlators in higher dimensions. The Gaussian smoothing extends to this case as well, with the complex coefficient of $G_N\tilde\omega^2$ having a positive real part as in the canonical ensemble.} In general one might have expected the matrix elements to contribute to the $G_N\tilde\omega^2$ term as well --- it would be interesting to understand if they play a role in higher dimensions.

\subsection{Stokes lines near $t=t_c$}
\label{sec:neartcstokes}
Having described the smoothing of the $t_c$ singularity, in this subsection we will discuss some finer aspects of the analytic structure of the correlator near $t_c$. 

\subsubsection*{Collision points of the Gaussian action}

\begin{figure}[t]
    \centering
    \includegraphics[width=\linewidth]{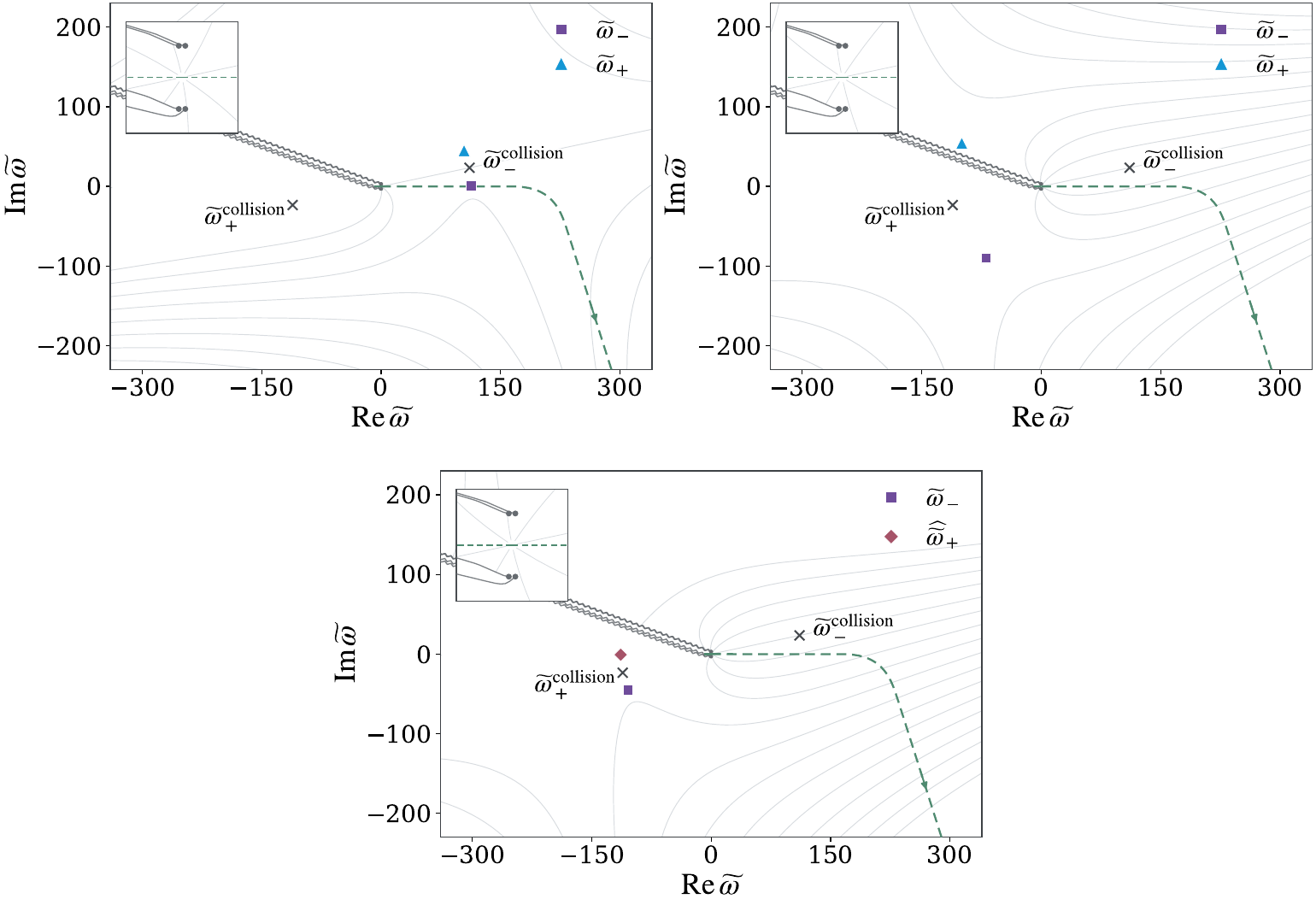}
    \caption{This figure depicts the collision between saddles that occurs at the two collision points $\tilde\omega_+^{\rm collision}$ and $\tilde\omega_-^{\rm collision}$. We have moved the branch cuts compared to Figure \ref{fig:gaussian_saddles} so that these two collision points may be pictured on the same sheet. For $t$ near $t_-^{\rm collision}$, the two saddles colliding at $\tilde\omega_-^{\rm collision}$ are shown in the top left figure. As we tune $t$ towards $t_+^{\rm collision}$, one of these saddles hits a branch cut and moves onto another sheet, which prevents it from meeting the other saddle on the same logarithmic sheet (depicted in the top right panel). For $t$ near $t_+^{\rm collision}$, the contributing saddle (given by $\tilde\omega_-$ in the above figure) then collides with a third saddle (denoted $\hat{\tilde\omega}_+$) that comes into the above picture from another sheet of the logarithmic Riemann surface. The two saddles are pictured close to the collision point in the bottom panel.}
    \label{fig:saddle_collision}
\end{figure}

First note that the Gaussian action in \eqref{eq:omega2_model} describing $G_{1,0}(t)$ develops two points at which two of its infinite number of  saddle points collide. These collisions happen when the argument of the square root $\Delta$ in \eqref{eq:quad_saddles} goes to zero, which happens for the values
\begin{align}
    t^{\text{collision}}_\pm = t_c \pm 4 i \sqrt{\mu G_N \gamma(\beta) }.
\end{align}
In the $\tilde\omega$ plane these correspond to the points
\begin{align}
    \tilde\omega_\pm^{\text{collision}}=\mp \sqrt{\frac{\mu}{G_N\gamma(\beta)}}.
\end{align}
To understand the saddles participating in the collision, it is useful to start at $t_-^{\rm collision}$ and track the colliding saddles. When $t$ equals this value, the picture in the $\tilde\omega$ plane involves the collision of the $\tilde\omega_+$ and $\tilde\omega_-$ saddles of Figure \ref{fig:gaussian_saddles} at   $\tilde\omega_-^{\text{collision}}$. As we then track $t$ to $t_+^{\text{collision}}$ in a straight line, these two saddle points travel to the point $\tilde\omega_+^{\text{collision}}$. However, at this endpoint, they lie on \textit{different} sheets of the logarithmic Riemann surface. Instead of colliding with each other, they collide with each other's respective image saddles. For example, the contributing saddle pictured in Figure \ref{fig:gaussian_saddles} interacts with two other saddles in this way. At $\tilde\omega_-^{\text{collision}}$ it collides with the $\tilde\omega_+$ saddle, and at $\tilde\omega_+^{\text{collision}}$ it collides with an image of that saddle on another logarithmic sheet. This image tracks the same path as that of the $\tilde\omega_+$ saddle on the other sheet, with its action differing by the monodromy of the $\sim\log\tilde\omega$ term. The entire collision process is pictured in Figure \ref{fig:saddle_collision}. In the rest of this section, we are interested in the dynamics of the three saddles identified above.

\subsubsection*{The local Airy behavior of the smoothed out $t_c$ singularity}

Note that the identified collision points are locally Airy-like. To see this, note that if we expand the action around them using $\tilde\omega=\tilde\omega_\pm^{\text{collision}}+x$ and $t=t^{\text{collision}}_\pm+\sqrt{G_N}\delta t$, the first and second derivatives vanish. Locally around these points, the argument of the exponential in $G_{1,0}(t)$ (referred to as ``the exponent" in the following) then behaves as
\begin{align}
&\mathcal{J}_{(1,0)}\!\left(
    \tilde\omega_\pm^{\text{collision}}+x,\,
    t_\pm^{\text{collision}}+\sqrt{G_N}\,\delta t
\right)
\nonumber\\
&\qquad=
\mathcal{I}_{(1,0)}^\pm
-i\tilde\omega_\pm^{\text{collision}}\sqrt{G_N}\,\delta t
-i\sqrt{G_N}\,\delta t\,x
+\frac{2\mu}{3(\tilde\omega_\pm^{\text{collision}})^3}\,x^3.
\end{align}
where we defined
\begin{align}
    \mathcal J_{(1,0)}(\tilde\omega,t)= \mathcal{I}_{(1,0)}(\tilde\omega,\bar E_*)-i\tilde\omega\left(t+i\frac{\beta}{2}\right)-\frac{\beta\bar E_*(\tilde\omega)}{G_N}.
\end{align}
On rescaling $x$, this gives the form of the action in the Airy function integral, multiplied by a term independent of $u$: 
\begin{align}
    Z=\text{independent of $u$}\cdot\int\d u \, e^{i\lambda_\pm(u^3/3-u)},\quad \lambda_\pm^2=\mp\frac{i\sqrt{\mu}(\delta t)^3}{2\left[\gamma(\beta)\right]^{3/2}}.
\end{align}
This observation will help us study the dynamics of the two saddle points of the quadratic action (and hence of the action corresponding to $G_{1,0}(t)$) involved in each collision. Note that the saddle point approximation here is controlled by $\lambda_\pm$ being large. This means that $|\delta t|$ (or the distance from any of the collision points divided by $\sqrt{G_N}$) has to be much greater than $\mu^{-1/6}\sqrt{ |\gamma(\beta)|}$. For large $\mu$, this constraint can be satisfied in the region between the collision points in Figure \ref{fig:local_stokes}, which gets larger with larger $\mu$. This region is in an $\mathcal{O}(\sqrt{G_N})$ vicinity of $t=t_c$. Even though large $\mu$ controls the saddle point approximation here, we need to consider small $G_N$ so that the Gaussian approximation is justified.

\subsubsection*{The local structure of Stokes lines}

\begin{figure}[t]
    \centering
    \includegraphics[width=0.4\linewidth]{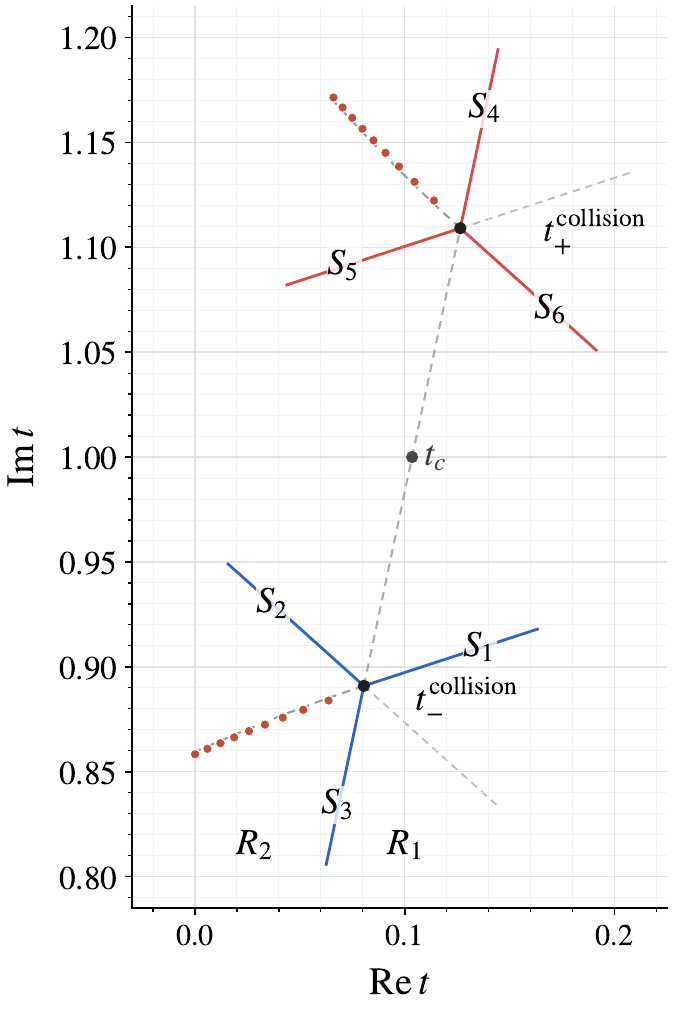}
    \caption{After backreaction, the singularity at $t=t_c$ gets smoothed out into an intricate analytic structure. In an $\mathcal{O}(\sqrt{G_N})$ region around $t_c$, the correlator is described by the Gaussian approximation in \eqref{eq:omega2_model}. Only one saddle point of the Gaussian action contributes at $t_c$. $t_\pm^{\text{collision}}$ indicate Airy-like collision points where this saddle collides with two other saddles, one at each collision point (see Figure \ref{fig:saddle_collision}). The dashed lines indicate anti-Stokes lines and $S_i$ indicate Stokes lines that dictate the interplay of these saddles. In regions where two saddles contribute, they destructively interfere to give rise to lines of zeroes (red dots) close to the anti-Stokes lines. These lines of zeroes approach the anti-Stokes lines as we increase $\mu$. In the limit where we take $G_N\to 0$ with $\mu G_N$ fixed to a small number, these lines of zeroes condense to form branch cuts (see \cite{Faulkner:2017hll}). This plot uses $\beta=2$, $G_N=10^{-3}$ and $\mu=10$.}
    \label{fig:local_stokes}
\end{figure}

Having justified the saddle point approximation, we can now study how the saddle points interact by understanding their Stokes and anti-Stokes lines. The Airy function form implies that locally, each collision point is the source of three Stokes lines and three anti-Stokes lines. 
Figure \ref{fig:local_stokes} illustrates these lines for the quadratic action governing the resolution of the $t_c$ singularity. 

On a Stokes line, the imaginary parts of the exponents evaluated on the two saddles participating in the collision are equal. When this happens, saddle points can join or leave the contour, depending on which saddle points are already on the contour and the orientation of the steepest descent contours of these saddles with respect to each other. The simplest case is when only one saddle is on the contour at some starting value of our parameter (which in this case is $t$). If the exponent evaluated on the saddle point already on the contour has a larger real part, then on crossing a Stokes line the other saddle joins the contour.  Its contribution is subleading. If the exponent evaluated on the saddle already on the contour has a smaller real part, then the other saddle point does not start to contribute. These rules are explained very clearly in \cite{Witten:2010cx}.

Let us see what these rules imply for the plot in Figure \ref{fig:local_stokes}. As we outlined in Section \ref{sec:first_correction}, only one saddle of the quadratic action contributes when $t=t_c$. This is the minus saddle $\tilde\omega_-$ of \eqref{eq:quad_saddles}. This means that in the diamond shaped region in Figure \ref{fig:local_stokes} that encloses $t_c$, only one saddle contributes. On the Stokes line $S_2$, the exponent evaluated on this saddle has a larger real part compared to that of the other saddle participating in the collision at $t_-^{\rm collision}$. When we cross this Stokes line, the other saddle joins the contour, so that both saddles contribute in the region $R_2$. On the other hand, on the Stokes line $S_1$, the exponent evaluated on the saddle that contributes inside the diamond-shaped region has a smaller real part compared to the action of the other saddle. The other saddle therefore does not join the contour when we cross this line, and only one saddle contributes in the region $R_1$. We can also similarly argue that a saddle joins the contour on crossing the Stokes line $S_5$, but not on crossing $S_6$. This is the third saddle point involved in this picture, which collides with the $t_c$ saddle at $t_+^{\rm collision}$.

\subsection{Global structure of Stokes lines in the complex $t$ plane}
\label{sec:stokes}
We now  study the global structure of Stokes lines of $G_{1,0}(t)$ in the complex $t$ plane.   We focus on the saddle points we've already identified in the region near $t=t_c$. We will find an interesting Stokes phenomenon near the coincident point singularity at $t=0$ which picks up the contribution of the geodesic with turning point close to the black hole singularity. As we discuss in Section \ref{sec:cft}, this phenomenon appears to be a universal feature of the smoothing out of singularities and appears in various conformal field theory quantities as well.

\begin{figure}[t]
    \centering
    \includegraphics[width=0.9\linewidth]{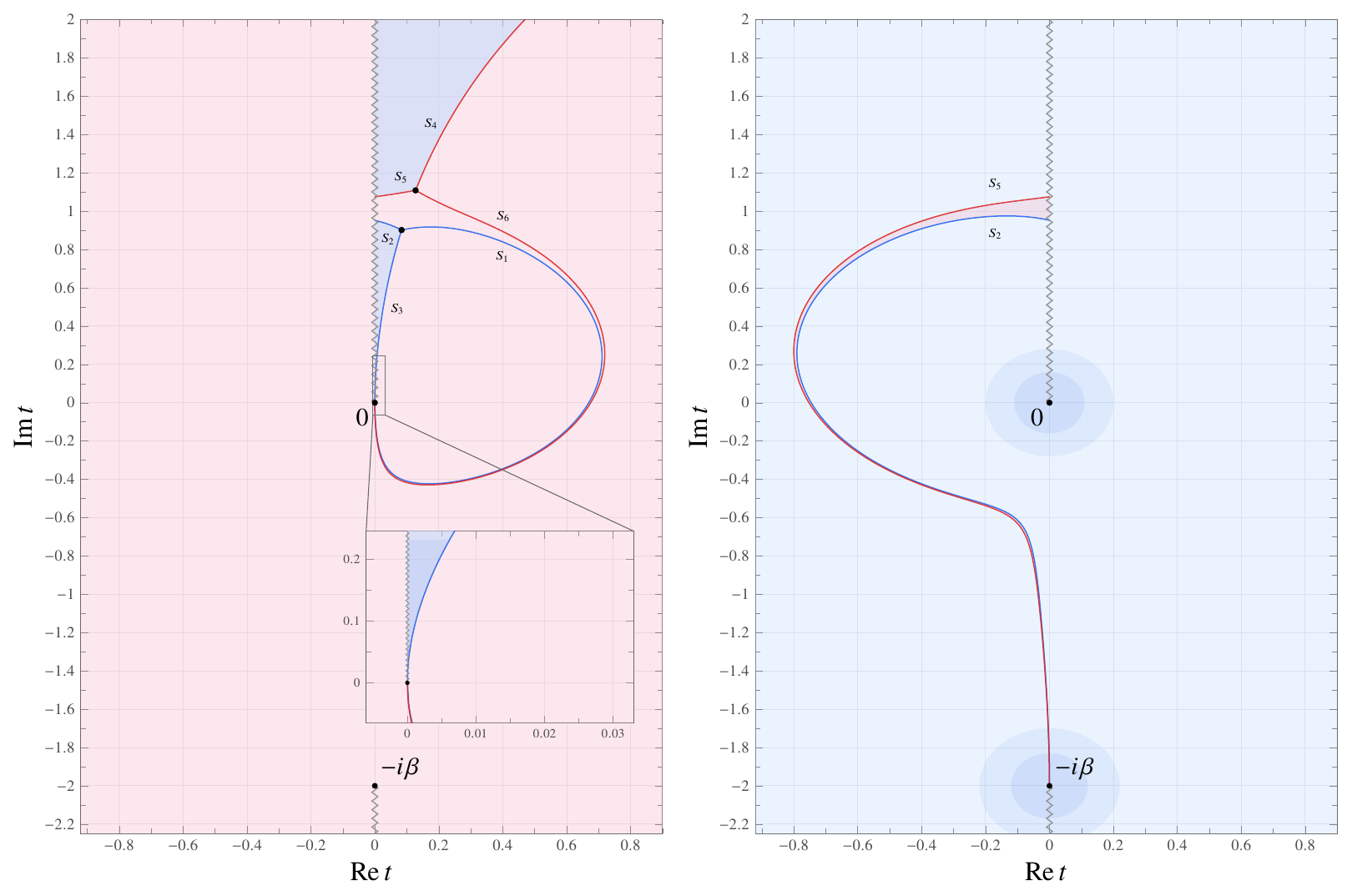}

    \caption{The global structure of the Stokes lines  of $G_{1,0}(t)$ involving the participating saddles  in the complex $t$ plane. The figure on the left includes the diamond-shaped region of Figure \ref{fig:local_stokes}. We choose a branch cut extending vertically upwards from $t=0$, so that some of the Stokes lines starting from the collision points hit this branch cut and evolve to another sheet. This second sheet is depicted in the right figure. The region shaded in pink is where one saddle point contributes, and the region shaded in blue is where two saddle points contribute. In total there are three saddles at play in this picture; two of them are images of each other on different logarithmic sheets. As we approach $t=0$ or $t=-i\beta$ in the blue region, one of these saddle points represents a bouncing geodesic with turning point close to the black hole singularity. This is depicted by the shaded disk regions; as the shading gets darker, the turning point gets closer to the singularity. The parameters used for this figure are $\beta=2$, $G_N=10^{-3}$ and $\mu=9$.}
    \label{fig:global_stokes}
\end{figure}

\begin{figure}[t]
    \centering
    \includegraphics[width=0.8\linewidth]{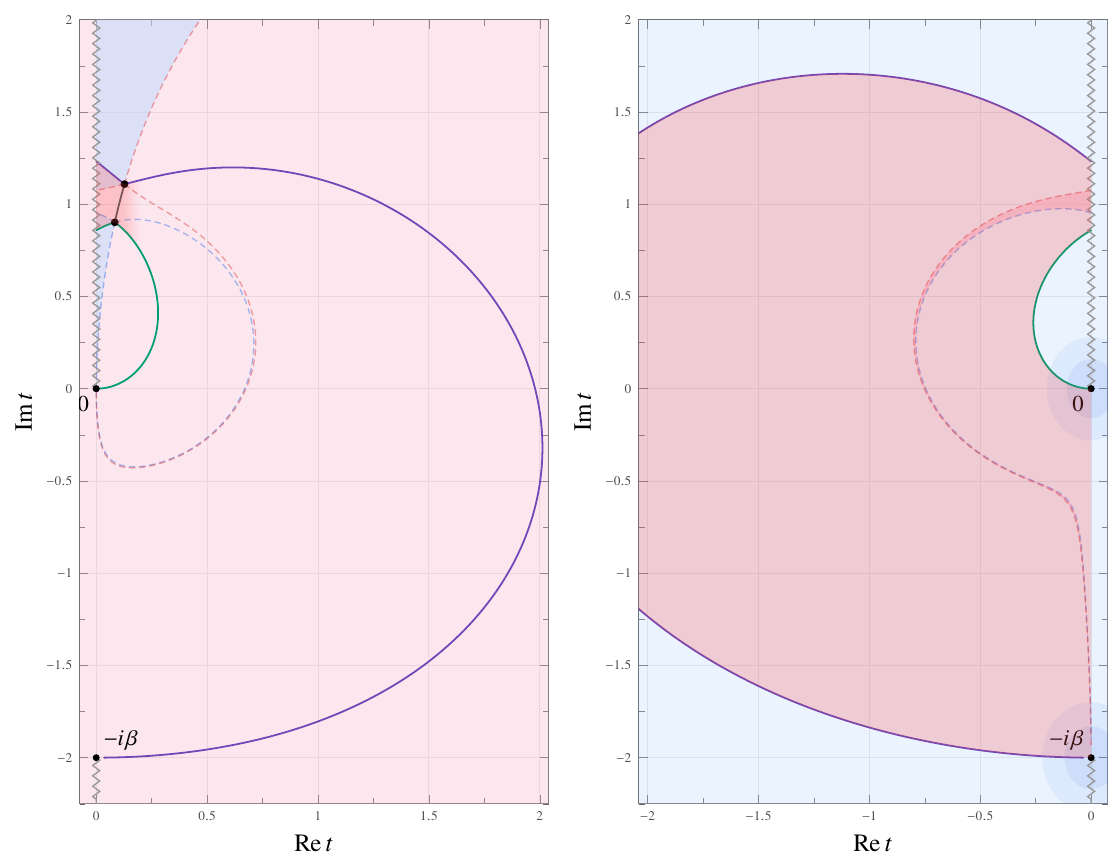}
    \caption{The global structure of anti-Stokes lines of $G_{1,0}(t)$ involving the participating saddles in the complex $t$ plane. The two figures here depict two sheets of the same function, similar to Figure \ref{fig:global_stokes}. In the right figure, one of the anti-Stokes lines exits from view, and then re-enters near the bottom before heading to $t=-i\beta$. Consider the green anti-Stokes line in the region of this plot where two saddles contribute. To the right of this line, the bouncing geodesic saddle is not the dominant saddle. To the left of this line, however, it becomes dominant, with its contribution to $G_{1,0}(t)$ scaling like $e^{1/G_N}$. Thus the correlator is large in the region extending from $t=t_c$ to $t=0$ where this saddle contributes and dominates. This is the shaded region between the green and blue anti-Stokes lines. Because of the $e^{1/G_N}$ scaling of the bouncing geodesic saddle, the correlator in this region does not have a $G_N\to 0$ limit. Instead in this limit, the anti-Stokes lines in the blue regions become ``boundaries" beyond which we have to analytically continue to obtain the usual probe limit answer. In other words, the $G_N\to 0$ limit does not commute with analytic continuation of the correlator outside the physical strip. The parameters used for this figure are $\beta=2$, $G_N=10^{-3}$ and $\mu=9$.}
    \label{fig:anti_stokes}
\end{figure}

We begin by asking how the Stokes and anti-Stokes lines corresponding to the participating saddles\footnote{There are other saddles that contribute to $G(t)$ in this region. These saddle points are subleading near $t=t_c$ and near the coincident point singularities.  We discuss some of these saddles in Appendix \ref{sec:additionalsaddles}.} described in Section \ref{sec:neartcstokes} evolve as we exit the local region in the vicinity of $t_c$.\footnote{We continue the labeling of the Stokes lines by $S_i$  to the global case as well.} Figure \ref{fig:global_stokes} shows the evolution of the Stokes lines computed numerically for the $1/\phi^3$ potential, and Figure \ref{fig:anti_stokes} shows the anti-Stokes lines.\footnote{To compute this evolution, we first locate the collision points in $(\bar E,\tilde\omega)$ above by solving $(\partial_{\bar E}\mathcal I=\beta/G_N$, $\partial_{\tilde\omega}\mathcal I=i\left(t+\frac{i\beta}{2}\right)$, $J=\det \partial^2_{\bar E,\tilde\omega} \mathcal I=0$). At a collision, the direction along which the saddles merge becomes a zero mode of the quadratic action, which is why the Hessian $\partial^2_{\bar E,\tilde\omega}$ has a vanishing eigenvalue (implying $J=0$). The local Stokes and anti-Stokes lines are initialized according to the local Airy form (which may be found by finding the null directions of $\partial^2_{\bar E,\tilde\omega} \mathcal I$). These rays are then continued using a predictor-corrector method, in which we track the participating saddles and impose that the difference in the real or imaginary parts of their exponents is zero. Note that to obtain continuous paths, we move continuously onto other sheets whenever branch cuts are encountered in the $(\bar E,\tilde\omega)$ plane.}
We observe that the lines $S_1$ and $S_6$ approach $t=0$ from the negative imaginary direction, while the line $S_3$ approaches $t=0$ from the positive imaginary direction. These are pictured on the left in Figure \ref{fig:global_stokes}. The Stokes lines $S_2$ and $S_5$ hit a branch cut and cross onto another sheet. They then approach $t=-i\beta$ from the positive imaginary direction. Their evolution is plotted on the right in Figure \ref{fig:global_stokes}. Thus the region $R_2$ where two saddles contribute (the blue shaded region) extends all the way to the coincident  points $t=0$ and $t=-i\beta$, and beyond.

\begin{figure}[t]
    \centering
    \includegraphics[width=0.45\linewidth]{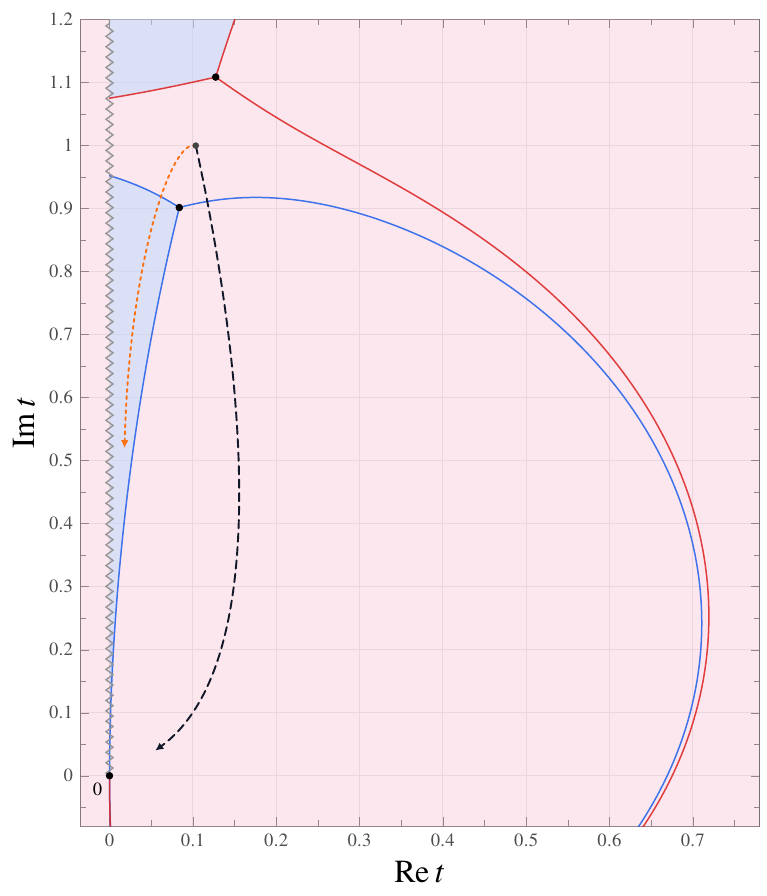}
    \caption{In the diamond-shaped region between the two collision points (drawn in black), only one saddle contributes at small $G_N$ (for a discussion of a subleading saddle see Appendix \ref{sec:additionalsaddles}). This saddle evolves to a geodesic with an $\mathcal{O}(1)$ energy along the black path. Along the orange path, it evolves to a geodesic with high energy and turning point close to the singularity. The parameters used for this figure are $\beta=2$, $G_N=10^{-3}$ and $\mu=9$.}
    \label{fig:stokes_paths}
\end{figure}

The global plot also tells us the fate of the backreacted bouncing geodesic saddle that contributes at $t=t_c$ as we leave the region in the $\mathcal{O}(\sqrt{G_N})$ vicinity of $t_c$ and the Gaussian approximation outlined in the previous section starts to fail. As $t$ changes, the location of this saddle in the $\tilde\omega$ plane evolves continuously.
If we follow a path in $t$ that moves to the point $t=0$ on crossing the Stokes line $S_1$ (see the path in black in Figure \ref{fig:stokes_paths}), the $t_c$ saddle moves to an $\mathcal{O}(1)$ point in $\tilde\omega$. 
On the other hand, if our path crosses the Stokes line $S_2$ and approaches $t=0$ (see the orange path in Figure \ref{fig:stokes_paths}), something interesting happens. 
On numerically tracking the saddle point in the $\tilde\omega$ plane, we find that it runs off to infinite $\tilde\omega$ without circling around any branch points. As we emphasized at the end of Section \ref{sec:backreaction}, this means that its turning point approaches the singularity. In other words, information about the singularity gets encoded in the vicinity of the coincident point singularities (whose divergences comes from the $G_{0,0}(t)$ integral).

 We reach the same conclusion if we 
start at the endpoint of the path in black in Figure \ref{fig:stokes_paths}. This point in $t$ corresponds to a saddle with $\tilde\omega\sim \mathcal{O}(1)$. On crossing the Stokes line $S_3$, this saddle participates in a Stokes phenomenon with the second saddle, which then starts to contribute. This second saddle point approaches infinite $\tilde\omega$ (and infinite $\bar E$) for $t\to 0$ and has turning point close to the singularity. We see that whether or not the $t_c$ saddle evolves to a saddle with turning point close to the singularity is path-dependent.
In subsequent sections, we will analyze the Stokes phenomenon that occurs across the Stokes line $S_3$ in more detail from an analytical approach.
Before we do so, it is useful to note that when the saddle point with turning point close to the singularity reaches large $\tilde\omega\sim 1/G_N$, the saddle point approximation is  controlled by the overall $1/G_N$ in the action. This is in contrast to the probe limit bouncing geodesic, which requires large $\mu$.

The reader might also wonder why the point $t=0$ is branched. To see this, consider evaluating the correlator at $t=-i\epsilon$ along two different paths.  If we start at $t=t_c$ and approach $t=-i\epsilon$ along a path that crosses $S_1$ and $S_6$, we find that one saddle contributes at the end. If we approach it along a path that only crosses $S_2$, we find that two saddles contribute at the end, summing up to give a different answer to the integral. This means that our endpoints must be on different sheets. In other words, the defining integration contour does not come back to itself on winding around $t=0$ --- the basic reason this happens is that the $(\bar E,\ \tilde\omega)$ plane is branched.

Note that we can also find a similar structure of Stokes lines in ensembles where we fix $\bar E$ or the energies $E_+$ or $E_-$ involved in the matrix elements. In the first ensemble, two Stokes lines instead of one extend down to $t=0$ from the $\Im t>0$ direction. However, because of the finite integration range in $\tilde\omega$, the saddle point corresponding to the bouncing geodesic stops contributing before we reach $t=0$. In the second ensemble, the structure of Stokes lines is qualitatively similar to the canonical ensemble, with one Stokes line going down to $t=0$ from the $\Im t>0$ direction.

\section{Imprint of the bouncing geodesic near the coincident  points}
\label{sec:OPE_sing}
In the previous section we found that a saddle point representing a geodesic with turning point close to the black hole singularity contributes near $t=0$ and $t=-i\beta$ in certain parts of the Riemann surface of $G(t)$. In this section, we will derive this from a local analysis near the coincident  points. We will show how the function develops an essential singularity at $t=0$ and $t=-i\beta$. This is a phenomenon that occurs in pure JT gravity at finite $G_N$ as well --- the geodesic with turning point close to the singularity produces an essential singularity with a different coefficient in the exponential.\footnote{In the case of the exponential dilaton potential the signal of the bouncing geodesic is more distinctive.  See Appendix \ref{sec:exp_potential}.}

\subsection{High energy saddles approach the coincident  points}
\label{sec:no_sing}
First, let us show that for each integral comprising $G(t)$ in \eqref{eq:correlator_omega},  a backreacted saddle that has turning point close to the singularity (and hence infinite $\tilde\omega$) may only contribute near the coincident  points $t=0$ or $t=-i\beta$. We will start with the first winding integral $G_{1,0}(t)$, which gives rise to the $t_c$ singularity. Recall that in the probe limit, this singularity is caused by a saddle point of this integral running off to infinite $\tilde{\omega}$ (see \eqref{eq:probe_tc_saddle}). This represents a complex geodesic becoming null and bouncing off the singularity. We would like to study how the equation for this saddle point changes at finite $G_N$. We will work in terms of the rescaled variable $\omega=\tilde\omega G_N$.

At finite $G_N$, the first winding integral changes from the probe form \eqref{eq:probe_bg}:
\begin{align}
     G_{1,0}(t)\sim\int^{\infty} \d\tilde\omega\exp\left[-i\tilde\omega t+i \tilde\omega t_c+
    2\mu
    \log\tilde\omega
    \right],\quad t_c=\frac{i\beta}{2}+\frac{\tilde\beta}{2}
\end{align}
to 
\begin{align}
\label{eq:G10_exact}
    G_{1,0}(t)\sim \int_{0}^\infty \d\bar E \int^{2\bar E}\d\omega\ \exp\left[-\frac{\beta\bar E}{G_N}-\frac{i\omega}{G_N}\left(t+\frac{i\beta}{2}\right)+\mathcal{I}_{(1,0)}\right],
\end{align}
where $\mathcal{I}_{(1,0)}$ is the on-shell action for the backreacted complex geodesic. It is given by \eqref{eq:S_reln}, and repeated here for convenience: 
\begin{align}
    \nonumber \mathcal{I}_{(1,0)}&=\frac{3\pi\phi_h(E_-)}{G_N}+ \frac{\pi \tilde \phi_h(E_-)}{G_N} -\frac{\pi \phi_h(E_+)}{G_N} - \frac{\pi \tilde \phi_h(E_+)}{G_N}  + (I_+^{\rm{OPE}} - I_-^{\rm{OPE}}) -i\pi \mu, \\
    I_\pm^{\rm{OPE}}&=\frac{1}{G_N} \int_{\phi_t}^{\infty} \frac{d\phi \, \phi\, U(\phi) P_\pm}{(W(\phi) - E_\pm) \sqrt{W(\phi) - E_\pm -P_\pm^2}}.
    \label{eq:another_action}
\end{align}
Because we have explicitly included the monodromy contributions
the contour from $\phi_t$ to $\infty$ in the above integral corresponds to that for the OPE geodesic, see Figure \ref{fig:OPE}.
Since this is now a two-dimensional integral, we obtain two saddle point equations. The first is obtained by varying with respect to $\bar E$, and fixes the inverse temperature to $\beta$. The second is obtained by varying with respect to $\omega$, and tells us the average time separation between the endpoints of the geodesic. They are given by
\begin{align}
    \beta=G_N\frac{\partial \mathcal{I}_{(1,0)}}{\partial\bar E}, \quad i\left(t+\frac{i\beta}{2}\right)=G_N\frac{\partial \mathcal{I}_{(1,0)}}{\partial\omega}.
\end{align}
Some algebra shows that these equations simplify to the following expressions:
\begin{align}
\label{eq:beta_eq}
    \frac{\beta}{2}&=-\pi \phi_h'(E_+)-\pi \tilde\phi_h'(E_+)+3\pi \phi_h'(E_-)+\pi \tilde\phi_h'(E_-)+ (T_+-T_-),\\
    \label{eq:t_eq}
    it-\frac{\beta}{2}&=-\pi \phi_h'(E_+)-\pi \tilde\phi_h'(E_+)-3\pi \phi_h'(E_-)-\pi \tilde\phi_h'(E_-)+(T_++T_-),\\
    T_\pm&=\int_{\phi_t}^{\infty} d\phi \frac{P_\pm}{(W(\phi)-E_\pm)\sqrt{W(\phi)-E_\pm-P_\pm^2}}.
\end{align}
Here we used the relations $E_-=2\bar E-\omega$ and $E_+=2\bar E+\omega$. To satisfy these saddle point equations at large $\omega$, we hold $E_-$ fixed and let $E_+$ go to infinity. This implies that both $|\bar E|$ and $|\omega|$ approach infinity. All the terms labeled with a $+$ in \eqref{eq:beta_eq} and \eqref{eq:t_eq} go to zero, so that we find
\begin{align}
    it-\frac{\beta}{2}\to-\frac{\beta}{2}\implies t\to 0, \quad E_+\to\infty \text{ and }E_-\text{ fixed}.
\end{align}
Therefore, a large $\omega$ saddle point of this type contributes only near $t=0$, implying that the associated singularity can only occur at this point.\footnote{As noted at the end of Section \ref{sec:stokes}, the sharpness of this saddle point is controlled by making the overall factor of $1/G_N$ in front of the action large.} 
Note that this argument is true for any winding $G_{n,m}(t)$. To see this, note that the integrals $T_+$ and $T_-$ inherit the same branch points as $I_+$ and $I_-$, given in \eqref{eq:bp_1} and \eqref{eq:bp_2}. The integrands for other windings are generated by winding around these branch points and adding the resulting monodromies to the action in \eqref{eq:another_action}. Doing such a winding changes $T_\pm$ by linear combinations of $\pi \phi_h'(E_\pm)$ and $\pi \tilde\phi_h'(E_\pm)$. Since these integrals appear in the combination $T_+-T_-$ in \eqref{eq:beta_eq} and $T_+ + T_-$ in \eqref{eq:t_eq}, the same argument as above shows that a large $\omega$ saddle point contributes only near $t=0$.

\begin{figure}[t]
    \centering
    \includegraphics[width=0.7\linewidth]{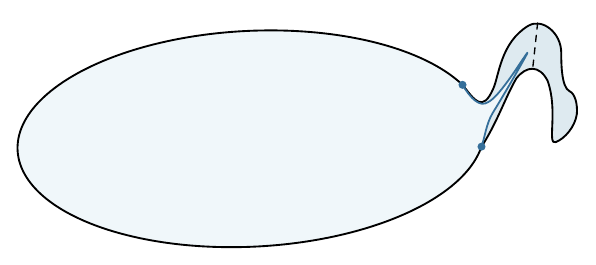}
    \caption{This figure depicts a geodesic with turning point very close to the singularity at finite $G_N$. For it to be a saddle point with a fixed temperature $\beta$, the mass of the black hole on one side of the geodesic goes to infinity. Thus the black hole effectively has infinite temperature on that side, and the spacing between the endpoints of the geodesic approaches $\beta$, the thermal image of zero. If the geodesic endpoints are on the other side of the fold, the spacing goes to zero.}
    \label{fig:high_T_geodesic}
\end{figure}

For windings $G_{-n,-m}(t)$ that contribute at large negative $\omega$, we instead hold $E_+$ fixed and let $E_-$ go to infinity.
This implies that all the terms of the type labeled with a $-$ in \eqref{eq:beta_eq} and \eqref{eq:t_eq} go to zero, so that 
\begin{align}
    it-\frac{\beta}{2}\to\frac{\beta}{2}\implies t\to -i\beta, \quad E_-\to\infty \text{ and }E_+\text{ fixed}.
\end{align}
Therefore the associated singularity can only occur at the point $t=-i\beta$, which is the thermal image of the $t=0$ point. The physics at play here is that the black hole solution on one side of the geodesic has infinite energy as $|\omega|\to \infty$. Since the inverse temperature of an infinitely massive black hole in AdS goes to zero, the part of the thermal circle on this side of the geodesic has almost zero length, while the part on the other side has length that is approximately $\beta$ (see Figure \ref{fig:high_T_geodesic}). This means that the endpoints of the geodesic effectively have Euclidean time separation that either equals zero or $\beta$.

In the next subsection, we will study the behavior of the action evaluated on these large $\omega$ saddles in more detail. We will also identify the regions near $t=0$ where these large $\omega$ saddles contribute, and see that the answer matches the expectation from Section \ref{sec:stokes}.

\subsection{Analytic structure of the correlator near the coincident  points}
\label{sec:essential}
In the previous subsection, we explained why backreacted geometries involving geodesics with turning point close to the singularity may only contribute as saddles for $t\to 0$ or $t \to -i\beta$. Whether or not these saddles actually contribute needs more thought. The analysis in Section \ref{sec:stokes} gave numerical evidence that for $G_{1,0}(t)$, these saddles do not contribute in the physical strip reviewed in Section \ref{sec:intro}. Instead, one picks up these saddles at a Stokes line that extends in the direction of increasing $\Im t$. In this section, we will show this analytically by systematically continuing outside the physical strip.

Let us start at the point $t=-i\epsilon$ with $\epsilon$ small, which lies within the strip of analyticity. At this starting point, we know that the contour of integration for the full correlator is $\bar E \in (0,\infty)$ and $\omega\in (-2\bar E,2\bar E)$. It is convenient to switch to the coordinates defined by
\begin{align}
    s_+= \sqrt{E_+}, \quad s_-= \sqrt{E_-},
\end{align}
so that the contour of integration runs from $0$ to $\infty$ in both $s_+$ and $s_-$. In the region near $t=0$, the results of the last subsection say that the integrand corresponding to $G_{1,0}(t)$ possesses a large $\omega$ saddle point for $s_+\to\infty$ with $s_-=s_-^*$ fixed. With these asymptotics, the leading behavior of the logarithm of the integrand for $G_{1,0}(t)$ in \eqref{eq:G10_exact} is 
\begin{align}
\label{eq:splus_integrand}
    \mathcal{J}_{(1,0)}\approx\mathcal{J}_{(1,0)}^+ = -\frac{it}{2 G_N}s_+^2 -\frac{\pi}{G_N}s_++2\mu \log(s_+^2)+\log (s_+).
\end{align}
Here we dropped $\mathcal{O}(1)$ terms that are bounded as $s_+\to\infty$. The last term comes from the Jacobian from changing variables to $s_+$ and $s_-$. For simplicity we will consider large $\mu$ and drop this last term from our analysis. At small $G_N$ when the saddle point approximation in $s_-$ is good, the full two-dimensional integral over the large $s_+$ region is well-approximated by the one-dimensional $s_+$ integral of $e^{\mathcal{J}_{(1,0)}^+}$ over this region. As we will see, this contribution tells us about the way in which the backreacted bouncing geodesic saddle affects the answer.

\begin{figure}[t]
    \centering
    \includegraphics[width=0.5\linewidth]{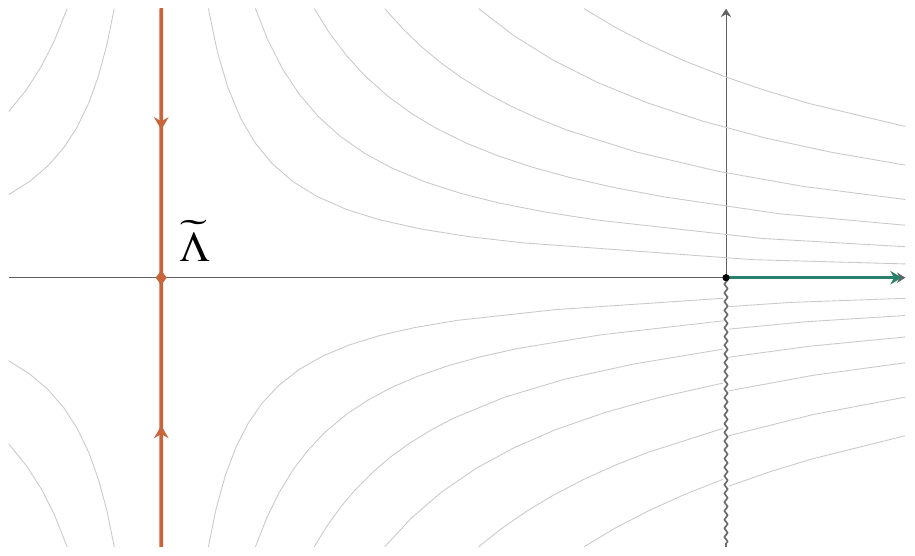}
    \caption{When $t=-i\epsilon$, the saddle point that corresponds to a geodesic with turning point close to the singularity resides at large negative $s_+$, where $s_+^2=2\bar E+\omega$. Its steepest ascent contour is pictured in orange. Since this contour doesn't intersect the defining contour of the integral $\int_0^\infty \d s_+ e^{\mathcal{J}_{(1,0)}^+(s_+)}$ (the postive real axis, in green), this saddle does not contribute. Following the notation in \cite{AliAhmad:2026wem}, we have labelled this saddle as $\tilde\Lambda$.}
    \label{fig:zero_theta}
\end{figure}

Now, for $t=-i\epsilon$, the leading saddle point for large $s_+$ is given by
\begin{align}\label{eq:saddle_sp}
    -\frac{\epsilon}{G_N} s_+ - \frac{\pi}{G_N}\approx0 \implies s_+ \approx -\frac{\pi}{\epsilon}.
\end{align}
This is the saddle point with turning point close to the singularity.
For positive $\epsilon$, even though this saddle point approaches infinite $\omega$ and $\bar E$, it lies at large \textit{negative} $s_+$ and does not contribute. To see this, first note that it does not contribute to the one-dimensional integral $\int_0^\infty \d s_+ e^{\mathcal{J}_{(1,0)}^+(s_+)}$ (as explained in Figure \ref{fig:zero_theta}) and hence to the integral over the large $s_+$ region of $e^{\mathcal{J}_{(1,0)}^+(s_+)}$. Its contribution also cannot arise from the remaining part of the two-dimensional integral. If it were to contribute, it would add a large non-analytic part to the answer:
\begin{align}
    \mathcal{J}_{(1,0)}^+ \sim \frac{\pi^2}{2 G_N\epsilon} +4\mu\log\left(\frac{1}{\epsilon}\right) \implies e^{\mathcal{J}_{(1,0)}^+ } \sim \left(\frac{\pi}{\epsilon}\right)^{4\mu}\exp\left[\frac{\pi^2}{2 G_N\epsilon}\right] \sim t^{-4\mu}\exp\left[-\frac{i\pi^2}{2 G_Nt}\right] .
\end{align}
This has an essential singularity at $t=0$, along with a generically branched prefactor.
However, the integral over $s_+$ and $s_-$ not in the asymptotic region considered above is holomorphic at $t=0$.\footnote{To see this, consider $G_{1,0}(t)$ integrated up to some cutoff in $s_+$:
\begin{align}
    G_{<R}(t)=\int_0^R \d s_+ \int \d s_-\, \mathcal{A}(s_+,s_-)e^{-\frac{it}{2 G_N} s_+^2}e^{\frac{it-\beta}{2 G_N}s_-^2}.
\end{align}
Then for $|\Im t|<\delta<\beta$ we have
\begin{align}
    \Big|e^{-\frac{it}{2 G_N} s_+^2}e^{\frac{it-\beta}{2 G_N}s_-^2}\Big|\leq e^{\frac{\delta R^2}{2 G_N}}e^{-\frac{(\beta-\delta)}{2G_N}s_-^2}.
\end{align}
Since the factor $\mathcal{A}(s_+,s_-)$ doesn't grow fast enough to overcome this Gaussian, the above integral converges uniformly in an open neighborhood of $t=0$, and hence is holomorphic at $t=0$. \label{fn:int}
} So it cannot produce the factor above. The entire two-dimensional integral therefore does not receive a contribution from this saddle point.\footnote{As we note in Appendix \ref{sec:JT}, similar contributions arise in the case of JT gravity. That case is another check of this saddle point not contributing for $t=-i\epsilon$.}

This small and intermediate $s_+$ region where the turning point is not close to the singularity still dominates the integral. To get the full answer we have to include the corrections from this part of the integral, for which the $s_-$ direction also becomes important.

\begin{figure}[t]
    \centering
    \includegraphics[width=0.35\linewidth]{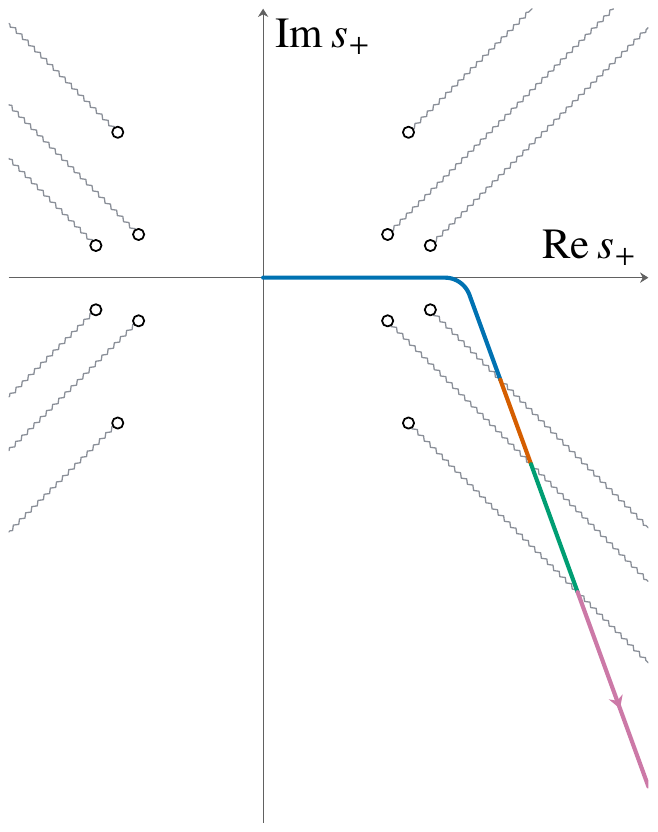}
    \caption{To analytically continue $G_{1,0}(t)$ outside the strip of analyticity, we rotate the $s_+$ contour into the lower half plane. As we do so, we go smoothly through any branch cuts encountered. A change in color of the contour in this figure indicates that the contour has passed onto another sheet. Note that the analytic structure pictured here is for the full integrand $e^{\mathcal J_{(1,0)}}$.}
    \label{fig:contour_deformation}
\end{figure}

\begin{figure}[t]
    \centering
    \includegraphics[width=\linewidth]{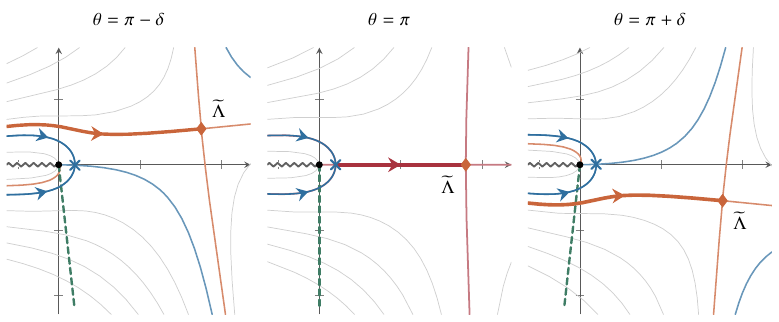}
    \caption{A Stokes phenomenon occurs in the integral $\int_0^\infty \d s_+ e^{\mathcal{J}_{\tilde\Lambda}^+}$ at $\theta=\pi$. The contours depicted here are lines of constant imaginary part of the action, and arrows indicate the direction in which the real part of the action and hence the size of the integrand decreases.  The defining contour is denoted by the green dashed line starting at the origin.  At $\theta=\pi-\delta$ (left panel), only one saddle point (labelled by a cross) contributes to the integral.  The steepest descent contour is the blue line starting at the origin and bending down to the right.  At $\theta=\pi$, a line of constant imaginary part of the action (here a part of the steepest descent contour) joins this saddle and the saddle $\tilde\Lambda$, corresponding to the bouncing geodesic. This is the Stokes phenomenon. At a larger value of $\theta$ (right panel), both saddles contribute to the integral; the steepest ascent contour of $\tilde\Lambda$ now intersects the defining contour. The steepest descent contour now has two components: the blue line leaving the origin and bending up to the right and the nearly vertical orange line passing through $\tilde\Lambda$. Note that we have moved the location of the branch cut relative to Figure \ref{fig:zero_theta} for clarity.}
    \label{fig:simple_stokes}
\end{figure}

We now want to see how $\int^\infty \d s_+ e^{\mathcal{J}_{(1,0)}^+(s_+)} $ changes as we rotate the phase of $t$. Write $t=-i\epsilon e^{i\theta}$, with $\theta=0$ representing $t=-i\epsilon$.
Beyond $\theta=\pi/2$, we exit the physical strip where the integral over $\bar E$ and $\omega$ is convergent, and need to analytically continue the answer. For this reason, as we vary $\theta$, 
we rotate the $s_+$ contour such that asymptotically we have
\begin{align}
    s_+ = e^{-i\theta/2}r,\quad r\in\mathbb{R}^+.
\end{align}
While doing so we go smoothly through the branch cuts in the complex $s_+$ plane, see Figure \ref{fig:contour_deformation} for a snapshot at fixed $s_-$. This is the contour rotation inherited by the integral including only the first $G_N$ correction in \eqref{eq:rotated}.
As $\theta$ varies, the saddle point equation in \eqref{eq:saddle_sp} changes and hence the saddle with turning point close to the singularity also rotates:
\begin{align}
    -\frac{\epsilon e^{i\theta}}{G_N}s_+-\frac{\pi}{G_N}\approx 0 \implies s_+ \approx -\frac{\pi}{\epsilon}e^{-i\theta}.
\end{align}
Now, a Stokes phenomenon happens when $\theta=\pi$: the saddle point above starts to contribute to the integral $\int_{\mathcal C_\theta} \d s_+ e^{\mathcal{J}_{(1,0)}^+}$.  Before $\theta$ reaches $\pi$, the saddle point that contributes to $\int_{\mathcal C_\theta} \d s_+ e^{\mathcal{J}_{(1,0)}^+}$ is given by including the effects of the logarithmic term in $\mathcal{J}_{(1,0)}^+$ in the saddle point equation:
\begin{align}
     -\frac{\epsilon e^{i\theta}}{G_N}s_+-\frac{\pi}{G_N}+\frac{4\mu}{s_+}\approx 0 \implies s_+^*=\frac{-\pi \pm \sqrt{\pi^2+16\mu G_N \epsilon e^{i\theta}}}{2\epsilon e^{i\theta}}.
\end{align}
The contributing saddle corresponds to the plus sign, and is located in the small $s_+$ region.
At $\theta=\pi$, the steepest descent contour from this saddle point ends at the large $s_+$ saddle with turning point close to the singularity (which corresponds to the minus sign above); see Figure \ref{fig:simple_stokes}. Hence, the latter saddle  starts to contribute.\footnote{Note that keeping the $\mathcal{O}(1)$ terms in the action in \eqref{eq:splus_integrand} corrects the location of the Stokes phenomenon so that it happens at $\theta=\pi-\mathcal{O}(\epsilon)$. }

Since this saddle point contribution comes from the large $s_+$ region\footnote{Similar to the argument in footnote \ref{fn:int}, the integral over the region where $s_+$ is not large cannot produce a factor that gives an essential singularity.} in $\int_{\mathcal C_\theta} \d s_+ e^{\mathcal{J}_{(1,0)}^+}$, it is reflected in the answer to our full two-dimensional integral and hence contributes in the full problem. This matches the prediction using the global Stokes lines derived in Section \ref{sec:stokes}, where the saddle with turning point close to the singularity starts to contribute on crossing the Stokes line $S_3$ emanating from the origin. 

We can also study how this saddle changes the functional form of the answer. The contribution of this saddle to the correlator is given by
\begin{align}
\label{eq:essential}
    e^{\mathcal{J}_{(1,0)}^+}\sim \left(\frac{\pi}{\epsilon e^{i\theta}}\right)^{4\mu}\exp\left[\frac{\pi^2}{2 G_N\epsilon e^{i\theta}}\right].
\end{align}
At $\theta=\pi$ when it is picked up, this contribution is exponentially small. The exponential becomes purely oscillatory as we rotate $\theta$ further to $3\pi/2$, and then makes an exponentially large contribution for $\theta>3\pi/2$. A naive interpretation of this behavior might be that $G(t)$ contains a large signal of the curvature divergence near the black hole singularity. In the next section, by summing over different windings in \eqref{eq:saddle_sum}, we will show why this is not the case.

\section{The sum over windings and a natural boundary}
\label{sec:windings}
So far we have focused our attention on the $(0,0)$ and $(1,0)$ windings in the integrals that form the correlator $G(t)$ (see \eqref{eq:sum2}):
\begin{align}
\label{eq:another_sum}
G(t)=G_{0,0}(t)
+\sum_{\substack{n,m\geq0\\n+m\geq1}}
\bigl[G_{n,m}(t)+G_{-n,-m}(t)\bigr].
\end{align}
We discussed the coincident point  singularity in $G_{0,0}(t)$, as well as the smoothing out of the bouncing geodesic singularity that arises from $G_{1,0}(t)$. After the smoothing out, the saddle with turning point close to the black hole singularity contributes on crossing a Stokes line near the coincident point singularity at $t=0$. However, other windings in the above sum lead to singularities in the probe limit as well. These are located at $t=t_{n,m}$ (see \eqref{eq:tnm}) and the corresponding KMS images. The arguments of Sections \ref{sec:smoothing} and \ref{sec:OPE_sing} also hold for these singularities, so that we expect similar behavior for them. The discussion in the rest of this section applies to these windings and excludes $(n,m)=(0,0)$. For simplicity we also focus on the windings corresponding to $G_{n,m}(t)$ with nonnegative $n$ and $m$.

\begin{figure}[t]
    \centering
    \includegraphics[width=0.7\linewidth]{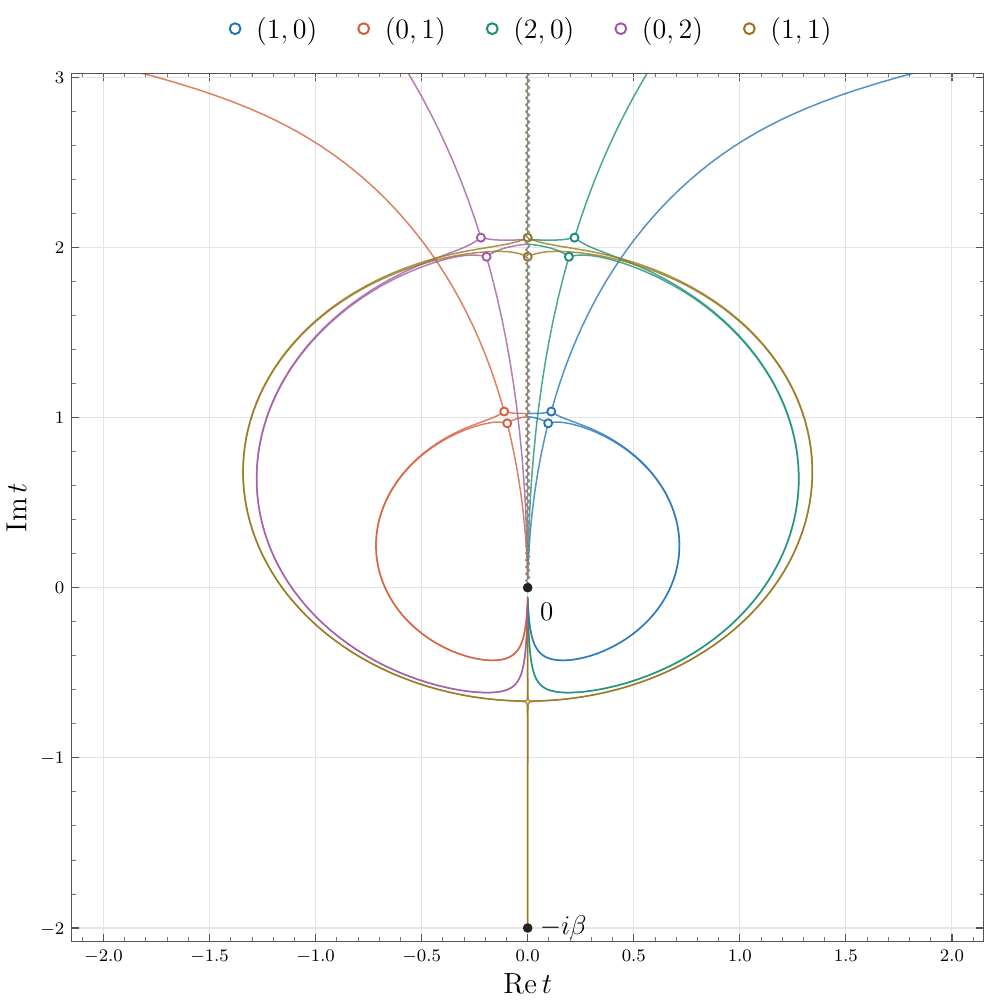}
    \caption{The generalization of the structure of Stokes lines depicted in Figure \ref{fig:global_stokes}. Each singularity forming the lattice of Figure \ref{fig:lattice} gets replaced by two Airy-like collision points, which source three Stokes lines each. These Stokes lines either travel to infinity or to the coincident  points $t=0$ and $t=-i\beta$. The color of a Stokes line labels the set of saddles that it is valid for, so that five sets of saddle points are at play in this figure. Note that the set labeled by $(1,1)$ behaves differently from the rest. It corresponds to two sets of collision points, which lie at the location of the chosen branch cut. One set is reached on approaching the branch cut from the left, and another set on approaching the branch cut from the right. As a result, the number of Stokes lines is also doubled. The parameters used in this plot are $\beta=2$, $G_N=10^{-3}$ and $\mu=1$.}
    \label{fig:lattice_stokes}
\end{figure}

This has two main implications for the structure of $G(t)$ in the complex $t$ plane. First, this means that the smoothing out of each singularity is locally modelled by a Gaussian integral of the form
\begin{align}
\label{eq:omega2_model_nm}
    G_{n,m}(t)\sim \int^\infty \d \tilde\omega \exp\left[-i\tilde\omega t + i \tilde\omega t_{n,m} +2\mu \log \tilde\omega + G_N \tilde\omega^2 \gamma_{n,m}(\beta) \right].
\end{align}
Here $\gamma_{n,m}(\beta)$ is given by\footnote{Here we used the fact that the $\tilde\omega$-dependent part of the on-shell action for a  winding goes from \eqref{eq:S_reln} to
\begin{align}
    \mathcal{I}_{(n,m)}=\mathcal{I}_{\text{OPE}}+(n+m)\left(\frac{\pi \phi_h(E_-)}{G_N}  -\frac{\pi \phi_h(E_+)}{G_N}\right)+(n-m)\left( \frac{\pi \tilde \phi_h(E_-)}{G_N} - \frac{\pi \tilde \phi_h(E_+)}{G_N}\right)-i\pi\mu(n-m)
\end{align}
for general $n$ and $m$.
This monodromy may be computed using the exact expression for the action in Appendix \ref{sec:phi2_formulas}. Winding clockwise around the lower right $\tilde\omega_1$ branch point in \eqref{eq:bp_1} increases $n$, while winding counterclockwise around the upper right $\tilde\omega_1$ branch point increases $m$.}
\begin{align}
    \gamma_{n,m}(\beta)
=
\pi \phi_h''(2\bar{E}_0)
\left[
1-
\left(
1+n+m
+
(n-m)
\frac{
\tilde{\phi}_h''(2\bar{E}_0)
}{
\phi_h''(2\bar{E}_0)
}
\right)^2
\right]
\end{align}
for general $n$ and $m$.
Note that this $(n,m)$-dependence of $\gamma_{n,m}(\beta)$ arises entirely from the saddle point in the $\bar E$ direction receiving a correction of the form $\bar E_*=\bar E_0+\# G_N\tilde\omega$ away from the probe limit.\footnote{Recall that near the smoothed-out singularity, certain cross sections of $G(t)$ look like peaks with a finite width (see Figure \ref{fig:resolution}). For large $n$ and $m$, this width gets larger as we increase $n$ and $m$. Moreover, when $n$ or $m$ are $\mathcal{O}(1/\sqrt{G_N})$, the saddle points of the $\tilde\omega$ integral instead lie at $\tilde\omega\sim \mathcal{O}(1)$ instead of $\tilde\omega\sim \mathcal{O}(1/\sqrt{G_N})$. This means that we cannot approximate the action relevant for $G_{n,m}(t)$ as a Gaussian in this case---its exact logarithmic dependence must be retained.}
The Gaussian smoothing above leads to the local Airy structure of Figure \ref{fig:local_stokes} in the vicinity of each smoothed-out singularity, which results in a lattice of Airy-like collision points in the complex $t$ plane. We can again numerically find the global behavior of the Stokes lines that emanate from these collision points, see Figure \ref{fig:lattice_stokes}. As in the case of the first $t_c$ singularity in Section \ref{sec:stokes}, Stokes lines either head to infinity or meet the coincident  points $t=0$ and $t=-i\beta$.

This naturally leads to the second implication for the analytic structure of $G(t)$. The results of Section \ref{sec:essential} say that upon crossing a Stokes line near $t=0$ with $\Im t>0$, each of the above integrals starts to receive a contribution from a saddle point that causes an essential singularity. This was shown for the first winding $G_{1,0}(t)$ in \eqref{eq:essential}. For an arbitrary winding, the relevant one-dimensional integral is instead given by
\begin{align}
\label{eq:nm_integrand}
    \int_0^\infty\, \d s_+ e^{\mathcal{J}_{{(n,m)}}^+},\quad \mathcal{J}_{{(n,m)}}^+ = -\frac{it}{2 G_N}s_+^2 -\frac{\pi}{G_N}(n+m)s_++2\mu \log(s_+^2)+\log (s_+).
\end{align}
So, the contribution of the saddle point causing the essential singularity becomes (for large $\mu$)
\begin{align}
    e^{\mathcal{J}_{(n,m)}^+}\sim t^{-4\mu} \exp\left[-\frac{i\pi^2(n+m)^2}{2 G_N t}\right].
\end{align}
Parameterizing $t=-i\epsilon e^{i\theta}$, any such saddle is picked up when $\theta=\pi$.
The full correlator is then given by an infinite sum over all of these saddles. It is interesting to consider what happens when we rotate $\theta$ further to $3\pi/2$. Similar to the discussion at the end of Section \ref{sec:essential}, all the exponentials in this sum are oscillatory at $\theta=3\pi/2$. At this stage, the sum over saddles has the form
\begin{align}
\label{eq:exp_bdry_sum}
    \sum_{n,m\geq 0,n+m\geq 1}(-\epsilon)^{-4\mu} \exp\left[\frac{i\pi^2(n+m)^2}{2 G_N \epsilon}\right].
\end{align}
\begin{figure}[t]
    \centering
    \includegraphics[width=0.8\linewidth]{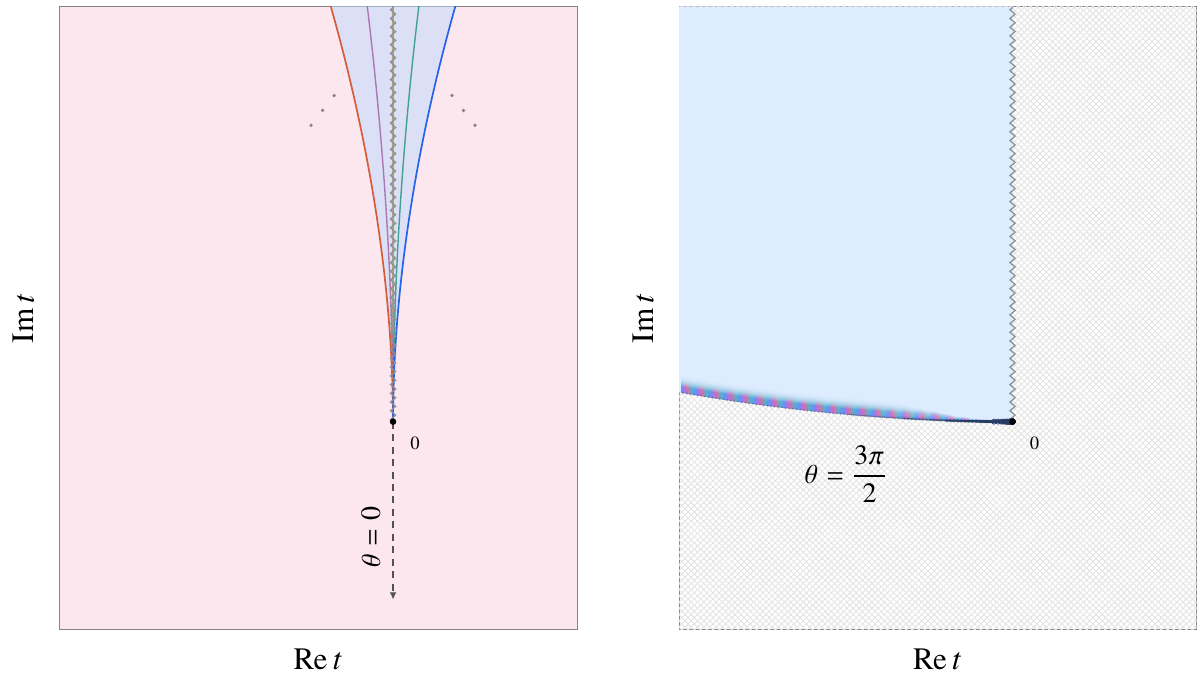}
    \caption{In the full correlator $G(t)$, an infinite number of saddles get switched on across an infinite number of Stokes lines emanating from $t=0$. The two figures above show two branches of the function $G(t)$ stitched across the branch cut on the positive imaginary axis. With $\theta=0$ lying on the negative imaginary axis on the first sheet, these saddles start to contribute when $\theta$ is almost $\pi$. They then add up to form a natural boundary of analytic continuation when $\theta$ is close to $3\pi/2$ (see right figure), which lies on the second sheet.}
    \label{fig:nat_bdry}
\end{figure}

The $(n+m)^2$ dependence of the argument of the exponential in each term in this sum indicates the existence of a natural boundary of analytic continuation\footnote{See \cite{Mizera:2022dko} for a nice pedagogical discussion of natural boundaries.} at $\theta=3\pi/2$.\footnote{The corrections to the action in \eqref{eq:nm_integrand} that are $\mathcal{O}(1)$ as $t\to 0$ also depend on $n$ and $m$. These make the natural boundary bend into the positive imaginary direction, correcting $\theta$ to $3\pi/2+\mathcal{O}(\epsilon)$. } Using the fact that the radius of convergence of the above series is one, this is implied by the Fabry gap theorem, and means that there is a dense set of singularities on the $\theta=3\pi/2$ axis. This natural boundary prevents us from accessing the region where the contribution of each saddle becomes exponentially large. See Figure \ref{fig:nat_bdry} for an illustration of this phenomenon.\footnote{Note that this sum may also include contributions from geodesic saddles obtained by winding around the $\tilde\omega_2$ branch points of \eqref{eq:bp_2}, depending on whether they contribute. Since these branch points only exist for finite $G_N$, these contributions disappear in the probe limit and do not cause any singularities. Taking them into account changes the overall coefficients of the exponentials in \eqref{eq:exp_bdry_sum} by $\mathcal{O}(1)$ constants, so that the Fabry gap theorem still applies to give a natural boundary.} Since we have derived this natural boundary from a local analysis near $t=0$, its existence is not affected by the Gaussian approximation breaking down for very high winding number.

 Note that this putative natural boundary is not a distinctive signal of the curvature divergence near the black hole singularity. Indeed, in Appendix \ref{sec:JT} we show that it is expected to exist in pure JT gravity as well; we prove that it exists for $\Delta=1/2$ without doing a saddle point approximation. Instead, the signal of the black hole singularity gets enmeshed in the fine structure of this natural boundary.

 This raises a question about higher dimensional holographic models where the bulk has a CFT dual. How does the power of locality and the  convergent OPE expansion around $t=0$ it implies affect the above conclusions?  The approach of the high energy saddle points to $t=0$ in our model arose from the fact that a high energy probe injects large amounts of energy into the black hole, and that the inverse temperature of the resulting infinite mass black hole in AdS goes to zero. So, it is reasonable to expect that such saddles would approach $t=0$ in higher dimensions as well. Assuming this is true, how does the behavior we have described here manifest itself in the dual CFT? What aspects of it are consistent with the existence of a convergent operator product expansion? In the next section we will address some of these questions in the context of a simple model.

\section{Seeing the Stokes web in 2D CFT vacuum blocks}
\label{sec:cft}
In this section we will turn our attention to another type of classical singularity that is resolved by quantum effects; the ``forbidden" singularities in AdS$_3$/CFT$_2$, which were first introduced and resolved in \cite{Fitzpatrick:2016ive,Faulkner:2017hll}.\footnote{We thank Ahmed Almheiri for emphasizing this analogy to us, and for pointing out the smoothing of such forbidden singularities due to backreaction in JT gravity \cite{AliAhmad:2026wem,Yang:2018gdb}.} These arise in correlation functions of light operators measured in individual high energy microstates:
\begin{align}
\label{eq:HHLL}
    \mathcal{A}(z,\bar z)=\ev{\mathcal O_H(\infty) \mathcal O_L(1)\mathcal O_L(z,\bar z)\mathcal O_H(0)}.
\end{align}
At infinite central charge $c$, the Eigenstate Thermalization Hypothesis implies that this should equal the correlator of these light operators in a thermal state: 
\begin{align}
    \mathcal{A}(z,\bar z)\approx\mathcal{A}_\beta(z,\bar z)= \ev{\mathcal{O}_L(1)\mathcal O_L(z,\bar z)}_\beta.
\end{align}
Thermal correlators are periodic in Euclidean time, so they contain an infinite number of periodic images of the  OPE singularity.  However, at finite $c$ the Euclidean correlator \eqref{eq:HHLL} can only have singularities at coincident points. This means that the thermal images are ``forbidden" singularities, which must be resolved at finite central charge. 

In \cite{Fitzpatrick:2016ive}, the authors argued that the smoothing out of these singularities is dictated by a Gaussian model similar to that in Section \ref{sec:first_correction}. As a useful playground for understanding this smoothing out, they considered the Virasoro vacuum block (which contains the stress tensor/gravitational physics) contributing to a four-point function 
\begin{align}
    \ev{\mathcal O_{L}(\infty) \mathcal O_L(1)\mathcal O_{r,s}(z,\bar z)\mathcal O_{r,s}(0)} 
\end{align}
involving two heavy degenerate operators $\mathcal O_{r,s}$ and two light operators of dimension $h_L$. 
(Note that we have changed the placement of the heavy and light operators relative to \eqref{eq:HHLL} to align with the notation in \cite{Fitzpatrick:2016ive}.) Here the heavy operators have large negative scaling dimension proportional to $c$. By extending their analysis, we will see a striking similarity to the results of Sections \ref{sec:smoothing} and \ref{sec:OPE_sing}.  But in this situation the analog of the $t=0$ point has a convergent OPE  around it. This demonstrates that Stokes phenomena like those discussed in Section \ref{sec:OPE_sing} are mathematically compatible with a convergent OPE.

Let us focus on $(r,s)=(2,1)$. At finite $c$, the holomorphic  vacuum block has the useful Coulomb gas integral representation
\begin{align}
\label{eq:block_int}
    \mathcal V=\frac{\Gamma(2 b^2+2)}{\Gamma(b^2+1)^2}\,z^{-2 h_{(2,1)}}(1-z)^\gamma\int_0^1\d w\, e^{\mathcal I}, \quad \mathcal I=b^2\log[w(1-w)]-2\gamma\log (1-wz),
\end{align}
where $b$ is related to the central charge,  $c=1+6\left(b+\frac{1}{b}\right)^2$. Here we defined
\begin{align}
    2\gamma=b^2+1-\sqrt{(b^2+1)^2-4b^2 h_L}\sim 2h_L,\quad b\gg 1.
\end{align}
To justify the saddle point approximation, we will consider the regime where $h_L\gg 1$ but doesn't scale with $b$, which we also take large.
The points $z=0$ and $z=1$ are analogs of the coincident  points $t=-i\beta$ and $t=0$. When $0<z<1$, the above integral converges absolutely. As we take $z$ to a real value $z_*>1$ , a branch point of the integrand at $w=1/z$ approaches the defining contour and coincides with it. This leads to a divergent integral; to analytically continue the answer we continuously deform the defining contour to avoid the branch point singularity. We obtain two different contours depending on whether we approach $z_*>1$ from $\Im z<0$ or $\Im z>0$.
The contours give different answers, indicating a branch cut for the integral on the real $z$ axis starting at $z=1$. 

We now study this integral by saddle point.
The saddle point equation for $w$ is exactly solvable, and gives rise to saddle points located at:
\begin{align}
    &\frac{1}{w}-\frac{1}{1-w}+\frac{2h_L z}{b^2(1-wz)}=0\nonumber\\
    \implies & w_\pm(z)=
\frac{
2+z\left(1-\frac{2h_L}{b^2}\right)
\pm
\sqrt{
4(1-z)+z^2\left(1-\frac{2h_L}{b^2}\right)^2
}
}{
2z\left(2-\frac{2h_L}{b^2}\right)
}.
\label{eq:forbidden_saddles}
\end{align}
Note that the full set of saddles includes an infinite number of images of the two saddles identified above, which lie on other logarithmic sheets of the action.
If we approach $z_*>1$ from $\Im z<0$, the defining contour runs from $w=0$ to $w=1$, passing below the branch point singularity at $w=1/z$. Similarly, if we approach $z_*>1$ from $\Im z>0$, the defining contour lies above it. 
In both cases the contour may be smoothly deformed to the $w_-$ saddle above, whose continuation from $0<w<1$ lies below the real $w$  axis in the first case and above the real $w$  axis in the second case.  

Now, as $b\to\infty$, the above integral develops forbidden singularities at $z=2$. We encounter this singularity regardless of whether we approach $z=2$ from $\Im z>0$ or $\Im z<0$.
This happens because as $z\to 2$, both the $w_+$ and $w_-$ saddle points of the equation above along with the branch point singularity of the integrand at $w=1/z$ approach $w=1/2$. The actions of these saddle points diverge, giving rise to the singularities.

\begin{figure}[t]
    \centering
    \includegraphics[width=0.8\linewidth]{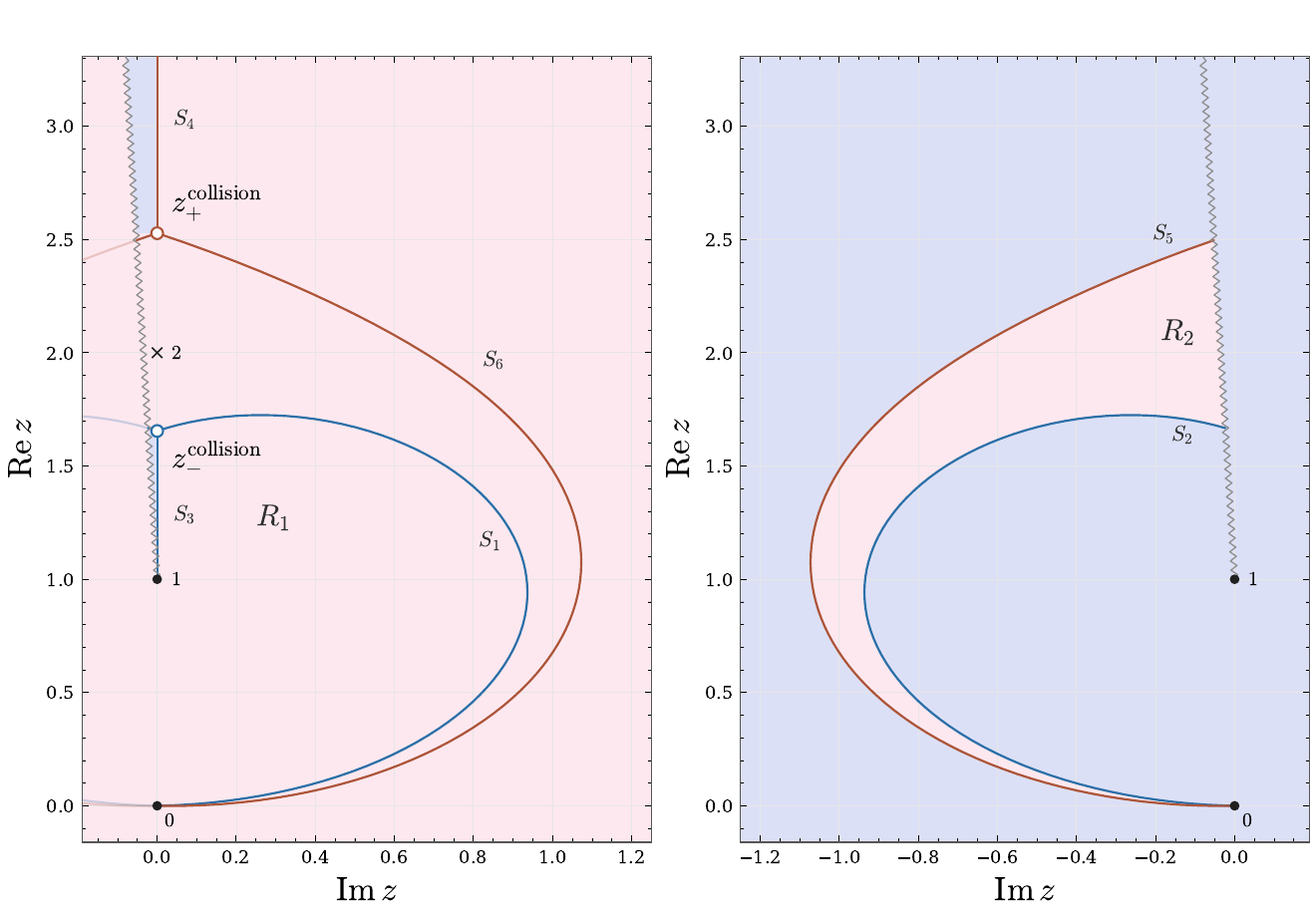}
    \caption{The global structure of the Stokes lines of the integral in \eqref{eq:block_int} in the complex $z$ plane. This structure is similar to that in Figure \ref{fig:global_stokes} (a related plot is given in \cite{Bissi:2024wur} in the large $c$ limit). In the large $c$ limit, forbidden singularities develop at $z=2$. We have rotated the branch cut emanating from $z=1$ slightly to the left to make one of these $z=2$ points visible. The left and right figures are two sheets of the same function stitched together across this branch cut. Only one saddle contributes in the regions shaded in pink; this includes the region enclosing $z=2$. Similar to the Stokes phenomenon described in Section \ref{sec:stokes}, a second saddle is picked up on crossing the Stokes line $S_3$ (the regions where two saddles contribute are shaded in blue). This saddle approaches the endpoint of the contour of integration as $z\to 1$. A forbidden singularity also occurs at large $c$ when we approach $z=2$ from the left of the branch cut in the left figure. The associated Stokes lines are pictured in a lighter color. The parameters used for this figure are $b=10$ and $h_L=1.1$.}
    \label{fig:f_stokes}
\end{figure}

At finite $b$ however, the situation is similar to that in Section \ref{sec:smoothing}. In the region $z=2+\mathcal{O}(1/b)$, the integral is dominated by the $\mathcal{O}(1/b)$  region around $w=1/2$. Setting $w=1/2+x$ and Taylor expanding  gives the Gaussian model
\begin{align}
    \int_0^1 \d w\, e^{\mathcal I}\sim\int_{-\infty}^\infty \d x\, 4^{-b^2}e^{-4b^2x^2}\left(-\frac{z-2}{4}-x\right)^{-2h_L}\sim \int_0^\infty \d p \,p^{2h_L-1} e^{\mp\frac{ip}{4}(z-2)-\frac{p^2}{16b^2}}.
\end{align}
The minus sign is used if we approach the point $z=2$ from $\Im z<0$, and the plus sign for $\Im z>0$.
To go from the left hand side to the right hand side of the above expression, we use the identity
\begin{align}
    \left(-x-\frac{\delta}{4}\pm i\epsilon\right)^{-2h_L}\propto \int_0^\infty \d p \, p^{2 h_L-1} e^{\mp i p(x+\delta/4)-\epsilon p},\quad \delta \in \mathbb{R},
\end{align}
and then do the $x$ integral.
Note we are able to extend the region of integration of the $x$ integral to $(-\infty,\infty)$ because $x\sim 1/b$ dominates the resulting integral as well. 
This Gaussian model reproduces the saddles in \eqref{eq:forbidden_saddles}, which collide in pairs at two points $w_\pm^\text{collision}$ lying in this $x\sim 1/b$ region. In particular, the contributing saddle for $z=2$ collides with two different saddles at the two different collision points. Similar to Section \ref{sec:stokes}, these collision points behave like turning points of the Airy function and source three Stokes lines each. One can again numerically solve for the global evolution of these Stokes lines, see Figure \ref{fig:f_stokes}. We see again that Stokes lines either approach infinity, or the points $z=0$ and $z=1$ that correspond to the OPE limit. In particular, one Stokes line approaches the point $z=1$ from the direction of the $z=2$ region; this is the analog of the Stokes line $S_3$ in Figures \ref{fig:local_stokes} and \ref{fig:global_stokes}.

The three saddles involved behave in a way that is similar to our analysis of the bouncing geodesic singularity. 
As in Section \ref{sec:first_correction}, in the region including $z=2$ in Figure \ref{fig:f_stokes}, only one saddle point contributes. The $z=2$ point noted in the figure lies to the right of the branch cut, and the contributing saddle is the $w_-$ saddle of \eqref{eq:forbidden_saddles}.
As we move from this region into region $R_1$ by crossing the Stokes line $S_1$, the real part of the action of the contributing saddle is smaller, so that the other saddle participating in the collision at $z=z_-^{\rm collision}$ does not get picked up. So, to the right of the Stokes line $S_3$, only one saddle contributes. 

On crossing $S_3$, however, the action of this other saddle has a smaller real part. This saddle gets picked up, so that two saddles contribute to the left of $S_3$. One of the saddles contributing in this region approaches the endpoint of the integration contour $w=1$. Near $z=1$, the contribution of this saddle behaves as
\begin{align}
\label{eq:z0}
    e^{\mathcal{I}}\sim (1-z)^{b^2-2h_L}.
\end{align}
(Note that this is multiplied by the additional factors in \eqref{eq:block_int} to obtain the contribution to the conformal block $\mathcal V$.) The more precise value of the exponent in the above equation is given by $b^2+1-2\gamma$, where the $+1$ comes from one loop effects.

This Stokes phenomenon is the analog of the one studied in Sections \ref{sec:stokes} and \ref{sec:essential}.
It provides a mathematical example of how this phenomenon can occur near a point with a convergent OPE-like expansion. Indeed, the block $\mathcal V$ may be written exactly in terms of hypergeometric functions, which near $z=1$ has the convergent Frobenius expansion 
\begin{align}
\mathcal V
\sim
z^{-2h_{(2,1)}}\left[
(1-z)^\gamma C_-(1-z)
+
 (1-z)^{b^2+1-\gamma} C_+(1-z)
\right],
\quad z\to1.
\end{align}
where $C_-, C_+$ are convergent power series in $1-z$ (these two terms are just two independent solutions of the hypergeometric differential equation). Note that the presence of a Stokes line ending at $z=1$ is not in tension with this convergent expansion because of ``Stokes smoothing" \cite{Berry1988}.\footnote{The connection between the forbidden singularity saddle point and the convergent Frobenius expansion is particularly clear in the Mellin-Barnes representation of he hypergeometric function \cite{paris2001asymptotics}.   Here deforming the contour in one direction picks up residues that form the terms in the Frobenius expansion.  Alternatively one can deform the contour into the steepest descent contour passing through the forbidden singularity saddle.}

One can ask whether this example teaches us any lessons about the fate of the bouncing geodesic singularity in higher dimensions.  In particular, if we are able to follow the bouncing geodesic saddle as in our 2D analysis what happens to it near the $t=0$ OPE point?  Although we can't give a concrete answer to this question, in the rest of this section we will outline some intermediate questions that may help understand the answer to it.

\begin{enumerate}
    \item In the example considered in this section the saddle point that joins the contour is associated with operators that contribute to the OPE. For instance, the behavior near $z=1$ in \eqref{eq:z0} is associated with the contribution of the operator $V_{\alpha_L+b/2}$\footnote{Here $\alpha_L=\gamma/b$.} to the OPE between $\mathcal{O}_{2,1}$ and the light operator (recall that $z=1$ is the heavy-light OPE point). This operator has large positive scaling dimension. 
    
    Similarly, 
    we can track this saddle further beyond the Stokes phenomenon near $z=1$. Recall that it joins the contour on crossing the Stokes line $S_3$ on the left hand side of Figure \ref{fig:f_stokes}. Let us now move $z$ towards $z=0$ along a path on the second sheet of the block (see the right hand side of Figure \ref{fig:f_stokes}). As we do so, the contribution of this saddle point grows, so that near $z=0$ its contribution behaves as
    \begin{align}
        e^{\mathcal I}\sim z^{-2b^2-1}.
    \end{align}
    Here the extra $-1$ is a one loop effect. Including the  $z^{-2h_{(2,1)}}$ prefactor in \eqref{eq:block_int}, the contribution of this saddle to the block $\mathcal{V}$ blows up as $z^{-b^2/2}\sim z^{h_{(3,1)}-2h_{(2,1)}}$, which is the behavior expected from the contribution of $\mathcal O_{3,1}$ to the $\mathcal{O}_{2,1}\times \mathcal{O}_{2,1}$ OPE (recall that $z=0$ is the heavy-heavy OPE point). 
    
    Based on this relation of saddles to operators contributing to the OPE, one might speculate that the infinite number of saddle points picked up near $t=0$ in the bouncing geodesic example correspond in some way to operators exchanged in the light-light OPE in the full CFT in the thermal state. Can this speculation be checked more concretely? 
    
    \item A downside of the  degenerate block example is that the Stokes phenomenon happens near the heavy-light OPE point instead of the light-light OPE point. Moreover, it lacks the presence of an infinite lattice of singularities. For this reason, it may be helpful to instead consider the more physical example of the large $c$ Virasoro vacuum block where the heavy operators are not degenerate and have large positive $h_H>c/24$. This block exhibits an infinite number of forbidden singularities---the authors of \cite{Fitzpatrick:2016ive} provide evidence that the Gaussian model of smoothing of is expected to apply here as well. Near the forbidden singularity at $z=2$, the vacuum block is expected to behave as
    \begin{align}
        \int_0^\infty \d p\, p^{2 h_L-1} \exp\left[-ip(z-2)-\frac{\sigma_n^2}{2b^2}p^2\right].
    \end{align} 
    An important distinction from the degenerate operator case is that $\sigma_n^2$ is expected to be negative here, and the contour for $p$ has to be rotated to define the integral.
    This implies that the two collision points surrounding the location of the forbidden singularity (see $z_+^{\rm collision}$ and $z_-^{\rm collision}$ in Figure \ref{fig:f_stokes}) lie along the imaginary $z$ direction instead. 
    
    By symmetry we expect two Stokes lines to extend from these points towards $z=1$, which here is the light-light OPE point (see Figure \ref{fig:expected_stokes}). It would be interesting to understand the global behavior of the saddle points colliding at the two points near $z=2$ in this case. Since the block possesses a lattice of smoothed-out singularities, each with their own structure of Stokes lines, the story here could be more subtle. In particular, if these Stokes lines have saddles in common, then parts of them could switch off at points where they cross \cite{Howls:2004}.

    \begin{figure}[t]
        \centering
        \includegraphics[width=0.3\linewidth]{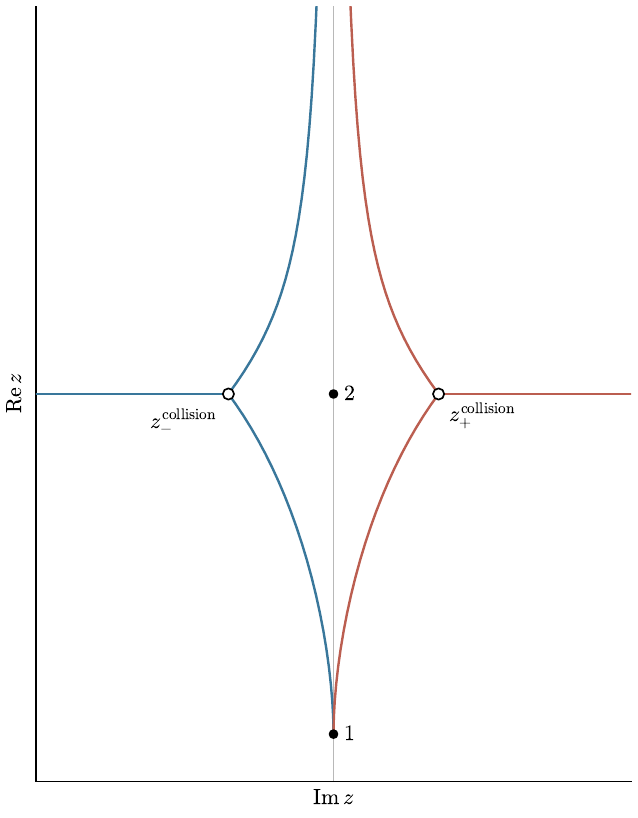}
        \caption{The expected pattern of Stokes lines for the large $c$ Virasoro vacuum block. Note that two Stokes lines asymptote to $z=1$ from above, instead of one.}
        \label{fig:expected_stokes}
    \end{figure}
    
    Non-vacuum blocks with intermediate $h_I>0$ each have their own forbidden singularities as well \cite{Fitzpatrick:2016ive}, and hence an associated structure of Stokes lines at finite but large $c$. How does adding up contributions from all blocks in a full correlator like \eqref{eq:HHLL} change the structure of Stokes lines? One place to explore this could be in the context of Liouville CFT, where the HHLL four-point function also exhibits forbidden singularities
    \cite{Balasubramanian:2017fan}. Since all the ingredients for computing the full four-point function are known in principle, one could try to generate the Stokes web numerically and study the result.  This example has the shortcoming, though, that it contains no vacuum block.
    
    \item Another way to approach the large $c$ vacuum block is by using the tools of resurgence, as considered in \cite{Benjamin:2023uib} (see also \cite{Bissi:2024wur}). This paper also analyzes an analogous extra saddle in a block with four degenerate external operators, and points out its relation to $\mathcal{O}_{3,1}$. It then extends this analysis to the general large $c$ vacuum block using Zamolodchikov's recursion relations to generate an asymptotic expansion for the block in $1/c$. The additional saddles appear as singularities in the Borel plane.  However, they find that these saddles correspond to the exchange of an infinite number of primaries with large \textit{negative} conformal dimension.\footnote{Note that this infinite number of saddles was first noticed in \cite{Fitzpatrick:2016mjq}, and also derived in \cite{Faulkner:2017hll} using the monodromy method.} This means that they do not appear as operators in the OPE of a unitary CFT. As the authors of \cite{Benjamin:2023uib} also emphasize, this is in contrast to the case of the degenerate block, where the additional saddle corresponds to an actual operator exchanged in the OPE. It would be interesting to understand the significance of this phenomenon for the bouncing geodesic problem.
    
    \item Another controlled setting that could provide insight  is the theory obtained by taking a kind of ``near-extremal" limit of a large $c$ 2D CFT \cite{Ghosh:2019rcj}. This limit is connected to the Schwarzian theory. This may be seen from the two-point function of light operators in the grand-canonical ensemble, which has the form
    \begin{align}
        G^{(\mathrm{CFT})}
\left(z,\bar z;\beta_L,\beta_R\right)
\sim
\left(\frac{c}{6}\right)^{-2h}
\left[
\frac{1}{\pi T_R}
\sinh\left(\pi T_R\bar z\right)
\right]^{-2\bar h}
G_h^{(\mathrm{Schw})}(-z),
    \end{align}
    where $G_h^{(\mathrm{Schw})}(-z)$ is the two-point function in the Schwarzian theory.
    Here $\beta_L=1/T_L$ and $\beta_R=1/T_R$ are linear combinations of the temperature and the chemical potential for  angular momentum, and $h$ and $\bar h$ are the holomorphic and anti-holomorphic conformal weights of the light operator. The above expression is valid in a regime where $T_L$ is $\mathcal{O}(c^{-1})$, and $T_R$ is very large. The Schwarzian theory also determines the JT gravity correlator of Appendix \ref{sec:JT}, which indicates that the above expression possesses a natural boundary. 
    The $G_h^{(\mathrm{Schw})}(-z)$ factor appears from the large $c$ limit of the torus Virasoro vacuum block --- studying what happens to this Schwarzian factor away from the large $c$ limit could help elucidate whether and how the natural boundary goes away in higher bulk dimensions.  
\end{enumerate}

\section{Preliminary analysis of loop corrections}
\label{sec:diagrams}
In this section we outline our preliminary understanding of loop corrections in $G_N$ that are not captured by backreaction.\footnote{We thank Simon Caron-Huot and Douglas Stanford for emphasizing the role of the diagrammatic point of view to us.}  We hope to give a more careful and complete treatment  in a follow-up paper \cite{BatraShenker:WIP}. 
Our tentative conclusions are as follows.

 For both of the dilaton potentials we have considered (see \eqref{eq:potentials}), local power counting suggests that the Gaussian model derived in Section \ref{sec:first_correction} is robust to loop corrections
in its regime of validity, $\tilde\omega\sim 1/G_N^k$, $k<1$, appropriate to $t$ near $t_c$. Outside of this region (corresponding to behavior near the OPE point $t=0$) the importance of the corrections depends on the potential. 

For the $1/\phi^3$ potential, corrections organize into an expansion in $G_N \tilde\omega$ (along with inverse powers of $\tilde\omega$). These do not compete with the $G_N\tilde\omega^2$ correction discussed in Section \ref{sec:first_correction} near $t=t_c$.  But these corrections are potentially large near the OPE points where $G_N \tilde\omega$ becomes large. 

For the $\exp(-\phi)$ potential, corrections superficially organize into an expansion in $G_N$ alone. Apart from not competing with the $G_N\tilde\omega^2$ term, these also do not affect the large $\tilde\omega$ behavior that determines the form of the correlator near $t=0$. However, there are possible infrared issues that need to be treated more carefully. 

We have tried to assess the superficial size of corrections by scaling arguments.  We have not attempted to choose field integration contours and boundary conditions, handle light cone singularities, deal with gauge-fixing or evaluate the sizes of individual diagrams.

We will work in conformal gauge, writing the metric as\footnote{Note that $e^{2\rho}$ here is negative, so that $z$ (the tortoise coordinate corresponding to $\phi$) is timelike.}
\begin{align}
\label{eq:conformal_metric}
    g_{\mu\nu}=e^{2\rho} \eta_{\mu\nu}\d x^\mu \d x^\nu=e^{2\rho}(-\d\tau^2+\d z^2).
\end{align}
We would like to understand the $G_N$ and $\tilde\omega$ scaling of Feynman diagrams that compute corrections to the action $\mathcal{I}$ that go beyond classical backreaction. These corrections come from fluctuations of the conformal factor $\rho\equiv \bar \rho+\sigma$, the dilaton $\phi\equiv \bar\phi+\varphi$ and the particle worldline $X^\mu$. Here $\tilde\omega$ plays the role of the energy of the point particle, whose turning point lies in the near-singularity region and hence scales with $\tilde\omega$.

In conformal gauge we may write the action as\footnote{Here we have gauge fixed the einbein $e$ in the point particle action to one.}
\begin{align}
    \mathcal{I}&=\mathcal{I}_{\rm{grav}}+\mathcal{I}_{\rm{pp}}+\mathcal{I}_{\rm{ghosts}},\\
    \label{eq:Igrav_conformal}\mathcal{I}_{\rm{grav}}&=\frac{1}{2G_N}
\int \d\tau \d z
\left[
-2\phi\,\hat\nabla^2\rho
+
e^{2\rho}U(\phi)
\right],\\
\mathcal{I}_{\rm pp}
&=
\frac12\int \d\lambda
\left[
e^{2\rho(X)}\delta_{\mu\nu}
\dot X^\mu\dot X^\nu
+\mu^2
\right],
\end{align}
where $\hat\nabla^2\equiv -\partial_\tau^2+\partial_z^2$.
As derived in Section \ref{sec:backreaction}, the classical saddle point solution for the metric is defined piecewise. Depending on the side of the particle worldline, in the near-singularity region we have
\begin{align}
    e^{2\bar\rho_\pm}=W(\bar\phi(z_\pm))-E_\pm,\quad W(\bar\phi(z_\pm))\approx -e^{-\bar\phi(z_\pm)} \text{ or } -\frac{1}{\bar\phi(z_\pm)^2}.
\end{align}
Here we will focus only on fluctuations of the conformal factor and the dilaton.\footnote{We study the effects of the fluctuations of the particle worldline in Appendix \ref{sec:feynman_rules} and show in a similar local scaling analysis that while the fixed geodesic picture breaks down near the singularity, fluctuations around it may be resummed so that they are well-controlled at large $\tilde\omega$.} 
Expanding around the saddle point above, we find that these fluctuations are governed by the action
\begin{align}
\mathcal I^{\mathrm{fluc}}&=\mathcal I_{\mathrm{grav}}^{\mathrm{fluc}}+\mathcal I_{\mathrm{pp}}^{\mathrm{fluc}},\\
\mathcal I_{\mathrm{grav}}^{\mathrm{fluc}}
&={}
\frac{1}{2G_N}
\sum_{\alpha=\pm}
\int_{\mathcal M_\alpha}
\d\tau_\alpha\,\d z_\alpha
\left[
2\partial_\mu\varphi_\alpha\,
\partial^\mu\sigma_\alpha
+e^{2\bar\rho_\alpha}
\sum_{p=2}^{\infty}\frac{1}{p!}
\sum_{n=0}^{p}
\binom{p}{n}
(2\sigma)^{p-n}\varphi^n
U^{(n)}(\bar\phi)
\right],\\
\mathcal I_{\mathrm{pp}}^{\mathrm{fluc}}
&=
\frac12
\int_{\gamma}\d\lambda\,
e^{2\bar\rho(\bar X)}
\eta_{\mu\nu}
\dot{\bar X}^{\mu}\dot{\bar X}^{\nu}
\sum_{k=2}^{\infty}
\frac{\bigl(2\sigma_\gamma\bigr)^k}{k!},
\quad
\sigma_\gamma\equiv\sigma\bigl(\bar X(\lambda)\bigr).
\label{eq:I_fluc_pp}
\end{align}
Here $U^{(n)}$ denotes the $n^{\rm th}$ derivative of $U(\phi)$.

We want to focus on the region near the turning point of the geodesic. This is the region where $\tau$ and $z$ are $\mathcal{O}(1/\tilde\omega^3)$ (for the $1/\phi^3$ potential) or $\mathcal{O}(1/\tilde\omega^2)$ (for the exponential potential). Moreover in this region, the metrics on the two sides of the geodesic are equal in the large $\tilde\omega$ regime that we are considering here.

Let us first consider the $1/\phi^3$ potential. It is simplest to go to coordinates $(\tilde\tau,\tilde z)$ where the vicinity of the singularity is at an $\mathcal{O}(1)$ location. To make propagators $\mathcal{O}(1)$, we also define $\sqrt{G_N}\tilde\sigma=\tilde\omega^{-1/2} \sigma$ and $\sqrt{G_N}\tilde\varphi=\tilde\omega^{1/2} \varphi$. After plugging in the way that $U^{(n)}(\bar\phi)$ and $e^{2\bar\rho_\alpha}$ scale with $\tilde\omega$ when $\tilde\tau$ and $\tilde z$ are $\mathcal{O}(1)$, and using the fact that $\d\tau/\d\lambda\sim \tilde\omega^{-1}$, we obtain an action
with terms that scale in the following way:
\begin{align}
\mathcal I_{\rm grav}^{\rm fluc}
\sim{}&
\frac12
\sum_{\alpha=\pm}
\int
\d\tilde\tau_\alpha\,
\d\tilde z_\alpha
\Bigg[
2\tilde\partial_\mu\tilde\varphi_\alpha\,
\tilde\partial^\mu\tilde\sigma_\alpha
+
\sum_{p=2}^{\infty}
\sum_{n=0}^{p}
(G_N\tilde\omega)^{p/2-1}
(2\tilde\sigma_\alpha)^{p-n}
\tilde\varphi_\alpha^n
\Bigg],\\
\mathcal I_{\rm pp}^{\rm fluc}
\sim{}&
\frac{1}{2\tilde\omega^2}
\int_{\tilde\gamma}
\d\tilde\tau
\sum_{k=2}^{\infty}
\frac{
(G_N\tilde\omega)^{k/2}
(2\tilde\sigma_\gamma)^k
}{k!}.
\end{align}
(Here we also used the mass shell condition $e^{2\bar\rho(\bar X)}
\eta_{\mu\nu}
\dot{\bar X}^{\mu}
\dot{\bar X}^{\nu}\sim \mathcal{O}(1)$.) From the above action it is clear, at least in terms of scaling, that the fluctuations around the saddle point organize into a series in $G_N\tilde\omega$. The only $G_N$ and $\tilde\omega$ dependencies come from the vertices, since propagators are $\mathcal{O}(1)$ if $\tilde\tau$ and $\tilde z$ are $\mathcal{O}(1)$.\footnote{Again, we assume that field integration contours can be appropriately rotated and boundary conditions can be chosen appropriately.} We need not consider regions where $\tilde\tau$ and $\tilde z$ are larger --- in any diagram contributing to corrections to the correlator, at least one vertex must be anchored on the particle worldline. Propagators that extend out to a larger region where  $\tilde\tau$ and $\tilde z$ scale with $\tilde\omega$ then decay exponentially with  $\tilde\omega$.

Now let us turn to the exponential potential. Following similar reasoning to the above, the action for the metric and dilaton fluctuations scales like
\begin{align}
\mathcal I_{\rm grav}^{\rm fluc}
\sim{}&
\frac{1}{2}
\sum_{\alpha=\pm}
\int
\d\tilde\tau_\alpha\,
\d\tilde z_\alpha
\Bigg[
2\tilde\partial_\mu\tilde\varphi_\alpha\,
\tilde\partial^\mu\tilde\sigma_\alpha
+
\sum_{p=2}^{\infty}
\frac{G_N^{p/2-1}}{p!}
\left(
2\tilde\sigma_\alpha-\tilde\varphi_\alpha
\right)^p
\Bigg],
\\
\mathcal I_{\rm pp}^{\rm fluc}
\sim{}&
-\frac{\mu^2}{2\tilde\omega}
\int_{\tilde\gamma}
\d\tilde\tau
\sum_{k=2}^{\infty}
\frac{
G_N^{k/2}
\left(2\tilde\sigma_\gamma\right)^k
}{k!}.
\label{eq:exponential-pp-fluc}
\end{align}
Here we defined $\sqrt{G_N}\tilde\sigma =\sigma$ and $\sqrt{G_N}\tilde\varphi =\varphi$.
At first sight this indicates that the fluctuations around the saddle point for the exponential potential organize into an expansion in $G_N$, and so are well-controlled at large $\tilde\omega$. However, the existence of a massless mode makes this case more subtle. In particular, the combination $\eta \equiv 2\tilde\sigma+\tilde\varphi$ has no mass term. This is easy to see by writing \eqref{eq:Igrav_conformal} in terms of a free mode and a timelike Liouville mode \cite{Kruthoff:2024gxc}:
\begin{align}
    \mathcal{I}_{\rm grav}\sim\frac{1}{2 G_N}\int \d\tau \d z \left[\frac{1}{4}(\partial \zeta)^2-\frac{1}{4}(\partial\chi)^2+e^{-\chi}\right],\ \zeta=\phi+2\rho, \ \chi=\phi-2\rho.
\end{align}
$\eta$ then represents the fluctuation around the classical value of $\zeta$.
This means that the propagators in which $\eta$ participates do not decay exponentially with $\tilde\omega$. For this reason, we cannot restrict vertices of diagrams involving $\eta$ to just the region where $\tilde\tau$ and $\tilde z$ are $\mathcal{O}(1)$, and need to understand enhancements from integration over the range $(\tilde\tau,\tilde z)\in (1,\tilde\omega^2)$. The upper limit of this range is the $\mathcal{O}(1)$ value of $\phi$ where the part proportional to $2\phi$ of $U(\phi)$ starts to compete with $e^{-\phi}$.

We have not analyzed this full problem, but we can argue that the infrared is well-controlled in a simpler system that has some of the essential features. The action describing this system is
\begin{align}
\label{eq:etamodel}
    \mathcal{I}^{\rm fluc}_\eta\sim  \int \d\tau\, \d z \,(\partial_\mu\eta)^2+\frac{1}{\tilde\omega}
\int_{\gamma}
\frac{\d z}{z}\,
e^{\sqrt{G_N}\eta},
\end{align}
where the $z$ and $\tau$ coordinates range from $0$ to an $\mathcal{O}(1)$ value in the kinetic term. In the second term, the $z$ coordinate ranges from $1/\tilde\omega^2$ to $1$, the lower limit being the turning point of the geodesic. The $\eta$  interactions only appear in the part of the action \eqref{eq:exponential-pp-fluc} that couples gravitational fluctuations to the particle worldline, which is why it suffices to consider only this part in the above action. The $dz/z$ coefficient in the second term reflects the saddle point form of the worldline action along the geodesic for the exponential potential.

We can now estimate the effective size of the interaction term in \eqref{eq:etamodel} by an RG argument. The scaling dimension of the operator $e^{2\sqrt{G_N}\eta}$ is small, $\sim \alpha G_N, ~\alpha \sim 1$. Since the ratio of the upper and lower $z$ limits is $\sim 1/\tilde\omega^2$ and $dz/z$ is scale invariant,  the size of the interaction term in \eqref{eq:etamodel} scales like $(\tilde\omega^{2 \alpha G_N})/\tilde\omega$. So the IR enhancements of this interaction do not overwhelm its small coefficient.

A proper analysis of the above problem requires patching the two geometries with energies $E_+$ and $E_-$ on both sides of the worldline, with $E_+\sim G_N\tilde\omega$ at large $\tilde\omega$. In the region where $\phi$ is of $\mathcal{O}(1)$, we can no longer approximate the two sets of tortoise coordinates $(\tau_+,z_+)$ and $(\tau_-,z_-)$ to be the same, and the range of the $z_+$ coordinate caps off at $1/(G_N\tilde\omega)$. We might not expect this to ruin the $\tilde\omega$ scaling above since this region shrinks at large $\tilde\omega$, but additional care is required to fully analyze this situation. In future work \cite{BatraShenker:WIP} we plan to study the details of this patching more carefully.

\section{Discussion}
\label{sec:disc}
In this paper we have analyzed the effect of gravitational backreaction on the bouncing geodesic singularity in 2D dilaton gravity.   We have found that these effects smooth out the divergence of the correlator at $t_c$, but they do not resolve the black hole singularity.
 Instead, they move the imprint of the black hole singularity to the OPE points in $G(t)$ at $t=0, - i \beta$. At these points the imprint is hidden under the OPE divergence, which persists at finite $G_N$.
 
 The imprint first appears at the OPE points \textit{outside} the physical strip, where the contribution of the saddle corresponding to the bouncing geodesic is picked up in a Stokes phenomenon. At this stage, this signal of the singularity is exponentially small (for $t\to 0$, it behaves as $\sim \exp(-{1}/{G_N t})$\,).\footnote{The exponential potential behavior has an additional correction, see \eqref{eq:appdcorr}.} It becomes order one as one continues around $t=0$ until one encounters a natural boundary.   We emphasize that the natural boundary is not a signature of the singularity -- it is already present in JT gravity, for example.   The signature is a more subtle one.  

The bulk picture of backreaction makes it clear that these effects are not enough to truly resolve the black hole singularity.  The inputs to the calculation are still classical geometries with divergent curvatures at the singularity. Instead, what this analysis does is to refocus the question about the imprint of the black hole singularity in the $t$ plane from the behavior near $t_c$ to the behavior near the OPE points. 

Gravitational backreaction is not the only quantum gravitational effect in these 2D models, though.    We have given a preliminary analysis of higher order perturbative corrections in $G_N$.  This analysis suggests that for the $\exp(- \phi)$ potential these corrections are subleading for all $t$ even near the OPE points.  This would indicate that the black hole singularity is not resolved in this model, at least to all orders in $G_N$.

For the $1/\phi^3$ potential (and others with power law behavior) perturbative corrections seem to go out of control at the OPE points.   2D dilaton gravity is a UV complete model of quantum gravity, at least at fixed topology, so the nature of the true finite $G_N$ correlator behavior near the OPE points is a well-posed, and interesting, question.  For this reason, it might  serve as a good setup for studying other more physically natural observables such as the measurements made by an infalling observer close to the singularity.

 One of the main virtues of such 2D gravity models is the ability to understand the role of more complicated topologies.  At first glance one might think that these effects would be enhanced near the black hole singularity.  In the $\exp(-\phi)$ potential, for example, the dilaton $\phi$ approaches $- \infty$ at the singularity. In this region the topological term in the action, $\int \phi R$, would strongly favor higher genus configurations localized there.  But to be on-shell in these theories a configuration must have $U(\phi)=0$ (or have a Killing vector -- see \eqref{eq:metric_eom}). This is far from the case for such higher genus geometries, so such near-singularity wormholes would be far off-shell and give highly suppressed contributions.\footnote{We thank Douglas Stanford for explaining this to us.} Since exponentially suppressed contributions do play a role in our story it would still be interesting to calculate the behavior of $G(t)$ on the handle-disk topology.

From the boundary perspective higher topologies encode the fluctuations in the random matrix ensemble of boundary Hamiltonians \cite{Saad:2019lba}.  A natural question to ask is what remains of the analytic structure discussed here if we take a single draw from the ensemble.\footnote{ We would imagine fixing the operator matrix elements using the disk bulk calculation along the lines of \cite{Jafferis:2022uhu,Iliesiu:2024cnh}.}  Such random series typically have natural boundaries at the edge of their radius of convergence.\footnote{Even a simple nonrandom example with discrete spectrum like the anharmonic oscillator has such a natural boundary. We thank Yiming Chen for pointing this out, building on discussions with Matthew Dodelson, Douglas Stanford and Zhenbin Yang.}   There is a wormhole way of seeing this.  First consider the partition function $Z(\beta) = \sum_n \exp(-\beta E_n)$ where $E_n$ are drawn from a (double-scaled) random matrix ensemble with JT type large $E$ density of states.   This series converges for $\Re \beta >0$, but has a natural boundary along the line $\Re \beta = 0$.  To see this compute the  variance of $Z$ in the ensemble, $\langle Z(\beta) Z(\beta^*)\rangle_c$.  At large $S_0$ the leading contribution is given by the double trumpet 
 \begin{equation}
     \langle Z(\beta) Z(\beta^*) \rangle_c =  \frac{\sqrt{\beta \beta^*}}{2 \pi (\beta + \beta^*)} .
 \end{equation}
 This diverges along the whole line $\Re \beta = 0$, suggesting that the typical series diverges there.  This suggests a natural boundary for the typical series.  We can apply a similar argument\footnote{developed in a discussion with Douglas Stanford.} to the variance of $G(t)$.   This diverges for $\Im t$ (strictly) outside of the physical strip, suggesting a natural boundary at the edge of the strip for a typical member of the ensemble.  Note that this natural boundary is much more severe than the one in the ensemble averaged case discussed in Section \ref{sec:windings}.  Here the natural boundary prevents any continuation outside the physical strip.

The most interesting question this work raises is to what extent these phenomena occur in higher dimensional holographic CFTs like $N=4$ SYM.  Here the convergence of the OPE expansion (a correlate of spatial locality) implies that we can continue in $t$ at least some distance outside the physical strip even without ensemble averaging. In fact Caron-Huot, Simmons-Duffin and Stanford \cite{DS_DSD_SCH:WIP} have been able to argue that in any CFT with a finite number of degrees of freedom there can be no singularity of $G(t)$ in the region of the $t$ plane that includes the $t_c$ point.\footnote{Zohar Komargodski informed us that he has a similar result.} If a bulk mechanism similar to what has been discussed here applies in these systems, it is natural to ask whether a subtle signature of the singularity behavior moves to the OPE points. 

In CFTs, though, the convergent operator product expansion rules out the more exotic behavior (essential singularities/natural boundary) found in the simpler models we discuss.  Could there be some imprint of the (resolved) singularity in the high order terms of the convergent OPE, and if so what is it? In this case could one also study the Gaussian smoothing of the singularity, and track the pattern of Stokes lines from points near $t=t_c$ to the OPE points? The forbidden singularity story for the conformal block discussed in Section \ref{sec:cft} shows that it is possible for such a singularity to smoothly deform into a subleading term in a convergent OPE.\footnote{With the caveat that this occurs in the heavy-light channel rather than the light-light one we are most interested in for the small $t$ behavior of $G(t)$. As we discussed in Section \ref{sec:cft}, for the large $c$ vacuum block the Stokes lines may extend to the light-light OPE channel instead.} A Stokes phenomenon of the kind we found that picks up the contribution of the bouncing geodesic  near $t=0$ can occur around a point with convergent OPE. 
A natural first step to address these questions would be to study backreaction in the higher dimensional context, perhaps building on \cite{Batra:WIP}.\footnote{A higher dimensional probe with simple 2D kinematics is a brane wrapped on the $S^3$ of AdS$_5$. Unfortunately the $r$ dependent mass of the effective 2D world line, $m(r) \sim r^3$, eliminates the nearly null bouncing geodesic.}

This work underlines the "UV-UV" nature of the bouncing geodesic probe \cite{Festuccia:2005pi} -- high energies in the boundary theory probe short distances in the bulk (rather than the long ones of the standard UV-IR correspondence).   This makes the role of bulk degrees of freedom necessary for the UV completion of gravity, like strings and branes, and their imprint on $G(t)$ an especially important target for study. Understanding the impact of stringy effects even at $N \to \infty$ is another natural first step.
We hope to explore the above questions further in future work.

\section*{Acknowledgements}
We are grateful to Ahmed Almheiri, Simon Caron-Huot, Yiming Chen, Matthew Dodelson, Alex Frenkel, Tom Hartman, Luca Iliesiu, Zohar Komargodski, Guanda Lin, Henry Lin, Hong Liu, Yuhang Liu, Alex Maloney, Alessio Miscioscia, Mukund Rangamani, Douglas Stanford,  Haifeng Tang, Wayne Weng and Zhenbin Yang for helpful discussions. We also thank Tom Hartman, Mukund Rangamani and Douglas Stanford for comments on the draft.
GB and SHS were supported in part by NSF Grant PHY-2310429. AL was supported by DOE grant DE0SC0025937 and the Templeton Foundation Agreement 41001491-013, \#63670. 
SHS thanks the Aspen
Center for Physics for hospitality, supported by NSF grant PHY-2210452. 
This work used and benefited from AI input. We thank OpenAI for providing ChatGPT Pro access.

\appendix

\section{$G(\tilde\omega)$ from the wave equation and the sum over windings}
\label{sec:qnms}
In this section we will derive an expression for the Wightman function $G(\tilde\omega)$ in frequency space in the probe approximation, using the wave equation. Our goal is to determine which saddles to include in the sum in \eqref{eq:saddle_sum} (and consequently \eqref{eq:sum2}) at high energies. We will do so by studying the asymptotics of $G(\tilde\omega)$ for large $\tilde\omega$, which includes an infinite set of non-perturbative corrections that correspond to the sum over these saddles.  We will follow  the method used in  \cite{Festuccia2007BlackHoleSingularities} for the AdS$_5$-Schwarzschild black hole, and focus on the $1/\phi^3$ potential.

Our starting point is the wave equation for a massive scalar $\psi$ in the bulk geometry in \eqref{eq:metric}:
\begin{align}
    \partial_\phi\left(f(\phi)\partial_\phi\psi\right)+\left(\frac{\tilde\omega^2}{f(\phi)}-m^2\right)\psi=0.
\end{align}
Similar to \cite{Festuccia2007BlackHoleSingularities}, we make the following coordinate and field redefinitions:
\begin{align}
    \phi(\rho)^2&=\phi_h^2\cosh^2\rho-\tilde\phi_h^2\sinh^2\rho,\nonumber\\
    \psi&=\frac{1}{(E^2+4)^{1/4}}\left(\frac{2\phi(\rho)}{\sinh(2\rho)}\right)^{1/2}u(\rho),
\end{align}
and define
\begin{align}
    \nu=\sqrt{m^2+\frac{1}{4}},\ h=\frac{\tilde\omega\beta}{\pi},\ \tilde h=\frac{\tilde\omega\tilde\beta}{\pi}.
\end{align}
This brings the wave equation to the form
\begin{align}\label{eq:wave_eq}
    u''(\rho)+\left(\frac{h^2+1}{4\sinh^2\rho}+\frac{\tilde h^2-1}{4\cosh^2\rho}-\nu^2+\frac{\phi_h^2+\tilde\phi_h^2}{4\phi(\rho)^2}+\frac{5}{4 \phi(\rho)^4}\right)u(\rho)=0.
\end{align}
Following the AdS$_5$-Schwarzschild case, we may drop the last two terms in the above equation in the regime of large $\tilde\omega$ and $\nu$, and large black hole mass:
\begin{align}
    h^2,\ \tilde h^2,\ \nu^2 \gg \Big|1+\frac{\tilde\phi_h^2}{\phi_h^2}\Big|,\ \Big|\frac{\tilde\phi_h^2}{\phi_h^2}\Big|.
\end{align}
Doing so also does not interfere with the imposition of boundary conditions at the horizon ($\rho=0$) or the AdS boundary ($\rho=\infty$), since the last two terms are much smaller than the first three terms for $\rho\to 0$ or $\rho\to\infty$. The remaining equation is then identical to the $p=0$ wave equation in AdS$_5$-Schwarzschild in a similar approximation.

$G(\tilde\omega)$ may then be computed by expressing it in terms of the Jost function $f(\tilde\omega)$:
\begin{align}
    G(\tilde\omega)=(2\nu)^2\frac{e^{\beta\tilde\omega}}{e^{\beta\tilde\omega}-1}\frac{2\tilde\omega}{f(\tilde\omega)f(-\tilde\omega)}.
\end{align}
$f(\tilde\omega)$ is constructed using the solutions to the wave equation in \eqref{eq:wave_eq}, and is proportional to their Wronskian. It encodes how a given linearly independent solution at the boundary connects to the ingoing and outgoing solutions at the horizon (see \cite{Festuccia2007BlackHoleSingularities} for more details). It follows that in this approximation, the Wightman function $G(\tilde\omega)$ has an expression identical to the AdS$_5$-Schwarzschild case. This takes the form of the product of four Gamma functions:
\begin{align}
 \label{eq:gamma_expr}   G(\tilde\omega)=\frac{e^{\beta\tilde\omega/2}}{\pi (\Gamma(\nu))^2}\left(\frac{2\pi}{|\mathcal B|}\right)^{2\nu} \Gamma\left(\frac{1+\nu}{2}\pm \frac{\tilde\omega \mathcal B}{4\pi}\right)\Gamma\left(\frac{1+\nu}{2}\pm \frac{\tilde\omega \bar{\mathcal B}}{4\pi}\right),
\end{align}
where 
\begin{align}
    \mathcal B=\tilde \beta+i\beta.
\end{align}
\begin{figure}[t]
    \centering
    \includegraphics[width=0.3\linewidth]{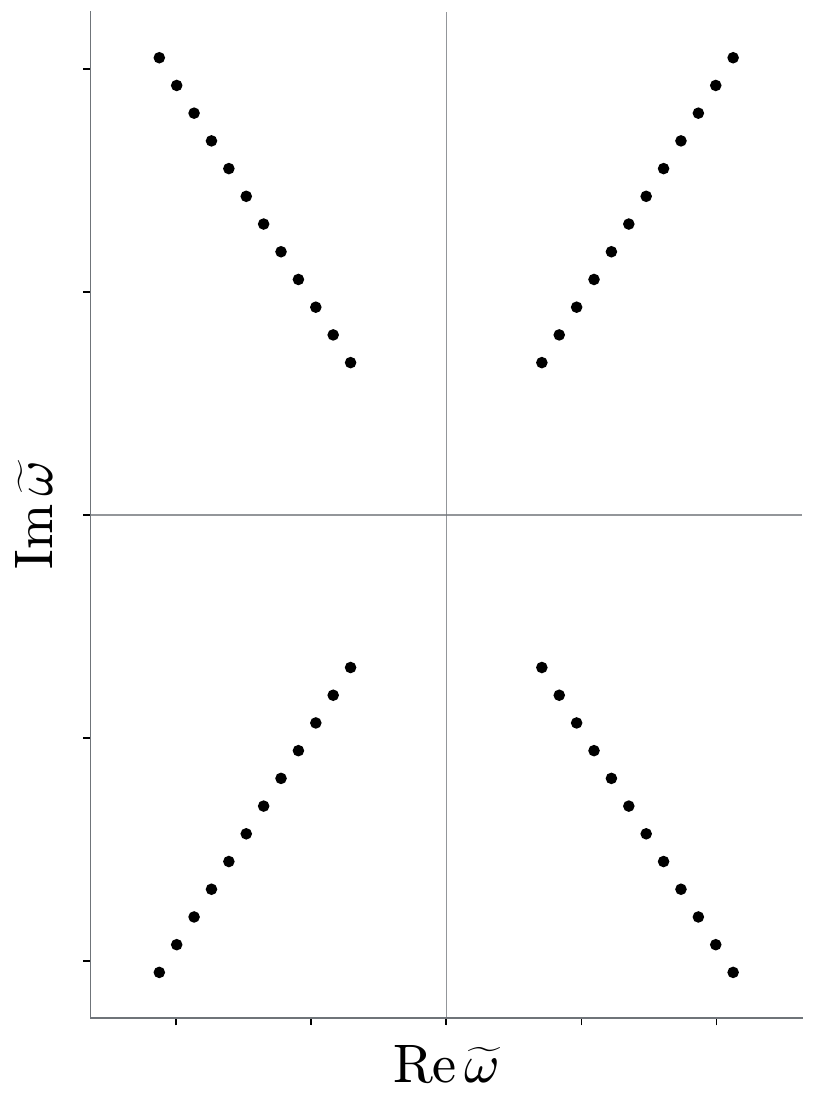}
    \caption{The quasinormal mode poles of the Wightman function $G(\tilde\omega)$ derived using the wave equation in \eqref{eq:gamma_expr}.}
    \label{fig:qnms}
\end{figure}

The notation $\Gamma(a \pm b)$ is an abbreviation for $\Gamma(a + b)\Gamma(a - b)$. The poles of $G(\tilde\omega)$ arising from the gamma functions in \eqref{eq:gamma_expr} correspond to the quasinormal mode spectrum, see Figure \ref{fig:qnms}. As noted in  \cite{Festuccia2007BlackHoleSingularities},  $G(\tilde\omega)$ has the asymptotic expansion\footnote{The importance of such an expansion has been emphasized in \cite{Afkhami-Jeddi:2025wra} where it is derived using WKB techniques.} for $\tilde\omega\to\infty$,
\begin{align}
\label{eq:gamma_asymptotic}
    G(\tilde\omega)\sim\frac{4\pi}{(\Gamma(\nu))^2}\left(\frac{\tilde\omega}{2}\right)^{2\nu}\left(1-e^{-i\pi\nu +i\tilde\omega \mathcal{B}/2}-e^{i\pi\nu -i\tilde\omega \bar{\mathcal{B}}/2}+\cdots\right)\left(1+\mathcal{O}(\tilde\omega^{-2})\right).
\end{align}
For $\tilde\omega\to -\infty$ we have instead
\begin{align}
    G(\tilde\omega)
    \sim
    \frac{4\pi}{(\Gamma(\nu))^2}
    e^{\beta\tilde\omega}
    \left(\frac{-\tilde\omega}{2}\right)^{2\nu}
    \left(
    1-e^{-i\pi\nu-i\tilde\omega\mathcal B/2}
     -e^{i\pi\nu+i\tilde\omega\bar{\mathcal B}/2}
     +\cdots
    \right)
    \left(1+\mathcal O(\tilde\omega^{-2})\right).
\end{align}
The exponentially small terms above give the asymptotic behavior of the extra saddles included in the sum in \eqref{eq:saddle_sum} at large positive and negative energies (see also \eqref{eq:sum2}); in \eqref{eq:probe_bg} we compute this asymptotic behavior for one winding and see that it matches the first exponentially small term in \eqref{eq:gamma_asymptotic}. It is straightforward to show this for all the other windings as well. Therefore these saddles must be summed over in \eqref{eq:saddle_sum} to get the full answer. Their Fourier transform leads to the lattice depicted in Figure \ref{fig:lattice}. As explained in Section \ref{sec:geodesics}, half of these windings correspond to complex geodesics with turning point near the singularity.

\section{Exact expressions for the $1/\phi^3$ potential}
\label{sec:phi2_formulas}
In this appendix we record the exact expression for the integral along the particle worldline used to compute the action in \eqref{eq:S_on_shell}:
\begin{align}
    I=\frac{1}{G_N} \int_{\phi_t}^{\infty} \frac{d\phi \, \phi\, U(\phi) P}{(W(\phi) - E) \sqrt{W(\phi) - E -P^2}},
\end{align}
with $W(\phi)=\phi^2-1/\phi^2$ and $U(\phi)=2\phi+2/\phi^3$. Here $\phi_t$ is the turning point near the boundary.
After subtracting a logarithmically divergent term that depends on the cutoff, it is useful to write the answer to this integral in the following way:
\begin{align}
\nonumber
I(E,P)
\doteq
-\frac{1}{2G_N}\Bigg\{&
\left[
P+i\left(\phi_h(E)+\widetilde{\phi}_h(E)\right)
\right]
\log\!\left[
P-\sqrt{-E-2i}
\right]
\\
\nonumber
+&
\left[
P-i\left(\phi_h(E)+\widetilde{\phi}_h(E)\right)
\right]
\log\!\left[
P+\sqrt{-E-2i}
\right]
\\
\nonumber
+&
\left[
P+i\left(\phi_h(E)-\widetilde{\phi}_h(E)\right)
\right]
\log\!\left[
P+\sqrt{-E+2i}
\right]
\\
+&
\left[
P-i\left(\phi_h(E)-\widetilde{\phi}_h(E)\right)
\right]
\log\!\left[
P-\sqrt{-E+2i}
\right]
\Bigg\},
\end{align}
which makes the monodromies around the branch points in \eqref{eq:probe_branches} manifest.
The on-shell actions $\mathcal{I}_{(0,0)}$ and $\mathcal{I}_{(1,0)}$ in \eqref{eq:S_on_shell} and \eqref{eq:S_reln} inherit these branch points.

\section{Additional saddle points}
\label{sec:additionalsaddles}
This paper focuses on the dynamics of three particular saddles --- those involved in the collisions of Section \ref{sec:neartcstokes}. Our analysis shows that these saddles determine the singular behavior in $G(t)$  at finite $G_N$, give the leading behavior near $t_c$ and explain how the bouncing geodesic saddle behaves in the complex $t$ plane. 
However, because of the infinitely branched structure of the integrand that gives rise to $G_{1,0}(t)$, an infinite number of saddle points exist. At generic values of $t$  other saddles contribute to $G_{1,0}(t)$. 

\begin{figure}[t]
    \centering
    \includegraphics[width=0.5\linewidth]{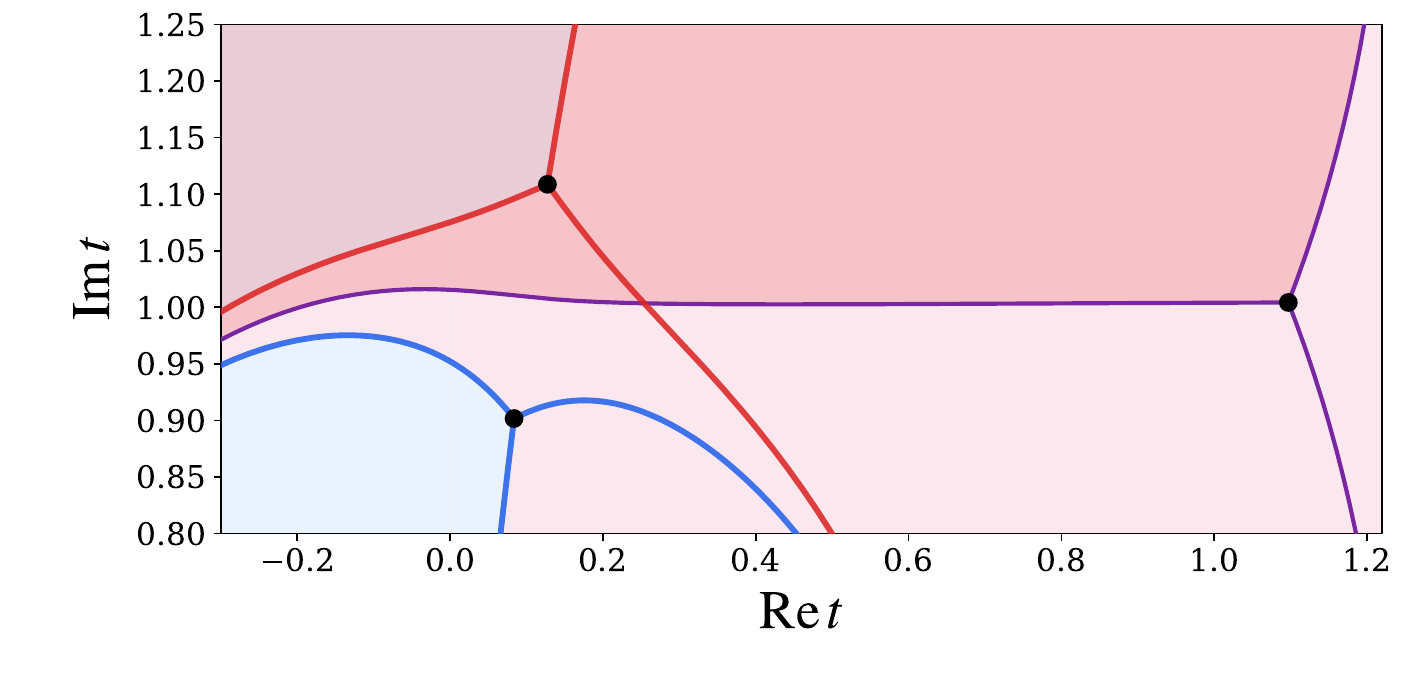}
    \caption{One of the saddles participating in the collisions around $t=t_c$ collides with a third saddle at the point pictured on the right of the figure. This saddle then contributes in the darker shaded region. In the vicinity of $t=t_c$, its contribution is subleading.}
    \label{fig:extra_saddle}
\end{figure}

As an illustration, at least one other saddle contributes in the darker shaded region in Figure \ref{fig:extra_saddle}. This saddle collides with one of the main participating saddles at an $\mathcal{O}(1)$ point in $t$ in an Airy-like collision.  Its contribution near $t_c$  is subleading compared to the saddles discussed in Section \ref{sec:neartcstokes}. 

We  did a substantial numerical and analytical search aided by OpenAI Astra for other saddles near the $t$ values most relevant to our analysis.   Near $t=t_c$, we found the one extra saddle just mentioned. Near $t=0$ we found no others that contribute besides those outlined in the main text. Near $t=-i\beta$ on the second sheet (pictured on the right in Figure \ref{fig:global_stokes}), we found one other contributing saddle whose contribution is parametrically subleading compared to the exponentially large bouncing geodesic saddle.   Despite this search we cannot exclude the possibility that other saddles make subleading contributions near these points.

\section{A natural boundary in JT gravity}
\label{sec:JT}
In this appendix we will explain why the natural boundary encountered in Section \ref{sec:windings} exists for pure JT gravity as well. We will then prove its existence for the case of $\Delta=1/2$ without using any saddle point approximation. Our starting point is the exact expression for the two point function. We follow the notation in \cite{Yang:2018gdb}:
\begin{align}
\label{eq:JT_correlator}\ev{O_1(u)O_2(0)}_{\text{QG}}=\frac{1}{\cal N}\int_0^\infty \d s_1 \d s_2\, \rho(s_1) \rho(s_2) e^{-\frac{s_1^2}{2}u-\frac{s_2^2}{2}(\beta-u)}\frac{|\Gamma(\Delta-i(s_1+s_2))\Gamma(\Delta+i(s_1-s_2))|^2}{2^{2\Delta+1}\Gamma(2\Delta)},
\end{align}
where $\rho(s)=\frac{s}{2\pi^2}\sinh(2\pi s)$. Here, the variable $u$ is the Euclidean time. Let us consider the integrand in the limit $s_1\to\infty$, $s_2$ fixed. We want to understand how the integral over this region affects the behavior of the correlator near $u=0$. First note the asymptotic expansion
\begin{align}
\nonumber\log \Gamma(\Delta+i y)
\sim&
\left(i y+\Delta-\frac{1}{2}\right)\log(i y)
-i y
+\frac{1}{2}\log(2\pi)
-\frac{1}{2}\log\!\left(
1-e^{-2\pi y+2\pi i\Delta}
\right)\\
&+\sum_{\nu=2}^{\infty}
\frac{(-1)^{\nu}B_{\nu}(\Delta)}
{\nu(\nu-1)(i y)^{\nu-1}},
\quad y\to+\infty .
\end{align}
This expansion is along the Stokes line of the above function; its form is taken from and studied more in \cite{Nemes:2013Hermite}, see their Eq. (42). It is an improved form of the Stirling expansion for the gamma function with a shifted argument.
Using this expansion one can then show
\begin{align}
\label{eq:asymptotic_log}
    |\Gamma(\Delta-i(s_1+s_2))\Gamma(\Delta+i(s_1-s_2))|^2&\sim
4\pi^2 e^{-2\pi s_1}s_1^{4\Delta-2}
\left(
1+
\mathcal{O}(s_1^{-2})
\right)
\left(
1+\sum_{k=1}^\infty c_k(\Delta,s_2) e^{-2k\pi s_1}
\right).
\end{align}
Plugging this back into the integral in \eqref{eq:JT_correlator}, we see that we obtain a sum over integrals that has the form of a sum over windings, like in \eqref{eq:another_sum}. Similar to the argument in Appendix \ref{sec:qnms}, the exponentially small terms in the expansion in \eqref{eq:asymptotic_log} give rise to these windings, and each winding is labeled by $k$.
The form of the integrand matches that in \eqref{eq:nm_integrand}. The difference is that we restrict to a subset of the values of $(n,m)$ summed over in Section \ref{sec:windings}, for which $n=m$.
Each integral then gives an essential singularity at $u=0$. The saddle point argument in Section \ref{sec:windings} then says that the sum over the windings starts to diverge when we rotate $u$ by $\theta=3\pi/2$ (with the starting point being $u=\epsilon$). Using the Fabry gap theorem, this strongly suggests the presence of a natural boundary at this location. 

\subsubsection*{The $\Delta=1/2$ case}
For $\Delta=1/2$, it is possible to show the existence of this natural boundary exactly, without using the saddle point approximation.\footnote{We thank ChatGPT 5.6 Sol for this argument.} This is because the gamma functions in \eqref{eq:JT_correlator} simplify so that the integral becomes
\begin{align}
    \label{eq:Delta_half}G(u)_{\Delta=1/2}=\ev{O_1(u)O_2(0)}_{\text{QG}}^{\Delta=1/2}\sim \int_0^\infty \d s_1 \d s_2 \,s_1 s_2\, e^{-\frac{u}{2}s_1^2-\frac{(\beta-u)}{2}s_2^2}\frac{\sinh 2\pi s_1 \sinh 2\pi s_2}{\cosh2\pi s_1+\cosh2\pi s_2}.
\end{align}
We can then use the identity
\begin{align}
    \int_0^\infty \d\omega\frac{\sin\omega s_1 \sin\omega s_2}{\cosh(\omega/2)}=2\pi \frac{\sinh \pi s_1 \sinh \pi s_2}{\cosh2\pi s_1+\cosh2\pi s_2}.
\end{align}
Plugging this into \eqref{eq:Delta_half} and exchanging the order of integration gives a one-dimensional integral:
\begin{align}
    G(u)_{\Delta=1/2}\sim\int_0^\infty \frac{\d \omega}{\cosh(\omega/2)}I_u(\omega) I_{\beta-u}(\omega),
\end{align}
where
\begin{align}
    \nonumber I_a(\omega)&:=\int_0^\infty \d s\, s\, e^{-\frac{a}{2}s^2}\cosh\pi s\sin \omega s\\
    &=\frac{1}{2}\sqrt{\frac{\pi}{2a^3}}
\left[
(\omega-i\pi)e^{-\frac{(\omega-i\pi)^2}{2a}}
+
(\omega+i\pi)e^{-\frac{(\omega+i\pi)^2}{2a}}
\right],\quad \Re a>0.
\end{align}
Now our task is to analyze this simpler integral. The Gaussian part of its asymptotic behavior in $\omega$ has the form
\begin{align}
    I_u(\omega) I_{\beta-u}(\omega)\sim \exp\left[-\frac{\beta \omega^2}{2 u(\beta-u)}\right].
\end{align}
For $|u|<<\beta$, 
the $\omega$ integral therefore diverges for $\Re u<0$. To analytically continue the answer to this region we have to rotate the $\omega$ contour. If we parameterize $u$ as $\epsilon e^{i\theta}$, then the rotated contour goes to infinity in the direction $\omega=e^{i\theta/2}x$, with $x\in \mathbb R_+$. Note that the complex $\omega$ plane has poles at the locations 
\begin{align}
    \omega_n=i\pi(2n+1).
\end{align}
These lie on the positive imaginary axis. As we rotate the contour in $\omega$ past $\theta=\pi$, we pick up the contributions from the residues of these poles. This is how the Stokes phenomenon of Section \ref{sec:OPE_sing} is encoded in these exact expressions. For $\theta>\pi$, $G(u)_{\Delta=1/2}$ is then given by the sum of an integral that converges in a neighborhood of $u=0$ and the residues of these poles. This sum over residues is proportional to
\begin{align}
    \sum(\text{residues})\sim-\frac{\pi}{2\,[u(\beta-u)]^{3/2}}
\sum_{n=1}^{\infty}
(-1)^n n(n+1)
\left[
e^{\frac{2\pi^2(2n+1)}{\beta-u}}
+
e^{\frac{2\pi^2(2n+1)}{u}}
\right]
e^{\frac{2\pi^2\beta n^2}{u(\beta-u)}}.
\end{align}
This sum possesses a natural boundary at
\begin{align}
\label{eq:nb_half}
    \Re \left(\frac{\beta}{u(\beta-u)}\right)=0,
\end{align}
where it first starts to diverge.
For small $u$ this is the 
$\theta=3\pi/2$ line. To show this natural boundary, we will show that the sum above develops a dense set of singularities on the curve in \eqref{eq:nb_half}. These singularities are located at values $u=u_0$ on the boundary that satisfy
\begin{align}
    \frac{2\pi^2\beta}{u_0(\beta-u_0)}=\frac{2\pi i p}{M},\quad p\in \mathbb{Z},\ M \in \mathbb{N}.
\end{align}
Since the values $u=u_0$ may be mapped to the roots of unity (right hand side of the equation above), they form a dense set. At these points we can define the periodic in $n$ quantity
\begin{align}
    c_n=(-1)^ne^{\frac{2\pi^2\beta n^2}{u_0(\beta-u_0)}},
\end{align}
with period $2M$ (or $M$ if $M$ is even). 
Since the $e^{\frac{2\pi^2 (2n+1)}{u}}$ term in the sum over residues decays exponentially with $n$ on the natural boundary, we can focus on proving that the following sum diverges as $u\to u_0$: 
\begin{align}
    \mathcal{S}=\sum_{n=1}^{\infty}
(-1)^n n(n+1)
e^{\frac{2\pi^2(2n+1)}{\beta-u}}
e^{\frac{2\pi^2\beta n^2}{u(\beta-u)}}.
\end{align}
Defining
\begin{align}
    h=\frac{2\pi^2\beta}{u_0(\beta-u_0)}-\frac{2\pi^2\beta}{u(\beta-u)},
\end{align}
this reduces to the sum
\begin{align}
  \mathcal{S} =e^{\frac{2\pi^2}{\beta-u(h)}} \sum_{n=1}^{\infty}
n(n+1) c_n
e^{\frac{4\pi^2n}{\beta-u(h)}}
e^{-hn^2}.
\end{align}
Now, since $c_n$ is periodic, we can write it in terms of a Fourier expansion, which gives
\begin{align}
    \mathcal{S}=e^{\frac{2\pi^2}{\beta-u(h)}}
\sum_{j=0}^{P-1}d_j
\sum_{n=1}^{\infty}
n(n+1)
\exp\left[
-hn^2+
\left(
\frac{4\pi^2}{\beta-u(h)}
+
\frac{2\pi ij}{P}
\right)n
\right].
\end{align}
Here $P$ is a period of the sequence $c_n$ -- we will choose $P=2M$.
Now, we will find a sequence $h_N\to 0$ with $\Re h_N>0$ along which the above sum becomes unbounded. Let us focus on a particular Fourier component $j=j_0$ and choose $h_N$ so that it satisfies 
\begin{align}
   h_N
=
\frac{1}{2N}\left[\frac{4\pi^2}{\beta-u(h_N)}
+
\frac{\pi i j_0}{M}\right]\approx \frac{1}{2N}
\left[
\frac{4\pi^2}{\beta-u_0}
+
\frac{\pi i j_0}{M}
\right]
+
O\left(\frac{1}{N^2}\right).
\end{align}
As $N\to\infty$, $h_N\to 0$ along a path defined by the above expression and we approach the natural boundary.
For the above choice of $h_N$, the saddle point of the sum over $n$ for $j_0$ is given by $n_{\text{saddle,}j_0}=N$. Evaluated on this saddle point, the magnitude of the contribution to the above sum from this saddle is
\begin{align}
    \exp\left[
2\pi^2(N+1)
\operatorname{Re}\left(\frac1{\beta-u(h_N)}\right)
\right].
\end{align}
This grows exponentially with $N$.
One can then check that the saddle points for other $j\neq j_0$ give contributions to the sum that are exponentially suppressed. Therefore as $h_N\to 0$ and $u\to u_0$ the above series diverges, giving rise to the natural boundary. Note that the contribution to the correlator from the integral over the rotated contour is holomorphic away from $u=0$, so that it cannot cancel this natural boundary.

\section{The exponential potential}
\label{sec:exp_potential}
In this appendix we will outline the computations in Sections \ref{sec:setup} to \ref{sec:windings} of the paper for the exponential potential of \eqref{eq:potentials}:
\begin{align}
    U(\phi)=2 \phi+e^{-\phi}.
\end{align}
This potential is of interest because of the existence of a simple defect expansion \cite{Maxfield:2020ale,Witten:2020wvy}, and because of the expectation that loop corrections are under control, see Section \ref{sec:diagrams}. Our goal is to highlight the universality of the local Gaussian correction in Section \ref{sec:first_correction}, and of the behavior near $t=0$ described in Section \ref{sec:OPE_sing}.

Let us first study the structure of the turning point equation for geodesics in the corresponding background. Recall this equation is given by
\begin{align}
\label{eq:exp_tp}
    W(\phi_t)-E-P^2=0\implies \phi_t^2-e^{-\phi_t}-E=P^2.
\end{align}
Unlike the case of the $1/\phi^3$ potential, this equation has infinitely many turning points in the near-singularity region. These are given by
\begin{align}
    \phi_t^\infty(P)\approx-\log(P^2+E)+i(2j+1)\pi.
\end{align}
Branch points of the function $\phi_t(P)$ occur when two of its solutions and hence two turning points coincide. At this location $\phi_t^*$, the derivative of $W(\phi_t)$ equals zero. This happens at infinitely many locations, parameterized by $k$:
\begin{align}
    \phi_{t,k}^*=\mathrm{LambertW}_k\left(-\frac{1}{2}\right),\quad k\in \mathbb{Z}.
\end{align}
Using \eqref{eq:exp_tp} this corresponds to an infinite number of branch points of $\phi_t(P)$ in the complex $P$ plane, located at $P_k^*$.
One can check that this is a square root branch point that exchanges two solutions, since locally we have
\begin{align}
    \phi_t-\phi_{t,k}^*\approx\pm \sqrt{\frac{P^2-{P_k^*}^2}{1+\phi_{t,k}^*}}.
\end{align}
To access a geodesic with turning point near the singularity, we can then follow a procedure
similar to that in Section \ref{sec:geodesics}. We can start at large $P$, so that the corresponding geodesic has turning point close to the AdS boundary. Then, we can wind around the branch point in $P$ that exchanges it with a given $\phi_t^\infty$, and go to large $P$ again.

\begin{figure}[t]
    \centering
    \includegraphics[width=0.6\linewidth]{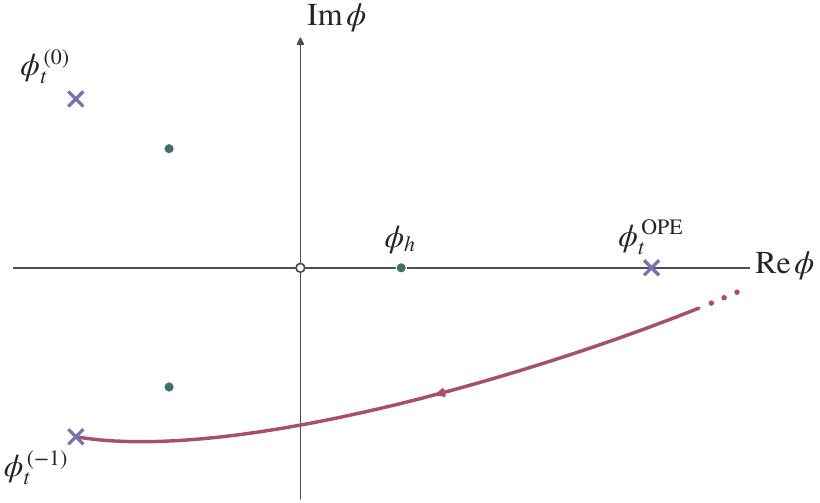}
    \caption{A contour in the complex $\phi$ plane that corresponds to a complex geodesic approaching the singularity for the exponential potential. The turning points are labelled by crosses, and some of the poles of \eqref{eq:exp_tc_action} corresponding to roots of the equation $W(\phi)=E$ are labelled by dots.}
    \label{fig:exp_potential}
\end{figure}

The contour for the geodesic corresponding to the $j=-1$ turning point near the singularity is pictured in Figure \ref{fig:exp_potential}. We are interested in the on-shell action evaluated on the backreacted geometry containing this geodesic, since that is used to compute the contribution from this geodesic to $G(t)$ (see \eqref{eq:saddle_sum}). The results of Section \ref{sec:backreaction} tell us that
\begin{align}
\label{eq:exp_tc_action}
    \mathcal{I}_{\tilde\Lambda}=\frac{2\pi \phi_h(E_-)}{G_N} + (I_+ - I_-), \quad
I_\pm=\frac{1}{G_N} \int_{\phi_t}^{\infty} \frac{d\phi \, \phi\, U(\phi) P_\pm}{(W(\phi) - E_\pm) \sqrt{W(\phi) - E_\pm -P_\pm^2}},
\end{align}
where the contour in the above integral for a given $E_+$ or $E_-$ is that in Figure \ref{fig:exp_potential}. 

Recall that the bouncing geodesic singularity arises from expanding the expression for $\mathcal{I}_{\tilde\Lambda}$ for small $G_N$. Using \eqref{eq:S_probe} and \eqref{eq:S_probe_2} with the above contour and doing this expansion gives
\begin{align}
    \mathcal{I}_{\tilde\Lambda}=\frac{2\pi \phi_h(2\bar E)}{G_N}
-2\pi\phi_h'(2\bar E)\tilde\omega
-
\mu
\ell_{\tilde\Lambda}(\tilde\omega, \bar E)
+
\tilde\omega \tau_{\tilde\Lambda}(\tilde\omega,\bar E).
\end{align}
Here, $\ell_{\tilde\Lambda}$ and $\tau_{\tilde\Lambda}$ are the regulated proper length and coordinate time elapsed along the complex geodesic we are considering. We also dropped unimportant additive constants. The large $\tilde\omega$ limit of $\tau_{\tilde\Lambda}$ sets $t_c$. The first correction due to backreaction is then obtained by expanding the action $\mathcal{I}_{\tilde\Lambda}$ to $\mathcal{O}(G_N)$. We will now argue that for the exponential potential, this first correction is of the form $G_N\tilde\omega^2$ as well, with subleading corrections forming the series in \eqref{eq:exp_form}:
\begin{align}
\label{eq:exp_series}
    \Delta  \mathcal{I}_{\tilde\Lambda}^{\text{probe}}\sim G_N\tilde\omega^2 + G_N^2\tilde\omega^3 + G_N^3 \tilde\omega^4+\cdots.
\end{align}
Recall that this form is valid when $\tilde\omega\sim 1/\sqrt{G_N}$ and hence is large.
As discussed in Section \ref{sec:first_correction}, the difference $I_+-I_-$ is even in $G_N$, so that the $G_N\tilde\omega^2$ correction only comes from the entropy term proportional to $2\pi \phi_h(E_-)$ by replacing $E_-=2\bar E-G_N \tilde\omega$ and Taylor expanding. Moreover, we can again write the difference $I_+-I_-$ in the form in \eqref{eq:I_diff}:
\begin{align}
I_+-I_-\sim \tilde\omega
\int_{\phi_t}^{\infty} d\phi\,
\phi\, U(\phi)\,
\frac{
2-\frac{\mu^2\left(W(\phi)-2\bar E\right)}{\tilde\omega^2}
}{
\left[\left(W(\phi)-2\bar E\right)^2-G_N^2\tilde\omega^2\right]
\sqrt{
-1+\frac{\mu^2\left(W(\phi)-2\bar E\right)}{\tilde\omega^2}
-\frac{\mu^4G_N^2}{4\tilde\omega^2}
}
}.
\end{align}
Here, we instead have $\phi_t\sim -2\log\tilde\omega-i\pi$. The contributions to this integral from near the turning point again go to zero at large $\tilde\omega$. The integral is dominated by the region where $\phi$ is $\mathcal{O}(1)$, and so at large $\tilde\omega$ with $\tilde\omega\sim 1/\sqrt{G_N}$ we get a series of the form \eqref{eq:exp_series}. The integral over the larger $\mathcal{O}(\log\tilde\omega)$ region inside the horizon in this case doesn't lead to extra enhancement.

This shows that the discussion of Section \ref{sec:smoothing} applies here as well, with the $t_c$ singularity smoothing into two Airy-like collision points. These collision points emit Stokes lines. It is harder to find the exact topology of Stokes lines in this case. However, the Stokes phenomenon near $t=0$ that picks up the contribution of the geodesic with turning point near the singularity is still present.

The reason is that as we increase $\tilde\omega$ beyond the regime where the Gaussian approximation is valid, so that it is at least of order $1/G_N$, the behavior of the action is similar to that of the $1/\phi^3$ potential in Section \ref{sec:OPE_sing}. An argument similar to that in Section \ref{sec:no_sing} using \eqref{eq:exp_tc_action} instead shows that a large $\omega\equiv G_N\tilde\omega$ saddle point contributes either near $t=0$ or $t=-i\beta$. To reach this saddle point, we take $E_-\to\infty$ with $E_+$ fixed or vice versa. In this regime, the form of the action $\mathcal{J}_{\tilde\Lambda}$ is also similar to that for the $1/\phi^3$ potential in \eqref{eq:splus_integrand}. For large $s_+=\sqrt{2\bar E+G_N\tilde\omega}$, fixing $s_-=s_-^*$, we have
\begin{align}
    \mathcal{J}_{\tilde\Lambda} \sim -\frac{it}{2 G_N}s_+^2 -\frac{\pi}{G_N}s_+-\frac{2i}{G_N}\log^2{s_+}+2\mu \log(s_+^2),
\end{align}
where $\mathcal{J}_{\tilde\Lambda}$ is the logarithm of the full integrand for large $\mu$.
The difference is the term proportional to $\log^2(s_+)$. This arises from the difference of integrals $I_+-I_-$ evaluated over the region inside the horizon from a value $\phi=\phi_c$ to an $\mathcal{O}(1)$ value of $\phi$ for which $|W(\phi)|$ is much larger than $|(s_-^*)^2|$. The value $\phi_c$ is determined by the condition $|W(\phi_c)|\sim s_+^2$, so that this region is the interval where $|(s_-^*)^2|<< |W(\phi)|<<|(s_+)^2|$. In this region, we can approximate $W(\phi)\approx -e^{-\phi}$ to obtain the integral
\begin{align}
    \d(I_+-I_-)\sim \d\phi \, \tilde\omega\,\phi\, e^{-\phi}\frac{2}{(-s_+^2) (-e^{-\phi})(-i)}.
\end{align}
Using the fact that $\phi_c\sim -2\log s_+$ and $\tilde\omega\sim s_+^2/(2 G_N)$ at large $\tilde\omega$ and fixed $s_-$, we obtain the $\propto \log^2 (s_+)$ behavior above. Note that the region that gives this behavior is separated from the turning point of the geodesic, which satisfies $\phi_t\sim -4 \log s_+$; it is a wide region inside the horizon.

The leading behavior for large $s_+$ is then determined by the first two terms in $\mathcal{J}_{\tilde\Lambda}$, so that the behavior of the integrand is similar to the case of the $1/\phi^3$ potential in Section \ref{sec:essential}. When $t=-i\epsilon$, the defining $s_+$ contour runs from $0$ to $\infty$, and the integral cannot receive a contribution from the saddle that lies at large negative $s_+$. As we then follow a path in $t$ given by a small circle around $t=0$, we rotate this defining contour as well. When we reach the positive imaginary $t$ axis, the asymptotic behavior of the action on the $s_+$ contour then gives rise to a term proportional to $\exp\left[-\frac{\pi^2}{2 G_N\epsilon }\right]$, indicating that the bouncing geodesic saddle starts to contribute. A key difference in this case, however, is the coefficient of this exponential behavior. The extra $\log^2s_+$ term changes this coefficient from a pure power law (see \eqref{eq:essential}), so that the contribution from the bouncing geodesic saddle becomes  
\begin{align}
\label{eq:appdcorr}
   e^{\mathcal J_{\tilde\Lambda}}\sim\left(\frac{\pi}{\epsilon}\right)^{4\mu} \exp\left[-\frac{2i}{G_N }\log^2\left(\frac{\pi}{\epsilon}\right)\right]\exp\left[-\frac{\pi^2}{2 G_N\epsilon }\right].
\end{align}
We see that the branch structure of the resulting Riemann surface becomes more complicated.

\section{Analysis of worldline fluctuations}
\label{sec:feynman_rules}
So far in our analysis the point particle has followed its classical trajectory. In this appendix, we derive the scaling of the coupling of particle fluctuations to fluctuations of the conformal factor, and show that fluctuations of the point particle in the near-singularity region grow as we increase $\tilde\omega$. However, these growing fluctuations may be resummed so that they don't compete with the large $\tilde\omega$ behavior derived in this paper. We set $\mu=1$ throughout this section, and work in the $\mathcal{O}(1)$ coordinates near the turning point outlined in Section \ref{sec:diagrams} (we also use the notation and definitions of that section).

Let us focus on the term in the point particle action with $k$ insertions of the $\sigma$ fluctuation (see \eqref{eq:I_fluc_pp}):
\begin{align}
\label{eq:Spp_k}
    \frac12\int \d\lambda
\left[
e^{2\bar\rho(X)} \frac{G_N^{k/2}}{k!}(2\sigma(X))^k\eta_{\mu\nu}
\dot X^\mu\dot X^\nu
\right].
\end{align}
Recall the dots represent derivatives with respect to $\lambda$.
First, consider the $1/\phi^3$ potential, so that $e^{2\bar\rho}\sim \tilde\omega^2$ and $\Delta\lambda \sim \tilde\omega^{-2}$. The mass shell constraint also
tells us that $\bar X^\mu \sim \tilde\omega^{-3}$. So, it is also helpful to rescale $Y^\mu=\tilde\omega^{3} X^\mu$. Making all the $\tilde\omega$ dependencies explicit, we may write the above action as
\begin{align}
    \label{eq:int_pp_action} \frac{G_N^{k/2}\tilde\omega^{k/2-2}}{2k!}\int \d\tilde\tau
\left[
\tilde{e^{2\bar\rho(Y)}}(2\tilde\sigma(Y))^k\delta_{\mu\nu}
{Y^\mu}'{Y^\nu}'
+1
\right],
\end{align}
where we defined the $\mathcal{O}(1)$ background conformal factor $\tilde{e^{2\bar\rho}}$.
Here, primes represent derivatives with respect to $\tilde\tau$. The propagator for particle fluctuations is obtained from the $k=0$ term.
To obtain an $\mathcal{O}(1)$ propagator, we expand the particle position as
\begin{align}
    Y^\mu = \bar Y^\mu +\tilde\omega\zeta^\mu.
\end{align}
To obtain the vertex coupling $k$ graviton legs and $r$ particle fluctuation legs, we then plug the above expression into the integrand in \eqref{eq:int_pp_action} and Taylor expand. Each term containing $r$ factors of $\zeta^\mu$ comes with a factor of $\tilde\omega^r$. The scaling of the resulting vertex is then
\begin{align}
    V_{\tilde\sigma^k \zeta^r}\sim (G_N\tilde\omega)^{k/2}\tilde\omega^{r-2}.
\end{align}
Note this vertex also involves derivatives of the $\tilde\sigma$ field.

For the exponential potential, following the same steps as above to make the $\tilde\omega$ dependence of \eqref{eq:Spp_k} explicit gives
\begin{align}
\frac{G_N^{k/2}}{2k!\,\tilde\omega}\int \d\tilde\tau
\left[
\tilde{e^{2\bar\rho(Y)}}(2\sigma(Y))^k\delta_{\mu\nu}
{Y^\mu}'{Y^\nu}'
+1
\right],
\end{align}
We instead expand the particle position as
\begin{align}
    Y^\mu = \bar Y^\mu +\tilde\omega^{1/2}\zeta^\mu.
\end{align}
This leads to the following scaling of the vertex:
\begin{align}
    V_{\tilde\sigma^k \zeta^r}\sim (G_N)^{k/2}\tilde\omega^{r/2-1}.
\end{align}
Contributions of these vertices to the diagrams we have considered in Section \ref{sec:diagrams} at first 
ruin the scaling that they point to, because these vertices contain arbitrarily high powers of $\tilde\omega$.\footnote{When $\tilde\omega$ is $\mathcal{O}(1)$, the effects of these corrections may be suppressed by taking $\mu$ large, which is how the saddle point approximation is controlled in the probe limit.} 
However, the perturbation series we are considering is no longer under control and a resummation of these diagrams is required to draw any conclusions. This breakdown tells us that we are expanding around the wrong background -- the point particle picture no longer holds in the near singularity region (see also \cite{Afkhami-Jeddi:2025wra} for a discussion of the breakdown of the WKB approximation near the singularity). In other words, the Compton wavelength of the particle exceeds the length scale associated to the curvature in the near-singularity region.

To resum all of the diagrams above that compete with the Gaussian approximation, we take a different approach that doesn't involve treating the matter field as a point particle. Recall we are computing the saddle point of the following path integral with some prescribed boundary conditions:
\begin{align}
    \int D\eta Dg D\phi\, e^{S_{\rm{grav}}[g,\phi]+S_{\rm{pp}}[g,\eta]}.
\end{align}
Here, $\eta$ represents the point particle. The idea is to replace part of the above integral by the propagator for the point particle with fixed energy as measured on the boundary:
\begin{align}
    \int D\eta\, e^{S_{\rm{pp}}[g,\eta]}=G_{\rm{pp}}^{\tilde\omega}[g].
\end{align}
In conformal gauge, we can then write the path integral as
\begin{align}
    \int D\sigma D\phi\, e^{S_{\rm{grav}}[\sigma,\phi]+\log G_{\rm{pp}}^{\tilde\omega}[\sigma]+S_{\rm{Liouville}}[\sigma]}.
\end{align}
The Liouville term comes from the ghosts from gauge fixing to conformal gauge. The effects of the matter particle are then encoded in effective $\sigma$ vertices generated by taking functional derivatives of $\log G_{\rm{pp}}^{\tilde\omega}[\sigma]$ with respect to $\sigma$. We will now understand the way these vertices scale with $\tilde\omega$ and $G_N$, and find that the resulting expansion they generate cannot compete with the Gaussian correction $G_N\tilde\omega^2$.

Let us first start with the propagator corresponding to the particle having fixed endpoints in $\tau$.
It may be computed using the wave equation; in conformal gauge, it is the inverse of the operator
\begin{align}
    \mathcal{D}_{\text{pp}}\equiv-\partial_\tau^2+\partial_z^2-\mu^2 e^{2\bar\rho(z)+2\sqrt{G_N}\sigma(\tau,z)}=\mathcal{D}_{\text{pp}}^{(0)}+\delta\mathcal D.
\end{align}
$\mathcal{D}_{\text{pp}}^{(0)}$ is the wave operator obtained on setting $\sigma=0$. Now, if we consider the propagator with fixed energy boundary conditions instead, the physical picture is that the incident matter wave with energy $\tilde\omega$ as measured on the boundary scatters against the potential created by the metric. The fluctuations of the metric $\sigma$ perturb the background potential and may then be treated in perturbation theory. 

The resulting series is then controlled by the energy of the incident wave being large compared to the magnitude of the perturbation of the potential, given that the perturbation acts over a region with $\mathcal{O}(1)$ width. We are again interested in the vicinity of the turning point of the geodesic, where 
 $\tau$ and $z$ are $\mathcal{O}(\tilde\omega^{-k})$. Let us first consider the $1/\phi^3$ potential, and work in the $\mathcal{O}(1)$ coordinates defined as $\tilde z\sim \tilde\omega^{3} z$ and $\tilde \tau\sim \tilde\omega^{3} \tau$. When the wave reaches the region where $\tilde\tau$ and $\tilde z$ are $\mathcal{O}(1)$, its locally measured energy is $\mathcal{O}(1)$. In these coordinates, the wave operator may be written as
\begin{align}
   \mathcal{D}_{\rm pp}\sim -\partial_{\tilde\tau}^2+\partial_{\tilde z}^2-\frac{\mu^2}{\tilde\omega^4} \tilde{e^{2\bar\rho}}-\frac{\mu^2}{\tilde\omega^4} \tilde{e^{2\bar\rho}}(e^{2\sqrt{G_N\tilde\omega}\tilde\sigma}-1).
\end{align}
Here we made all the $\tilde\omega$ dependence explicit by also using the $\mathcal{O}(1)$ perturbation $\tilde\sigma$ (which has $\mathcal{O}(1)$ propagators) and the $\mathcal{O}(1)$ background conformal factor $\tilde{e^{2\bar\rho}}$. In the Gaussian regime, when $\tilde\omega\sim 1/\sqrt{G_N}$, we can Taylor expand the exponential in the above expression. Each perturbation in the resulting series then has a height of $\mathcal{O}((G_N\tilde\omega)^{r/2}\tilde\omega^{-4})$. Comparing this to the energy of the wave gives the control parameter 
\begin{align}
    \epsilon_{1/\phi^3}=\frac{(G_N\tilde\omega)^{r/2}\tilde\omega^{-4}}{\tilde\omega^{-2}}=(G_N\tilde\omega)^{r/2}\tilde\omega^{-2}.
\end{align}
The $\tilde\sigma$ vertices generated by taking functional derivatives of $\log G_{\rm{pp}}^{\tilde\omega}$ hence also generate a series in the same parameter, for all values of $r$.  When $\tilde\omega\sim 1/\sqrt{G_N}$, none of these terms compete with the $G_N\tilde\omega^2$ of the Gaussian approximation.

For the exponential potential, a similar analysis leads to
\begin{align}
    \epsilon_{e^{-\phi}}=\frac{G_N^{r/2}}{\tilde\omega}.
\end{align}
The resulting perturbation series is well controlled even for large $\tilde\omega$, and so doesn't affect the Gaussian approximation or the large $\tilde\omega$ behavior dictating the correlator near $t=0$, as stated at the start of this section. Note that these conclusions are based on local estimates; a full analysis requires studying the infrared effects discussed in Section \ref{sec:diagrams}.

\bibliography{main}

\providecommand{\href}[2]{#2}\begingroup\raggedright\begin{thebibliography}{10}

\bibitem{Horowitz:2003he}
G.T.~Horowitz and J.M.~Maldacena, \emph{{The Black hole final state}}, \href{https://doi.org/10.1088/1126-6708/2004/02/008}{\emph{JHEP} {\bfseries 02} (2004) 008} [\href{https://arxiv.org/abs/hep-th/0310281}{{\ttfamily hep-th/0310281}}].

\bibitem{Silverstein:2005qf}
E.~Silverstein, \emph{{Dimensional mutation and spacelike singularities}}, \href{https://doi.org/10.1103/PhysRevD.73.086004}{\emph{Phys. Rev. D} {\bfseries 73} (2006) 086004} [\href{https://arxiv.org/abs/hep-th/0510044}{{\ttfamily hep-th/0510044}}].

\bibitem{Horowitz:2006mr}
G.T.~Horowitz and E.~Silverstein, \emph{{The Inside story: Quasilocal tachyons and black holes}}, \href{https://doi.org/10.1103/PhysRevD.73.064016}{\emph{Phys. Rev. D} {\bfseries 73} (2006) 064016} [\href{https://arxiv.org/abs/hep-th/0601032}{{\ttfamily hep-th/0601032}}].

\bibitem{Horowitz:2009wm}
G.~Horowitz, A.~Lawrence and E.~Silverstein, \emph{{Insightful D-branes}}, \href{https://doi.org/10.1088/1126-6708/2009/07/057}{\emph{JHEP} {\bfseries 07} (2009) 057} [\href{https://arxiv.org/abs/0904.3922}{{\ttfamily 0904.3922}}].

\bibitem{Itzhaki:2018rld}
N.~Itzhaki and L.~Liram, \emph{{A stringy glimpse into the black hole horizon}}, \href{https://doi.org/10.1007/JHEP04(2018)018}{\emph{JHEP} {\bfseries 04} (2018) 018} [\href{https://arxiv.org/abs/1801.04939}{{\ttfamily 1801.04939}}].

\bibitem{Frenkel:2020ysx}
A.~Frenkel, S.A.~Hartnoll, J.~Kruthoff and Z.D.~Shi, \emph{{Holographic flows from CFT to the Kasner universe}}, \href{https://doi.org/10.1007/JHEP08(2020)003}{\emph{JHEP} {\bfseries 08} (2020) 003} [\href{https://arxiv.org/abs/2004.01192}{{\ttfamily 2004.01192}}].

\bibitem{Grinberg:2020fdj}
M.~Grinberg and J.~Maldacena, \emph{{Proper time to the black hole singularity from thermal one-point functions}}, \href{https://doi.org/10.1007/JHEP03(2021)131}{\emph{JHEP} {\bfseries 03} (2021) 131} [\href{https://arxiv.org/abs/2011.01004}{{\ttfamily 2011.01004}}].

\bibitem{Leutheusser:2021qhd}
S.~Leutheusser and H.~Liu, \emph{{Causal connectability between quantum systems and the black hole interior in holographic duality}}, \href{https://doi.org/10.1103/PhysRevD.108.086019}{\emph{Phys. Rev. D} {\bfseries 108} (2023) 086019} [\href{https://arxiv.org/abs/2110.05497}{{\ttfamily 2110.05497}}].

\bibitem{Emparan:2021ewh}
R.~Emparan, D.~Licht, R.~Suzuki, M.~Toma{\v{s}}evi{\'c} and B.~Way, \emph{{Black tsunamis and naked singularities in AdS}}, \href{https://doi.org/10.1007/JHEP02(2022)090}{\emph{JHEP} {\bfseries 02} (2022) 090} [\href{https://arxiv.org/abs/2112.07967}{{\ttfamily 2112.07967}}].

\bibitem{Bousso:2022tdb}
R.~Bousso and A.~Shahbazi-Moghaddam, \emph{{Quantum singularities}}, \href{https://doi.org/10.1103/PhysRevD.107.066002}{\emph{Phys. Rev. D} {\bfseries 107} (2023) 066002} [\href{https://arxiv.org/abs/2206.07001}{{\ttfamily 2206.07001}}].

\bibitem{deBoer:2022zps}
J.~de~Boer, D.L.~Jafferis and L.~Lamprou, \emph{{On black hole interior reconstruction, singularities and the emergence of time}},  \href{https://arxiv.org/abs/2211.16512}{{\ttfamily 2211.16512}}.

\bibitem{DeClerck:2023fax}
M.~De~Clerck, S.A.~Hartnoll and J.E.~Santos, \emph{{Mixmaster chaos in an AdS black hole interior}}, \href{https://doi.org/10.1007/JHEP07(2024)202}{\emph{JHEP} {\bfseries 07} (2024) 202} [\href{https://arxiv.org/abs/2312.11622}{{\ttfamily 2312.11622}}].

\bibitem{Shahbazi-Moghaddam:2025fxi}
A.~Shahbazi-Moghaddam, \emph{{Constraints on the resolution of spacetime singularities}},  \href{https://arxiv.org/abs/2510.25927}{{\ttfamily 2510.25927}}.

\bibitem{Chakravarty:2025ncy}
J.~Chakravarty, \emph{{Imprint of the black hole interior on thermal four-point correlators}},  \href{https://arxiv.org/abs/2512.10912}{{\ttfamily 2512.10912}}.

\bibitem{Blacker:2026poc}
M.J.~Blacker and S.A.~Hartnoll, \emph{{Holographic Banners}},  \href{https://arxiv.org/abs/2604.02514}{{\ttfamily 2604.02514}}.

\bibitem{Maldacena:2001kr}
J.M.~Maldacena, \emph{{Eternal black holes in anti-de Sitter}}, \href{https://doi.org/10.1088/1126-6708/2003/04/021}{\emph{JHEP} {\bfseries 04} (2003) 021} [\href{https://arxiv.org/abs/hep-th/0106112}{{\ttfamily hep-th/0106112}}].

\bibitem{Louko:2000tp}
J.~Louko, D.~Marolf and S.F.~Ross, \emph{{On geodesic propagators and black hole holography}}, \href{https://doi.org/10.1103/PhysRevD.62.044041}{\emph{Phys. Rev. D} {\bfseries 62} (2000) 044041} [\href{https://arxiv.org/abs/hep-th/0002111}{{\ttfamily hep-th/0002111}}].

\bibitem{Kraus:2002iv}
P.~Kraus, H.~Ooguri and S.~Shenker, \emph{{Inside the horizon with AdS / CFT}}, \href{https://doi.org/10.1103/PhysRevD.67.124022}{\emph{Phys. Rev. D} {\bfseries 67} (2003) 124022} [\href{https://arxiv.org/abs/hep-th/0212277}{{\ttfamily hep-th/0212277}}].

\bibitem{Fidkowski:2003nf}
L.~Fidkowski, V.~Hubeny, M.~Kleban and S.~Shenker, \emph{{The Black hole singularity in AdS / CFT}}, \href{https://doi.org/10.1088/1126-6708/2004/02/014}{\emph{JHEP} {\bfseries 02} (2004) 014} [\href{https://arxiv.org/abs/hep-th/0306170}{{\ttfamily hep-th/0306170}}].

\bibitem{Festuccia:2005pi}
G.~Festuccia and H.~Liu, \emph{{Excursions beyond the horizon: Black hole singularities in Yang-Mills theories. I.}}, \href{https://doi.org/10.1088/1126-6708/2006/04/044}{\emph{JHEP} {\bfseries 04} (2006) 044} [\href{https://arxiv.org/abs/hep-th/0506202}{{\ttfamily hep-th/0506202}}].

\bibitem{Festuccia2007BlackHoleSingularities}
G.N.I.~Festuccia, \emph{Black Hole Singularities in the Framework of Gauge/String Duality}, Ph.D. thesis, Massachusetts Institute of Technology, Cambridge, Massachusetts, May, 2007.

\bibitem{Dodelson:2023vrw}
M.~Dodelson, C.~Iossa, R.~Karlsson and A.~Zhiboedov, \emph{{A thermal product formula}}, \href{https://doi.org/10.1007/JHEP01(2024)036}{\emph{JHEP} {\bfseries 01} (2024) 036} [\href{https://arxiv.org/abs/2304.12339}{{\ttfamily 2304.12339}}].

\bibitem{Ceplak:2024bja}
N.~{\v{C}}eplak, H.~Liu, A.~Parnachev and S.~Valach, \emph{{Black hole singularity from OPE}}, \href{https://doi.org/10.1007/JHEP10(2024)105}{\emph{JHEP} {\bfseries 10} (2024) 105} [\href{https://arxiv.org/abs/2404.17286}{{\ttfamily 2404.17286}}].

\bibitem{Dodelson:2025jff}
M.~Dodelson, C.~Iossa and R.~Karlsson, \emph{{Bouncing off a stringy singularity}},  \href{https://arxiv.org/abs/2511.09616}{{\ttfamily 2511.09616}}.

\bibitem{Afkhami-Jeddi:2025wra}
N.~Afkhami-Jeddi, S.~Caron-Huot, J.~Chakravarty and A.~Maloney, \emph{{Imprint of the black hole singularity on thermal two-point functions}},  \href{https://arxiv.org/abs/2510.21673}{{\ttfamily 2510.21673}}.

\bibitem{AliAhmad:2026wem}
S.~Ali~Ahmad, A.~Almheiri and S.~Lin, \emph{{Continuing past the inner horizon using WKB}},  \href{https://arxiv.org/abs/2601.02354}{{\ttfamily 2601.02354}}.

\bibitem{Fulling:1987otn}
S.A.~Fulling and S.N.M.~Ruijsenaars, \emph{{Temperature, periodicity and horizons}}, \href{https://doi.org/10.1016/0370-1573(87)90136-0}{\emph{Phys. Rept.} {\bfseries 152} (1987) 135}.

\bibitem{Grozdanov:2026cut}
S.~Grozdanov, S.~Valach and M.~Vrbica, \emph{{Bouncing geodesics, black hole singularities, and singularities of thermal correlators}}, \href{https://doi.org/10.1007/JHEP08(2026)022}{\emph{JHEP} {\bfseries 08} (2026) 022} [\href{https://arxiv.org/abs/2603.15598}{{\ttfamily 2603.15598}}].

\bibitem{Araya:2026shz}
I.J.~Araya, C.~Esper, Y.~Jia, M.~Kulaxizi and A.~Parnachev, \emph{{Bulkcone Singularities and Complex Geodesics}},  \href{https://arxiv.org/abs/2602.12893}{{\ttfamily 2602.12893}}.

\bibitem{Kaplan:2004qe}
J.~Kaplan, \emph{{Extracting data from behind horizons with the AdS / CFT correspondence}},  \href{https://arxiv.org/abs/hep-th/0402066}{{\ttfamily hep-th/0402066}}.

\bibitem{Hartnoll:2020fhc}
S.A.~Hartnoll, G.T.~Horowitz, J.~Kruthoff and J.E.~Santos, \emph{{Diving into a holographic superconductor}}, \href{https://doi.org/10.21468/SciPostPhys.10.1.009}{\emph{SciPost Phys.} {\bfseries 10} (2021) 009} [\href{https://arxiv.org/abs/2008.12786}{{\ttfamily 2008.12786}}].

\bibitem{Rodriguez-Gomez:2021pfh}
D.~Rodriguez-Gomez and J.G.~Russo, \emph{{Correlation functions in finite temperature CFT and black hole singularities}}, \href{https://doi.org/10.1007/JHEP06(2021)048}{\emph{JHEP} {\bfseries 06} (2021) 048} [\href{https://arxiv.org/abs/2102.11891}{{\ttfamily 2102.11891}}].

\bibitem{Horowitz:2023ury}
G.T.~Horowitz, H.~Leung, L.~Queimada and Y.~Zhao, \emph{{Boundary signature of singularity in the presence of a shock wave}}, \href{https://doi.org/10.21468/SciPostPhys.16.2.060}{\emph{SciPost Phys.} {\bfseries 16} (2024) 060} [\href{https://arxiv.org/abs/2310.03076}{{\ttfamily 2310.03076}}].

\bibitem{Giombi:2026kdz}
S.~Giombi, Y.-Z.~Li and J.~Shan, \emph{{Bouncing singularities and thermal correlators on line defects}}, \href{https://doi.org/10.1007/JHEP07(2026)170}{\emph{JHEP} {\bfseries 07} (2026) 170} [\href{https://arxiv.org/abs/2603.11012}{{\ttfamily 2603.11012}}].

\bibitem{Jia:2026ryl}
H.F.~Jia and M.~Rangamani, \emph{{Exact holographic thermal spectral functions: OPE, non-perturbative corrections, and black hole singularity}},  \href{https://arxiv.org/abs/2604.10803}{{\ttfamily 2604.10803}}.

\bibitem{Grozdanov:2026ktq}
S.~Grozdanov, V.~Movrin and S.~Valach, \emph{{Bouncing Geodesics, Singularities, and the Cavity Thermal Product Formula in Asymptotically Flat and de Sitter Black Holes}},  \href{https://arxiv.org/abs/2606.11297}{{\ttfamily 2606.11297}}.

\bibitem{Barrat:2026jfg}
J.~Barrat, D.N.~Bozkurt, E.~Marchetto, A.~Miscioscia and E.~Pomoni, \emph{{Analytic thermal bootstrap in momentum space: From thermal OPE to QNMs}},  \href{https://arxiv.org/abs/2607.24919}{{\ttfamily 2607.24919}}.

\bibitem{Arnaudo:2026axe}
P.~Arnaudo, C.~Iossa, R.~Karlsson and B.~Withers, \emph{{OPE = QNM}},  \href{https://arxiv.org/abs/2607.24909}{{\ttfamily 2607.24909}}.

\bibitem{Grozdanov:2026lnc}
S.~Grozdanov, V.~Movrin and S.~Valach, \emph{{Quasinormal modes as exterior probes of black hole interiors in our Universe}},  \href{https://arxiv.org/abs/2608.13643}{{\ttfamily 2608.13643}}.

\bibitem{Hartnoll:2026vhu}
S.A.~Hartnoll and A.~Zhiboedov, \emph{{Can one hear the shape of a black hole singularity?}},  \href{https://arxiv.org/abs/2609.07514}{{\ttfamily 2609.07514}}.

\bibitem{Horowitz:1989bv}
G.T.~Horowitz and A.R.~Steif, \emph{{Space-Time Singularities in String Theory}}, \href{https://doi.org/10.1103/PhysRevLett.64.260}{\emph{Phys. Rev. Lett.} {\bfseries 64} (1990) 260}.

\bibitem{Horowitz:1990sr}
G.T.~Horowitz and A.R.~Steif, \emph{{Strings in Strong Gravitational Fields}}, \href{https://doi.org/10.1103/PhysRevD.42.1950}{\emph{Phys. Rev. D} {\bfseries 42} (1990) 1950}.

\bibitem{Dodelson:2024atp}
M.~Dodelson, \emph{{Ringdown in the SYK model}}, \href{https://doi.org/10.21468/SciPostPhys.19.3.081}{\emph{SciPost Phys.} {\bfseries 19} (2025) 081} [\href{https://arxiv.org/abs/2408.05790}{{\ttfamily 2408.05790}}].

\bibitem{Buric:2026qsp}
I.~Buri{\'c}, C.-M.~Chang, I.~Gusev, E.~Helfenberger, A.~Parnachev and M.~Rangamani, \emph{{Thermal two-point functions in SYK and complex-time singularities}},  \href{https://arxiv.org/abs/2607.05258}{{\ttfamily 2607.05258}}.

\bibitem{Dodelsonetal:WIP}
M.~Dodelson, H.~Lin, H.~Xie and J.~Yeh, ``Work in progress.'' 2026.

\bibitem{Sarosi:2017ykf}
G.~S{\'a}rosi, \emph{{AdS$_{2}$ holography and the SYK model}}, \href{https://doi.org/10.22323/1.323.0001}{\emph{PoS} {\bfseries Modave2017} (2018) 001} [\href{https://arxiv.org/abs/1711.08482}{{\ttfamily 1711.08482}}].

\bibitem{Mertens:2022irh}
T.G.~Mertens and G.J.~Turiaci, \emph{{Solvable models of quantum black holes: a review on Jackiw{\textendash}Teitelboim gravity}}, \href{https://doi.org/10.1007/s41114-023-00046-1}{\emph{Living Rev. Rel.} {\bfseries 26} (2023) 4} [\href{https://arxiv.org/abs/2210.10846}{{\ttfamily 2210.10846}}].

\bibitem{Turiaci:2024cad}
G.J.~Turiaci, \emph{{Les Houches lectures on two-dimensional gravity and holography}}, \href{https://doi.org/10.21468/SciPostPhysLectNotes.113}{\emph{SciPost Phys. Lect. Notes} {\bfseries 113} (2026) 1} [\href{https://arxiv.org/abs/2412.09537}{{\ttfamily 2412.09537}}].

\bibitem{Grumiller:2002nm}
D.~Grumiller, W.~Kummer and D.V.~Vassilevich, \emph{{Dilaton gravity in two-dimensions}}, \href{https://doi.org/10.1016/S0370-1573(02)00267-3}{\emph{Phys. Rept.} {\bfseries 369} (2002) 327} [\href{https://arxiv.org/abs/hep-th/0204253}{{\ttfamily hep-th/0204253}}].

\bibitem{Witten:2020wvy}
E.~Witten, \emph{{Matrix Models and Deformations of JT Gravity}}, \href{https://doi.org/10.1098/rspa.2020.0582}{\emph{Proc. Roy. Soc. Lond. A} {\bfseries 476} (2020) 20200582} [\href{https://arxiv.org/abs/2006.13414}{{\ttfamily 2006.13414}}].

\bibitem{Maxfield:2020ale}
H.~Maxfield and G.J.~Turiaci, \emph{{The path integral of 3D gravity near extremality; or, JT gravity with defects as a matrix integral}}, \href{https://doi.org/10.1007/JHEP01(2021)118}{\emph{JHEP} {\bfseries 01} (2021) 118} [\href{https://arxiv.org/abs/2006.11317}{{\ttfamily 2006.11317}}].

\bibitem{Kruthoff:2024gxc}
J.~Kruthoff and A.~Levine, \emph{{Semi-classical dilaton gravity and the very blunt defect expansion}}, \href{https://doi.org/10.1007/JHEP07(2025)211}{\emph{JHEP} {\bfseries 07} (2025) 211} [\href{https://arxiv.org/abs/2402.10162}{{\ttfamily 2402.10162}}].

\bibitem{Bah:2022uyz}
I.~Bah, Y.~Chen and J.~Maldacena, \emph{{Estimating global charge violating amplitudes from wormholes}}, \href{https://doi.org/10.1007/JHEP04(2023)061}{\emph{JHEP} {\bfseries 04} (2023) 061} [\href{https://arxiv.org/abs/2212.08668}{{\ttfamily 2212.08668}}].

\bibitem{Ruan:2026talk}
S.~Ruan, ``{Smoothing the Bouncing-Geodesic Singularity in AdS/CFT}.'' Talk at \emph{Quantum Information in Quantum Gravity 2026 (QIQG 2026)}, Tsinghua University, 2026.

\bibitem{Fitzpatrick:2016ive}
A.L.~Fitzpatrick, J.~Kaplan, D.~Li and J.~Wang, \emph{{On information loss in AdS$_{3}$/CFT$_{2}$}}, \href{https://doi.org/10.1007/JHEP05(2016)109}{\emph{JHEP} {\bfseries 05} (2016) 109} [\href{https://arxiv.org/abs/1603.08925}{{\ttfamily 1603.08925}}].

\bibitem{Batra:WIP}
G.~Batra and D.~Stanford, ``Work in progress.'' 2026.

\bibitem{StanfordTangPrivateCommunication}
D.~Stanford and H.~Tang, \emph{Private communication},  Jan., 2026.

\bibitem{Maldacena:2015iua}
J.~Maldacena, D.~Simmons-Duffin and A.~Zhiboedov, \emph{{Looking for a bulk point}}, \href{https://doi.org/10.1007/JHEP01(2017)013}{\emph{JHEP} {\bfseries 01} (2017) 013} [\href{https://arxiv.org/abs/1509.03612}{{\ttfamily 1509.03612}}].

\bibitem{BatraShenker:WIP}
G.~Batra and S.H.~Shenker, ``Work in progress.'' 2026.

\bibitem{Faulkner:2017hll}
T.~Faulkner and H.~Wang, \emph{{Probing beyond ETH at large $c$}}, \href{https://doi.org/10.1007/JHEP06(2018)123}{\emph{JHEP} {\bfseries 06} (2018) 123} [\href{https://arxiv.org/abs/1712.03464}{{\ttfamily 1712.03464}}].

\bibitem{Witten:2010cx}
E.~Witten, \emph{{Analytic Continuation Of Chern-Simons Theory}}, {\emph{AMS/IP Stud. Adv. Math.} {\bfseries 50} (2011) 347} [\href{https://arxiv.org/abs/1001.2933}{{\ttfamily 1001.2933}}].

\bibitem{Mizera:2022dko}
S.~Mizera, \emph{{Natural boundaries for scattering amplitudes}}, \href{https://doi.org/10.21468/SciPostPhys.14.5.101}{\emph{SciPost Phys.} {\bfseries 14} (2023) 101} [\href{https://arxiv.org/abs/2210.11448}{{\ttfamily 2210.11448}}].

\bibitem{Yang:2018gdb}
Z.~Yang, \emph{{The Quantum Gravity Dynamics of Near Extremal Black Holes}}, \href{https://doi.org/10.1007/JHEP05(2019)205}{\emph{JHEP} {\bfseries 05} (2019) 205} [\href{https://arxiv.org/abs/1809.08647}{{\ttfamily 1809.08647}}].

\bibitem{Bissi:2024wur}
A.~Bissi, N.~Dondi, A.~Piazza, T.~Reis and M.~Serone, \emph{{On the 1/c expansion in 2d CFTs with degenerate operators}}, \href{https://doi.org/10.1007/JHEP03(2025)092}{\emph{JHEP} {\bfseries 03} (2025) 092} [\href{https://arxiv.org/abs/2412.04387}{{\ttfamily 2412.04387}}].

\bibitem{Berry1988}
M.V.~Berry, \emph{Stokes' phenomenon; smoothing a victorian discontinuity}, {\emph{Publications Mathématiques de l'IHÉS} {\bfseries 68} (1988) 211}.

\bibitem{paris2001asymptotics}
R.~Paris and D.~Kaminski, \emph{Asymptotics and Mellin-Barnes Integrals}, Encyclopedia of Mathematics and its Applications, Cambridge University Press (2001).

\bibitem{Howls:2004}
C.J.~Howls, P.J.~Langman and A.B.~Olde~Daalhuis, \emph{{On the higher-order Stokes phenomenon}}, \href{https://doi.org/10.1098/rspa.2004.1299}{\emph{Proc. Roy. Soc. Lond. A} {\bfseries 460} (2004) 2285}.

\bibitem{Balasubramanian:2017fan}
V.~Balasubramanian, A.~Bernamonti, B.~Craps, T.~De~Jonckheere and F.~Galli, \emph{{Heavy-Heavy-Light-Light correlators in Liouville theory}}, \href{https://doi.org/10.1007/JHEP08(2017)045}{\emph{JHEP} {\bfseries 08} (2017) 045} [\href{https://arxiv.org/abs/1705.08004}{{\ttfamily 1705.08004}}].

\bibitem{Benjamin:2023uib}
N.~Benjamin, S.~Collier, A.~Maloney and V.~Meruliya, \emph{{Resurgence, conformal blocks, and the sum over geometries in quantum gravity}}, \href{https://doi.org/10.1007/JHEP05(2023)166}{\emph{JHEP} {\bfseries 05} (2023) 166} [\href{https://arxiv.org/abs/2302.12851}{{\ttfamily 2302.12851}}].

\bibitem{Fitzpatrick:2016mjq}
A.L.~Fitzpatrick and J.~Kaplan, \emph{{On the Late-Time Behavior of Virasoro Blocks and a Classification of Semiclassical Saddles}}, \href{https://doi.org/10.1007/JHEP04(2017)072}{\emph{JHEP} {\bfseries 04} (2017) 072} [\href{https://arxiv.org/abs/1609.07153}{{\ttfamily 1609.07153}}].

\bibitem{Ghosh:2019rcj}
A.~Ghosh, H.~Maxfield and G.J.~Turiaci, \emph{{A universal Schwarzian sector in two-dimensional conformal field theories}}, \href{https://doi.org/10.1007/JHEP05(2020)104}{\emph{JHEP} {\bfseries 05} (2020) 104} [\href{https://arxiv.org/abs/1912.07654}{{\ttfamily 1912.07654}}].

\bibitem{Saad:2019lba}
P.~Saad, S.H.~Shenker and D.~Stanford, \emph{{JT gravity as a matrix integral}},  \href{https://arxiv.org/abs/1903.11115}{{\ttfamily 1903.11115}}.

\bibitem{Jafferis:2022uhu}
D.L.~Jafferis, D.K.~Kolchmeyer, B.~Mukhametzhanov and J.~Sonner, \emph{{Matrix Models for Eigenstate Thermalization}}, \href{https://doi.org/10.1103/PhysRevX.13.031033}{\emph{Phys. Rev. X} {\bfseries 13} (2023) 031033} [\href{https://arxiv.org/abs/2209.02130}{{\ttfamily 2209.02130}}].

\bibitem{Iliesiu:2024cnh}
L.V.~Iliesiu, A.~Levine, H.W.~Lin, H.~Maxfield and M.~Mezei, \emph{{On the non-perturbative bulk Hilbert space of JT gravity}}, \href{https://doi.org/10.1007/JHEP10(2024)220}{\emph{JHEP} {\bfseries 10} (2024) 220} [\href{https://arxiv.org/abs/2403.08696}{{\ttfamily 2403.08696}}].

\bibitem{DS_DSD_SCH:WIP}
S.~Caron-Huot, D.~Simmons-Duffin and D.~Stanford, ``Work in progress.'' 2026.

\bibitem{Nemes:2013Hermite}
G.~Nemes, \emph{{Error bounds and exponential improvement for Hermite's asymptotic expansion for the Gamma function}}, \href{https://doi.org/10.2298/AADM130124002N}{\emph{Appl. Anal. Discrete Math.} {\bfseries 7} (2013) 161}.

\end{thebibliography}\endgroup
\bibliographystyle{JHEP}
\end{document}